\documentclass[epjST]{svjour}
\usepackage{graphics}
\usepackage{rotating}
\def\d{\mbox{D$_{13}$(1520)}}
\def\s{\mbox{S$_{11}$(1535)}}

\def\f{\mbox{F$_{15}$(1680)}}
\def\ff{\mbox{F$_{37}$(1950)}}
\def\p{\mbox{P$_{33}$(1232)}}
\def\pp{\mbox{P$_{11}$(1440)}}
\begin{document}
\renewcommand{\textfraction}{0.00000000001}
\renewcommand{\floatpagefraction}{1.0}
\title{Photoproduction of mesons off nuclei}
\subtitle{Electromagnetic excitations of the neutron and meson-nucleus interactions}
\author{B. Krusche\inst{1}\fnmsep\thanks{\email{Bernd.Krusche@unibas.ch}} 
}
\institute{Department of Physics, University of Basel, Ch-4056 Basel, Switzerland
}
\abstract{
Recent results for the photoproduction of mesons off nuclei are reviewed.
These experiments have been performed for two major lines of research related 
to the properties of the strong interaction.  
The investigation of nucleon resonances requires light nuclei as targets 
for the extraction of the isospin composition of the electromagnetic 
excitations. This is done with quasi-free meson photoproduction off the bound
neutron and supplemented with the measurement of coherent photoproduction
reactions, serving as spin and/or isospin filters. Furthermore,
photoproduction from light and heavy nuclei is a very efficient
tool for the study of the interactions of mesons with nuclear matter and
the in-medium properties of hadrons. Experiments are currently 
rapidly developing due to the combination of high quality tagged 
(and polarized) photon beams with state-of-the-art 4$\pi$ detectors and
polarized targets.   
} 
\maketitle
\section{Introduction and overview of the research topics}
\label{intro}
During the last two decades, photoproduction of mesons has become a prime tool
for the study of the properties of the strong interaction in the regime where
this fundamental force cannot be treated with the methods of perturbation
theory. It has almost completely replaced meson induced reactions like pion
elastic scattering, which had previously dominated this field. On the
experimental side, this development was triggered by the large progress in
accelerator and detector technology which nowadays allows the measurement of
the cross sections of electromagnetically induced reactions routinely with 
comparable or even better precision than hadron induced reactions, although 
the latter typically have cross sections which are three orders of magnitude
larger.
The experiments are centered at high duty electron beams, in particular 
CEBAF at Jlab in Newport News, at the ELSA accelerator in Bonn, at the MAMI 
accelerator in Mainz, at SPring-8 in Osaka, at LNS at Tohoku University in Sendai, 
and, until recently, also at the GRAAL facility at the ESRF in Grenoble.

\begin{figure}[htb]
\resizebox{1.0\textwidth}{!}{%
\begin{turn}{90}
  \includegraphics{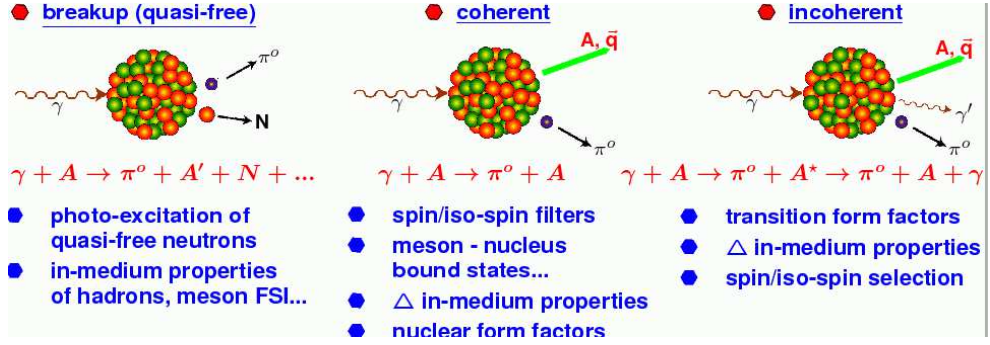}
\end{turn}
}
\caption{Different nuclear meson production processes for the example of single
$\pi^0$-photoproduction (see text).  
}
\label{fig:nucl}       
\end{figure}

The largest part of this program involves meson photo- or electro-production 
experiments off the free proton aiming at a better understanding of the
structure and the excitation spectrum of the nucleon. However, there is  
also a very active and growing program studying nuclear targets.
The physics topics of this program are diverse and the most important ones 
are summarized in Fig. \ref{fig:nucl}. 
   
Photoproduction of mesons off nuclei is mostly characterized by three types
of final state. In the case of breakup reactions, at least one nucleon is 
removed from the nucleus. As we will discuss later, this type of process
includes quasi-free meson production reactions, where the meson is
produced on one `participant' nucleon, while the rest of the nucleus acts as
`spectator'. This simple picture is mostly used for light nuclei, in particular
the deuteron, where final state interaction effects (FSI) play a minor role.
The study of meson photoproduction off the neutron can only be performed in 
this way. The methods for the extraction of elementary cross section of the 
neutron target and the recent results from such experiments are a central 
topic of this report.
However, breakup reactions are also used in the case of heavier nuclei to 
study in-medium properties of hadrons and FSI effects, e.g. exploring the 
scaling behavior of the reaction cross sections in dependence on the nuclear 
mass number. A recent review of the in-medium properties of scalar and vector
mesons derived from photoproduction and other reactions is given by Leupold,
Mosel, and Metag \cite{Leupold_10}; results for the in-medium properties of
nucleon resonances are summarized e.g. in \cite{Krusche_05}.
 
Coherent meson photoproduction is characterized by a final state with the
initial-state nucleus in its ground state. Exploring the spin and isospin
quantum numbers of the nucleus can be used to project out specific parts 
of the elementary reaction amplitudes. However, with a few exceptions, typical
cross sections are small and the experiments are so demanding that only
few results are available up to now. Coherent meson production is also used 
as a doorway reaction for the search of meson-nucleus bound states. As an 
example we will discuss tentative evidence for $\eta$-mesic states.
Coherent photoproduction of $\pi^0$ mesons was also used for the study of the
in-medium properties of the $\Delta$(1232) resonance 
(see e.g. \cite{Drechsel_99,Krusche_02}), and in a completely different context,
for the extraction of nuclear mass form factors \cite{Krusche_05a}.
 
Incoherent production finally denotes the process where the final-state nucleus
is excited (but otherwise identical with the initial state nucleus)
and de-excites typically by emission of $\gamma$-radiation. Such processes 
provide additional selection possibilities as spin- and isospin filters
(by selection of the final state quantum numbers), but are still almost unexplored 
due to the small reaction cross sections. Incoherent pion production can also
contribute to the study of $\Delta$-in-medium properties and give access to nuclear
transition form factors. The first precise experimental study for incoherent 
photoproduction of $\pi^0$ mesons to the 4.4 MeV excited state of $^{12}$C has been
recently reported by Tarbert et al. from the Crystal Ball/TAPS collaboration at MAMI
\cite{Tarbert_08}. 

The main emphasis of this article will be on the study of the isospin degree of
freedom of the nucleon excitation scheme via quasi-free meson photoproduction off
the neutron and coherent photoproduction of light nuclei and on meson-nucleus
interactions, in particular the search for meson-nucleus quasi-bound states.

\subsection{Nucleon resonances}
\label{sec:intro_res}

Understanding the properties of the strong interaction in the non-perturbative 
regime, where it gives rise to the observed hadron spectrum, is still a great 
challenge. One expects that, as in nuclear structure physics, the 
excitation spectrum of hadrons will reflect the main properties of the 
interaction, so that many efforts have been undertaken to study the excited states
of the nucleon. However, so far, the results are not satisfactory. On the theory 
side, the only direct connection between baryon properties and Quantum 
Chromodynamics (QCD) has been established through the numerical method of lattice 
gauge theory. During the last few years, the progress in this field has been tremendous 
for the ground state properties of hadrons \cite{Duerr_08}. But first results for 
excited states \cite{Bulava_10}, going beyond earlier quenched approximations 
\cite{Burch_06,Basak_07}, became available only very recently..  

\begin{figure}[thb]
\centerline{ 
\resizebox{0.30\textwidth}{!}{%
  \includegraphics{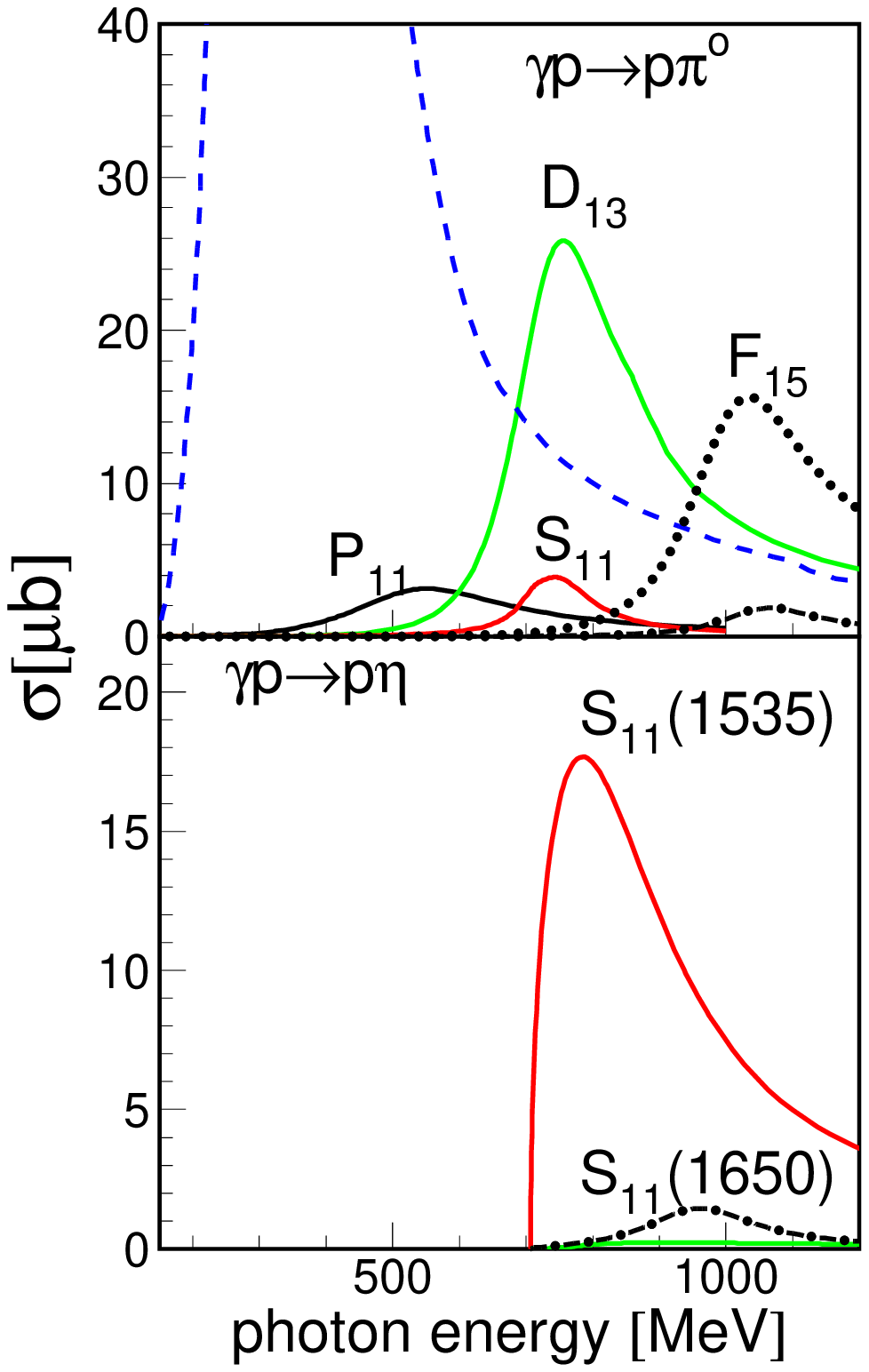} 
}
\resizebox{0.70\textwidth}{!}{%
  \includegraphics{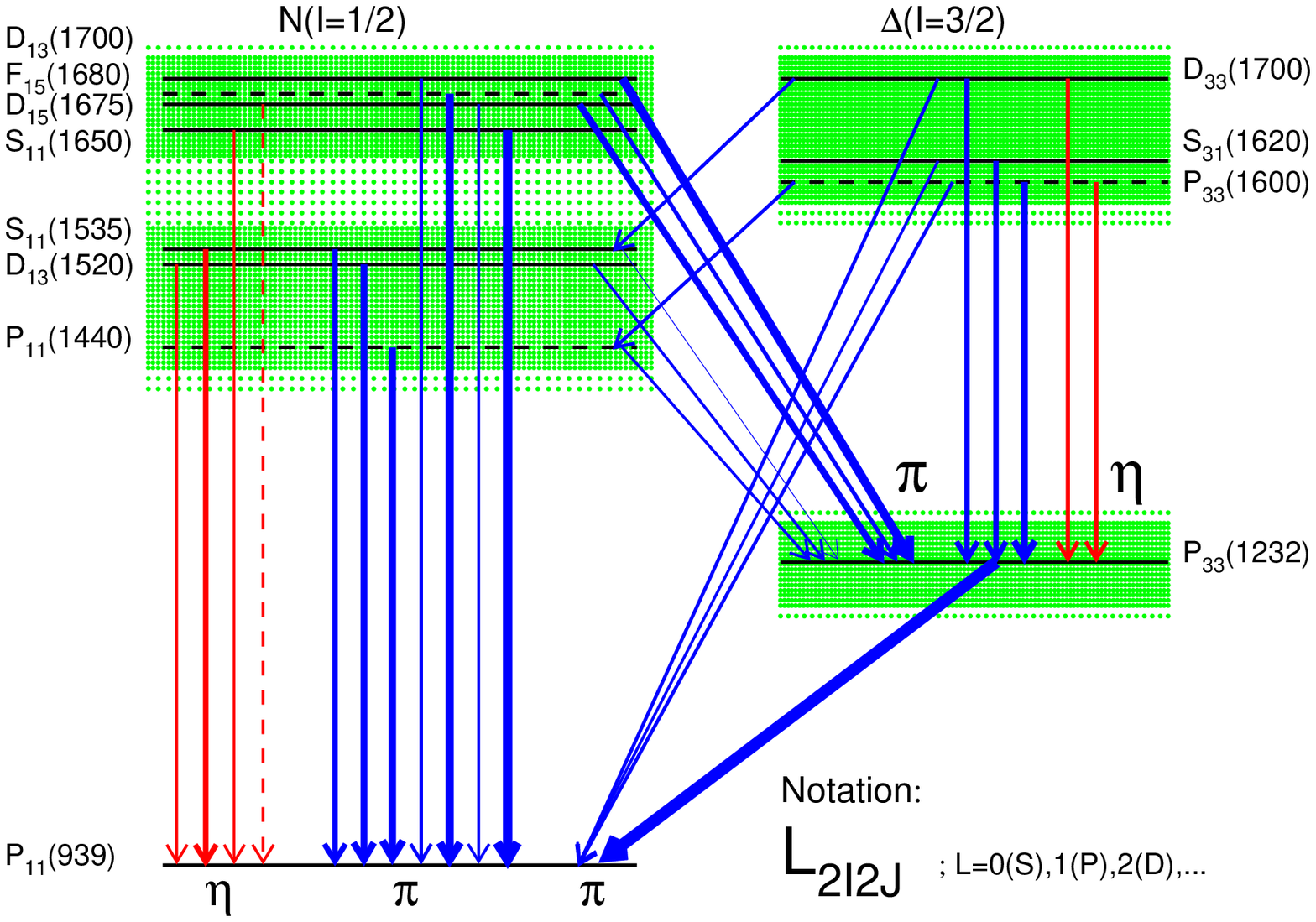}
}}
\caption{Left hand side: contributions of different nucleon resonances to
$\gamma p\rightarrow p\pi^0$ and $\gamma p\rightarrow p\eta$ (schematic).
Right hand side: low energy excitation scheme of the nucleon. Isospin $I=1/2$
$N^{\star}$ resonances left and isospin $I=3/2$ $\Delta$-resonances right. 
Typical decays are indicated. 
}
\label{fig:level}       
\end{figure}

In a more indirect way, experimental observations and QCD are connected via
QCD inspired quark models. However, in spite of their phenomenological successes, 
the basis of these models is still not well anchored.  
There is neither consent about the effective degrees of freedom nor about 
the residual quark - quark interactions (see e.g. Ref. \cite{Capstick_00,Klempt_10} 
for detailed reviews). So far, a comparison of the experimentally known excitation 
spectrum of the nucleon to such model predictions does not clearly favor any of the 
different models. In fact, in most cases even the ordering of some of the lowest 
lying excitations is not reproduced. In particular, the N(1440)P$_{11}$ (`Roper') 
resonance and the first excited $\Delta$, the P$_{33}$(1600), are notoriously 
problematic. Furthermore, even the models with the fewest effective degrees of 
freedom predict many more states than have been observed, which is known as the 
`missing resonance' problem. This problem is severe; only for very few 
combinations of quantum numbers has more than the lowest lying excited state
been found experimentally so far. Nevertheless, it could also be rooted in trivial
experimental bias. Most nucleon resonances have been established so far by elastic
pion scattering, so that states that decouple from $N\pi$ are suppressed.
Photon-induced reactions can avoid this bias when other final
states than single pion production are also investigated. This is the main motivation
for the world-wide experimental program for the study of photon-induced meson
production reactions off the nucleon. As an example, Fig. \ref{fig:level}
(left hand side) shows the contribution of different resonances to single $\pi^0$
and $\eta$-photoproduction at fairly low incident photon energies. While the 
extraction of the tiny contribution of the S$_{11}$(1535) resonance to pion
photoproduction requires complicated model analyses, this resonance completely 
dominates $\eta$-photoproduction in the threshold region 
\cite{Krusche_97,Krusche_95}, allowing significantly more detailed studies of this state.
Following the same approach, photoproduction of heavier mesons like the 
$\eta '$ has also been studied recently from the respective production threshold
\cite{Dugger_02,Williams_09,Crede_09}. However, here the situation turned out
to be less simple; first model analyses do not give conclusive results for the
contributing resonances. In this, as in many other cases, differential cross section
data alone do not sufficiently constrain the model analysis, which is the motivation
for the large current efforts to measure single and double polarization observables. 

Higher lying excited states may not only decay by single meson emission to the nucleon
ground state but also via an intermediate excited states, giving rise to double or even
triple meson production reactions. Making again the analogy to nuclear spectroscopy,
it is evident that spectral information based only on the `ground state' decays of 
excited states is very incomplete. As indicated in the nucleon `level-scheme' in
Fig. \ref{fig:level} already states at moderate excitation energies can have substantial
decay branching ratios to intermediate states and for higher excitation energies such
decay modes will become more probable. Therefore multi-meson production reactions
have also moved into the focus. Double pion production has recently been studied up to
incident photon energies of 1.8 GeV \cite{Assafiri_03,Sarantsev_08,Thoma_08}.
Also in this case polarization observables are urgently needed to constrain the model
analyses. First measurements of the beam-helicity asymmetry $I^{\odot}$ have revealed 
severe problems in the reaction models \cite{Strauch_05,Krambrich_09}. More recently,
also the $\eta\pi$-channel has been studied, which has
the additional advantage of isospin selectivity ($\eta$-mesons are only emitted in
$N^{\star}\rightarrow N^{(\star)}$ and $\Delta^{\star}\rightarrow \Delta^{(\star)}$
transitions). First results point to a dominant contribution of one resonance at
threshold \cite{Ajaka_08,Horn_08a,Kashevarov_09,Gutz_10,Kashevarov_10}, and possibly a 
parity doublet around $W\approx$1.9 GeV \cite{Horn_08b}. 

In summary, an intensive experimental program is currently under way in particular at
the CLAS facility at Jlab, the Crystal Barrel/TAPS experiment at ELSA, and the
Crystal Ball/TAPS experiment at MAMI to measure differential cross sections, single and
double polarization observables with polarized beams and polarized targets for many
different single and double meson production reactions off the proton. The last missing
degree of freedom in the experiments is the isospin dependence of the cross sections. 
  
The electromagnetic interaction does not conserve isospin. The electromagnetic 
transition operator $\hat{A}$ can be split in an isoscalar part $\hat{S}$ and an 
isovector part $\hat{V}$, giving rise to three independent matrix elements
\cite{Watson_52} in the notation $\langle I_f,I_{f3}|\hat{A}|I_i,I_{i3}\rangle$:
\begin{equation}
\label{eq:iso_1}
A^{IS} = \langle \frac{1}{2},\pm \frac{1}{2}|\hat{S}|\frac{1}{2},\pm 
      \frac{1}{2}\rangle ,
\;\;      
\mp A^{IV} = \langle \frac{1}{2},\pm \frac{1}{2}|\hat{V}|\frac{1}{2},\pm
      \frac{1}{2}\rangle ,
\;\;      
A^{V3} = \langle \frac{3}{2},\pm \frac{1}{2}|\hat{V}|\frac{1}{2},
        \pm\frac{1}{2}\rangle\;. 
\end{equation}
Photoproduction of isovector mesons like pions involves all three matrix elements,
while only $A^{IS}$ and $A^{IV}$ contribute in the case of isoscalar mesons like the
$\eta$. Nevertheless, in both cases at least one reaction on a neutron target must be
measured for a unique isospin decomposition of the multipole amplitudes
(see e.g. \cite{Krusche_03} for details). The situation for the excitation of nucleon
resonances is different for $N^{\star}$- and $\Delta$-states. The latter involve only
the $A^{V3}$ matrix element, so that they are excited identically on protons and
neutrons, while the combination of $A^{IS}$ and $A^{IV}$ components leads in general 
to different electromagnetic couplings for protons and neutrons in the excitation 
of $N^{\star}$ states. In the limit of SU(3) symmetry in extreme cases  
$\gamma NN^{\star}$ transitions may even be completely forbidden for the proton
but allowed for the neutron (Moorehouse selection rules \cite{Moorehouse_66}). 
Although due to the non-negligible spin-orbit mixing in the wave functions they are 
not strictly forbidden in more realistic models, they remain suppressed and can be 
better studied using neutron targets. 
  
\begin{figure}[htb]
\centerline{ \resizebox{1.0\textwidth}{!}{%
 \includegraphics{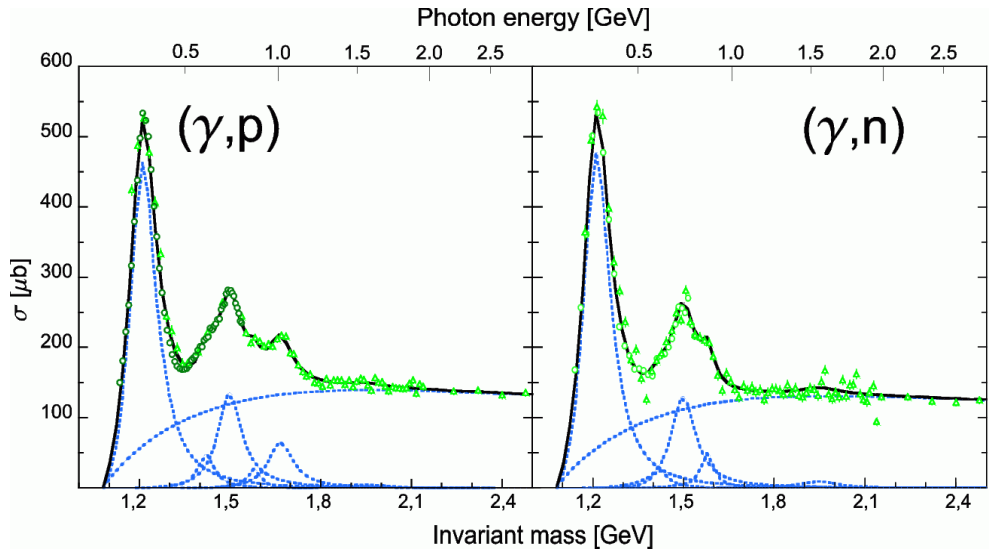}
 }}
\caption{Cross section for total photoabsorption on the proton (left hand side)
and the neutron (right hand side) \cite{Bianchi_96}. Points: 
measured data, curves: fit of Breit-Wigner shapes of nucleon 
resonances (\p, \pp, \d, \s, \f (only for proton), and \ff) and a 
smoothly varying background.
}
\label{fig:photoabs}       
\end{figure}

Already the measurement of total photoabsorption 
\cite{Bianchi_96,Armstrong_72,MacCormick_96}, reveals differences in 
the electromagnetic excitation of resonances on the proton and on the neutron 
(see Fig. \ref{fig:photoabs}).
The bump-like structures of the second and third resonance region are quite different,
but such inclusive measurements do not allow the extraction of detailed 
information about the isospin structure of the amplitudes. 

The investigation of exclusive meson production reactions off the neutron is still
much less advanced than for the proton case. The reason is that no free neutron
targets exist so that only quasi-free photoproduction off the neutron bound in light
nuclei and additionally coherent photoproduction from light nuclei can be used.
The use of quasi-free reactions off bound nucleons involves two complications.
Technically the experiments must detect the recoil nucleons. This is more or less
straightforward for protons but more challenging for neutrons, so that in particular
reactions with only neutral mesons in the final state are difficult to measure since
detection probabilities for neutrons and achievable energy resolution are at best
moderate. Furthermore, as is the case for all quasi-free measurements, the 
interpretation of the data must include nuclear effects like Fermi motion, 
re-scattering, and FSI, which leads to model dependent analyses and results.

\begin{figure}[thb]
\resizebox{1.0\textwidth}{!}{%
  \includegraphics{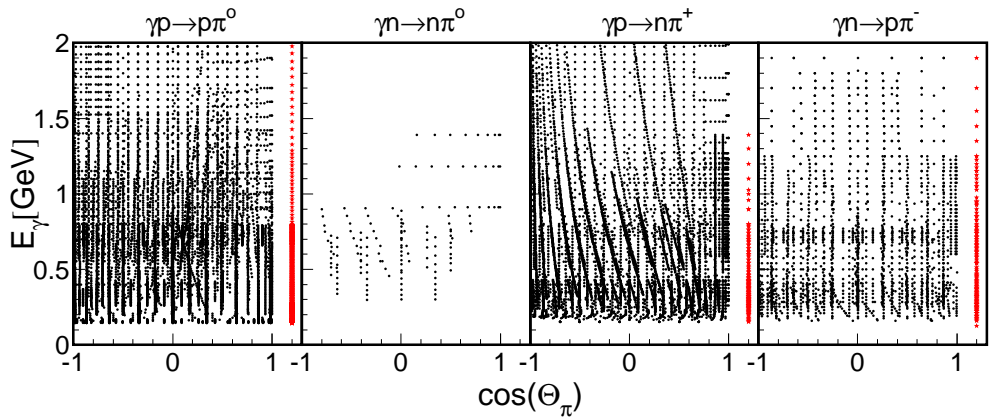}
}
\caption{Available data for angular distributions and total cross sections
(red stars at cos($\theta_{\pi}$) = 1.1) for the photoproduction of pions off 
the nucleon. 
}
\label{fig:piond}       
\end{figure}

As an example, the available entries from the SAID data base \cite{Said}
for angular distributions of pion production off the nucleon are compared 
for the different isospin channels in Fig. \ref{fig:piond}. The data base is
quite extensive for the two reactions off the free proton. Results for the 
$\gamma n\rightarrow p\pi^-$ final state are less abundant but also available 
for a fairly complete angular range below incident photon energies of 1 GeV,
while the fully neutral final state $n\pi^0$ is basically unexplored. This is
so because the reaction $\gamma d\rightarrow pp\pi^-$ with three charged 
particles in the final state can be easily studied. In fact most of these data 
go back to bubble chamber measurements from the early 1970's and have been 
analyzed with plane wave approximations assuming closure \cite{Benz_73}.
Only recently new data from the CLAS experiment  have been published for this 
channel \cite{Chen_09}.

Since in principle the measurement of three of the four isospin channels allows 
the complete determination of the isospin structure of the amplitudes, one might
ask at this point, whether the experimental efforts to measure also the fully 
neutral channel $n\pi^0$ are worthwhile. One answer is, that this trick works 
only for isovector mesons, while for the isoscalar ones like the $\eta$, 
$\eta '$, $\omega$,... photoproduction reactions with final state neutrons
cannot be avoided. The other answer is, that even for isovector mesons like
the pion the fully neutral final state carries important additional information
for partial wave analysis. This is so, because not only resonance excitations
contribute to the photoproduction reactions. Background terms like
meson-pole terms, Kroll-Rudermann-like terms, diffractive $t$-channel 
contributions etc. must also be considered and can contribute differently to the 
isospin channels.    

\begin{figure}[thb]
\centerline{ 
\resizebox{1.0\textwidth}{!}{%
  \includegraphics{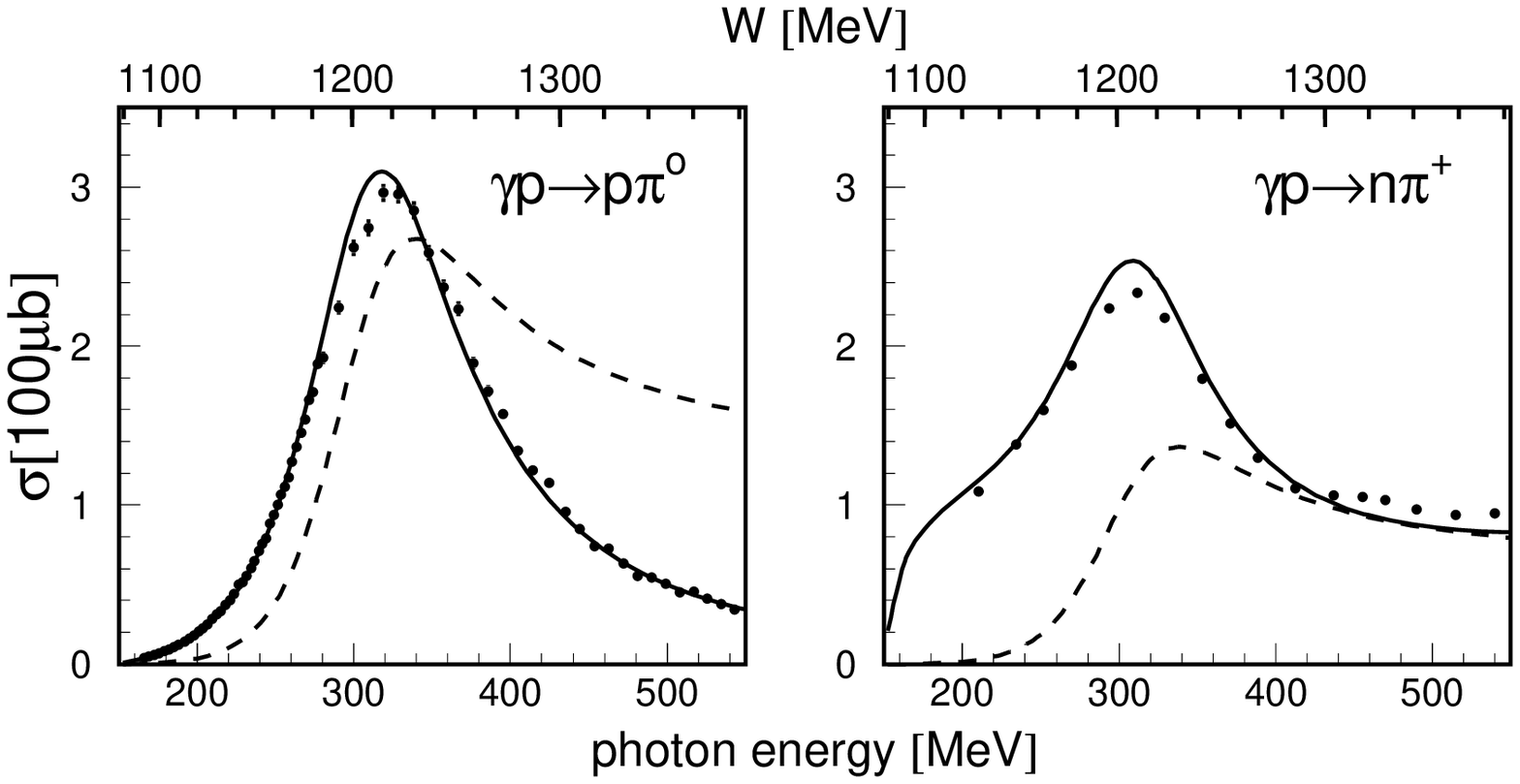} 
}}
\centerline{ 
\resizebox{1.0\textwidth}{!}{%
  \includegraphics{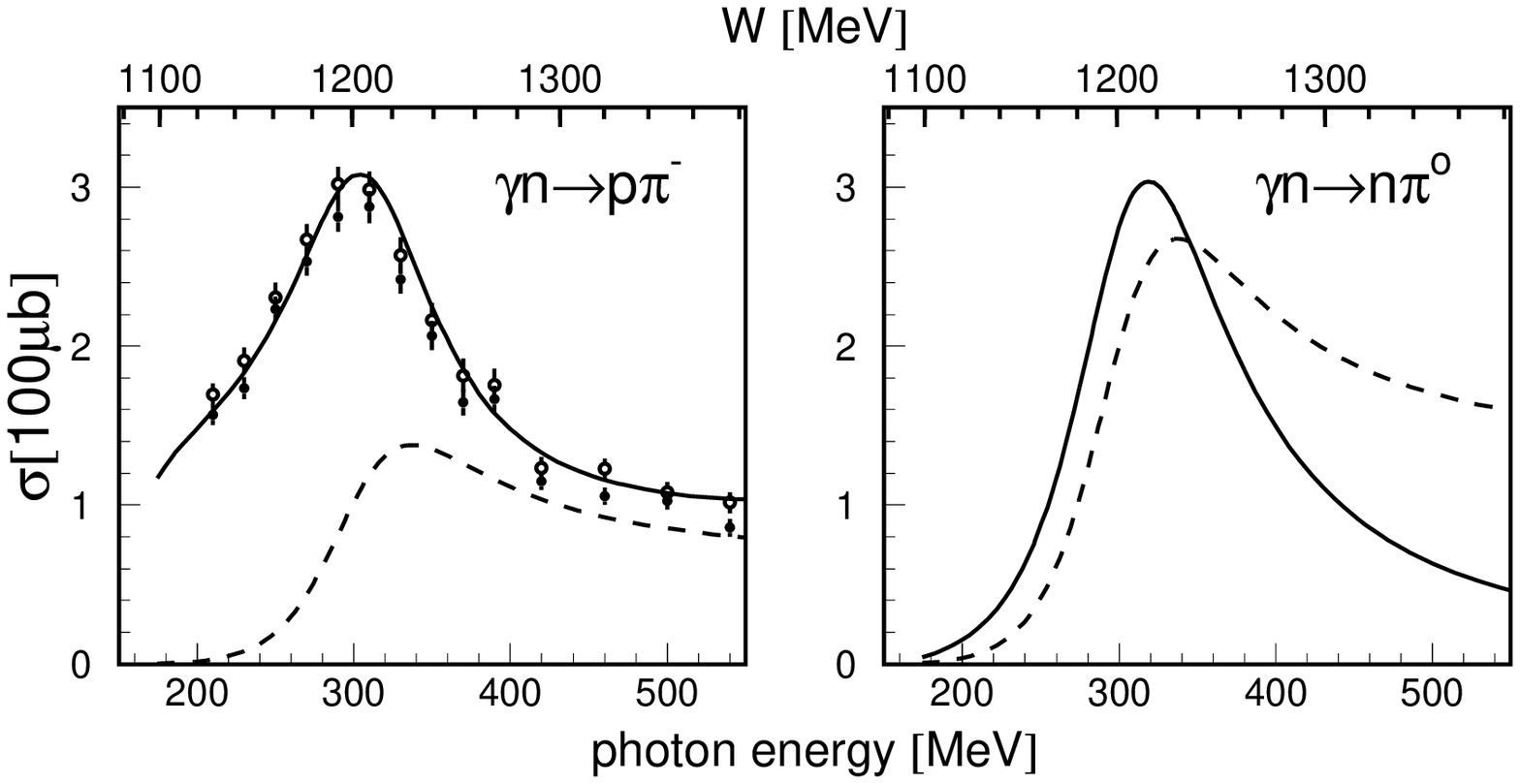}
}}
\caption{Pion production in the $\Delta$-resonance region. Data: $p\pi^0$
final state \cite{Fuchs_96,Krusche_99}, $n\pi^+$ final state \cite{Buechler_94},
$p\pi^-$ \cite{Benz_73}. Curves: MAID-model \cite{Drechsel_99b}, solid: full model,
dashed: only P$_{33}$(1232) resonance.
}
\label{fig:delta}       
\end{figure}

A very simple but instructive example is photoproduction of pions in the
$\Delta$-resonance region. Total cross sections are shown in Fig. \ref{fig:delta}.
The only nucleon resonance that contributes significantly in this energy range is 
the P$_{33}$(1232). It follows immediately from the isospin decomposition of  
the amplitude that as long as background terms are neglected (so that only
$A^{V3}$ contributes) the cross sections are related by
\begin{equation}
 \sigma(\gamma p\rightarrow p\pi^0) = 
 \sigma(\gamma n\rightarrow n\pi^0) = 
2\sigma(\gamma p\rightarrow n\pi^+) = 
2\sigma(\gamma n\rightarrow p\pi^-)\;\; .
\end{equation} 
But the measured cross sections show a completely different pattern due to the
contribution of the background terms, which are much more important for the
channels with charged pions since the photon does not couple directly to the
neutral pion. The contribution of the $\Delta$-resonance (dashed-curves) extracted 
by the MAID model \cite{Drechsel_99b} obeys of course the above relation since this is
built into the model. Even in this most simple case, additional experimental information 
for the $n\pi^0$ final state would be helpful in order to test the model. In more 
complicated cases, involving several overlapping $N^{\star}$ resonances where all three
matrix elements contribute, such information is almost indispensable.

In general, the extraction of resonance properties from photoproduction reactions 
involving neutral mesons profits from the above discussed suppression of background 
terms. However, quasi-free measurements of neutral mesons off bound nucleons
tend to suffer more strongly from FSI effects. Here, again pion production in the
$\Delta$ region is a nice example.  
\begin{figure}[thb]
\centerline{ 
\resizebox{0.8\textwidth}{!}{%
  \includegraphics{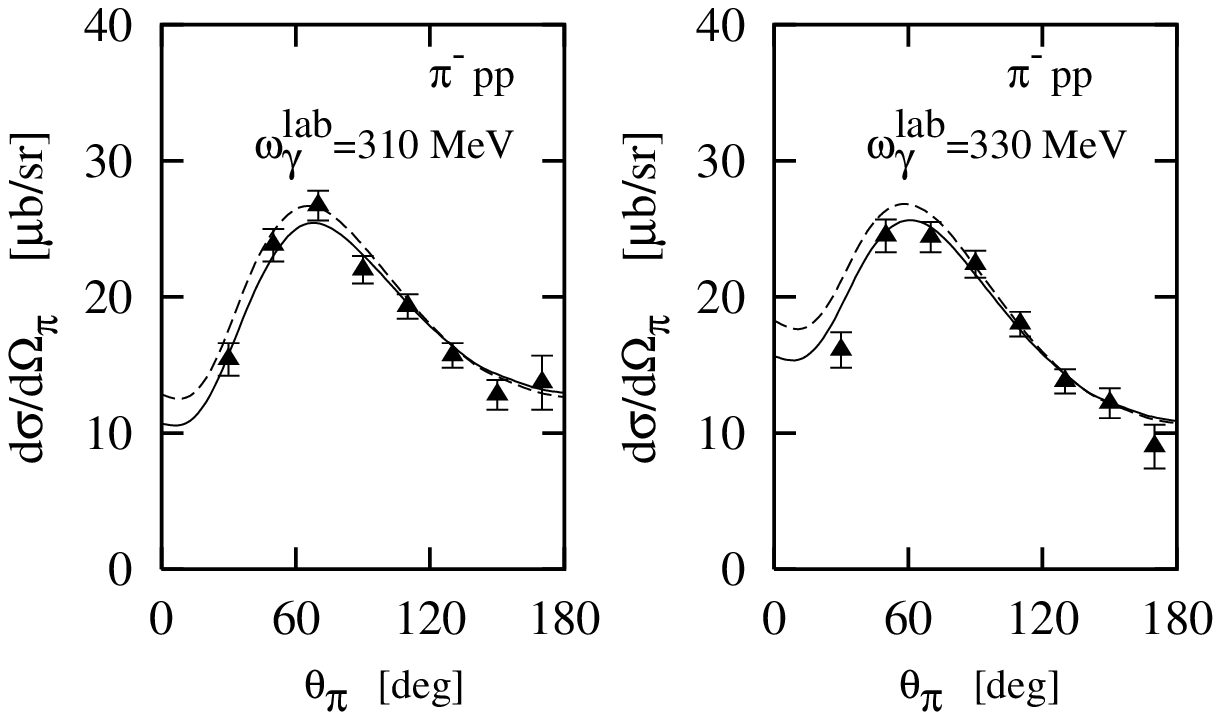} 
}}
\centerline{ 
\resizebox{0.8\textwidth}{!}{%
  \includegraphics{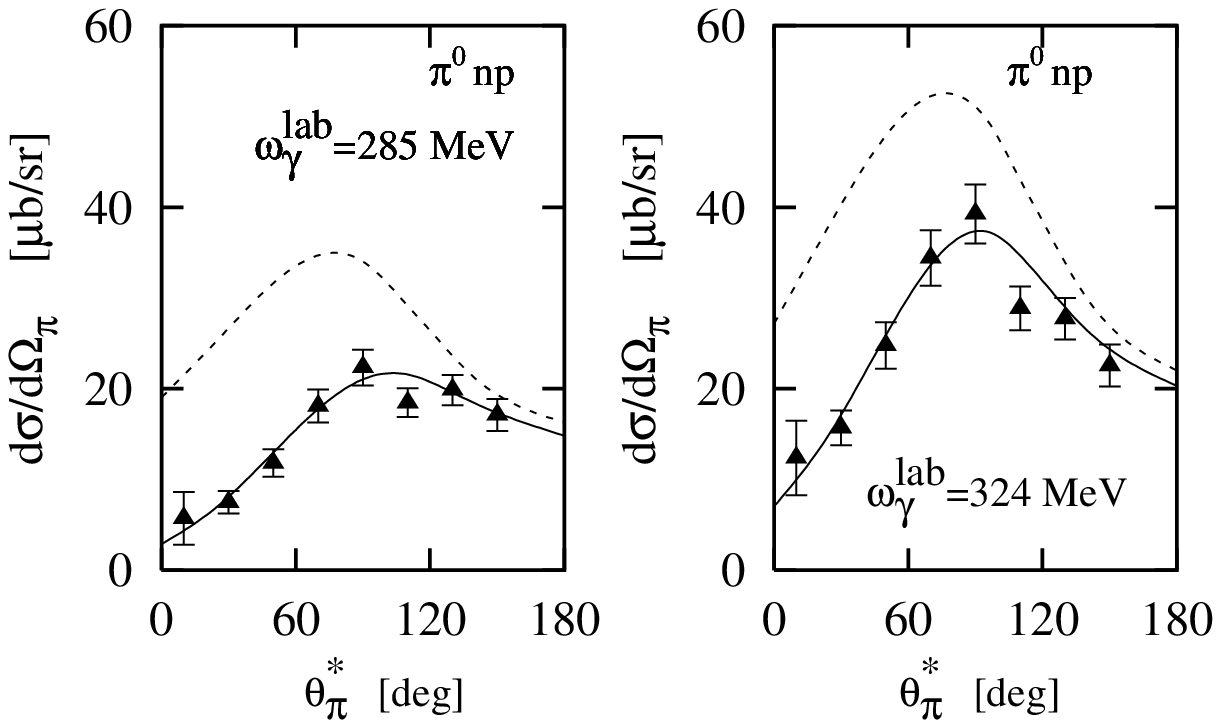}
}}
\caption{Angular distributions \cite{Darwish_03} for the reactions 
$\gamma d\rightarrow \pi^- pp$ (data from \cite{Benz_73}) and 
$\gamma d\rightarrow \pi^0 np$ (data from \cite{Krusche_99})
compared to model calculations from \cite{Darwish_03}, solid: full model,
dashed: without FSI.  
}
\label{fig:delta_diff}       
\end{figure}  
Figure \ref{fig:delta_diff} shows a few typical angular distributions for the 
$\gamma d\rightarrow \pi^- pp$ and $\gamma d\rightarrow \pi^0 np$ reactions close 
to the $\Delta$-peak. For the $\pi^0$-channel only the breakup reaction is 
considered, coherent contributions to the $\pi^0 d$ final state are removed with
kinematical cuts. The results are compared to a model calculation by
Darwish, Aren\"ovel, and Schwamb \cite{Darwish_03}, which takes into account FSI 
effects in all two-body subsystems (although it turns out that only $NN$ FSI 
is important). The large influence of FSI for the neutral final state is clearly 
visible. A major effect is that part of the reaction strength, in particular for 
pion forward angles for which the momentum mismatch between the two nucleons is
smallest, is re-distributed to the coherent channel. As a consequence, the sum of 
breakup and coherent contributions is a much better estimate for the sum of
the proton/neutron elementary cross sections than the quasi-free part 
(see \cite{Krusche_03} for details). Therefore, in this case a measurement of 
quasi-free cross sections in coincidence with recoil nucleons will produce 
model dependent results for the elementary cross sections. The inclusion
of such strong FSI effects in the models is far from trivial, results for pion
production from the work of Darwish et al. \cite{Darwish_03} are only available 
in the $\Delta$ resonance region. Recently, Tarasov et al. \cite{Tarasov_11} 
have presented results for $\gamma n \rightarrow p\pi^-$ from the 
deuteron up to incident photon energies of 3 GeV, but comparable calculations 
for the $n\pi^0$ final state are still missing.   

\begin{figure}[thb]
\centerline{ 
\resizebox{0.46\textwidth}{!}{%
  \includegraphics{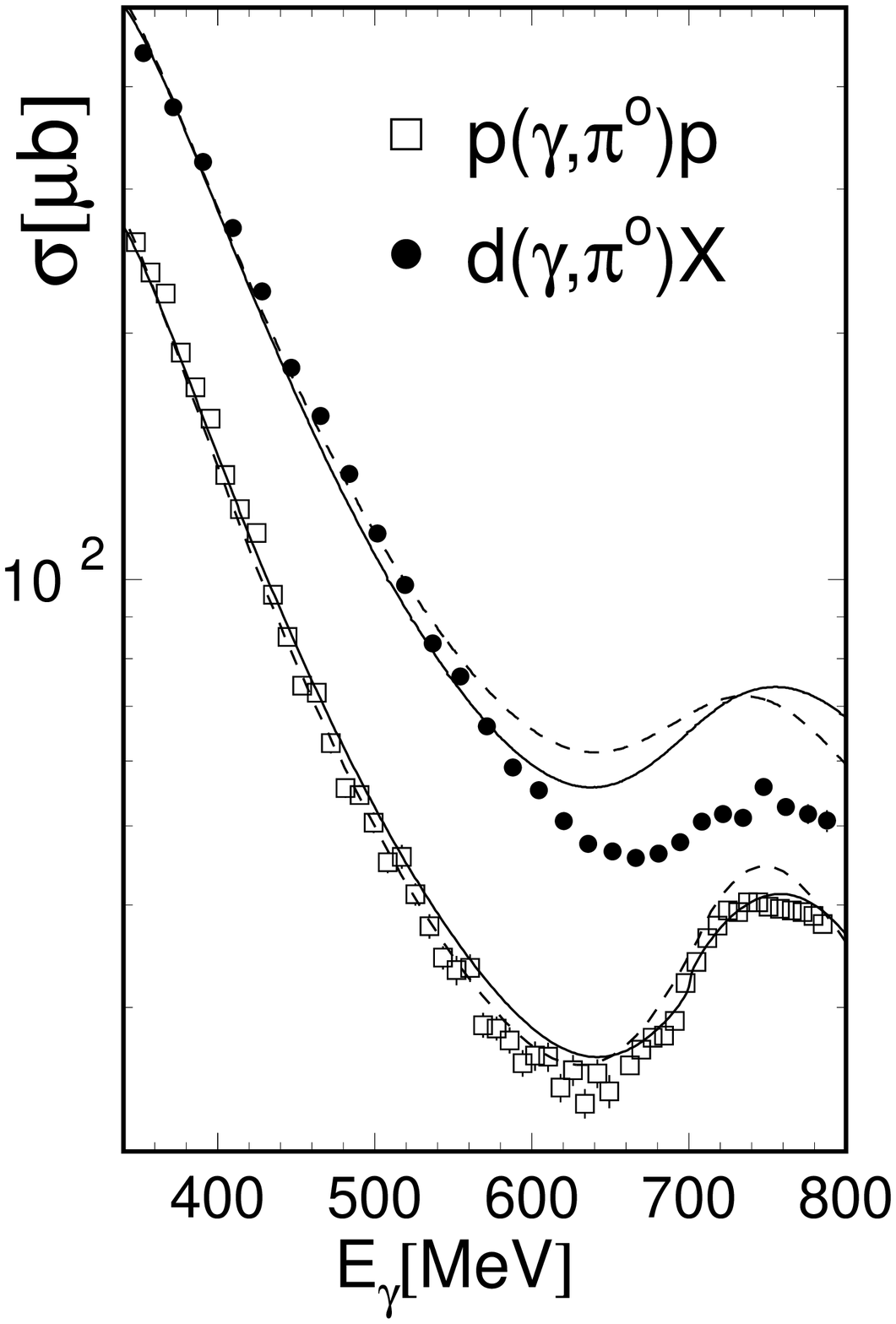} 
}
\resizebox{0.54\textwidth}{!}{%
  \includegraphics{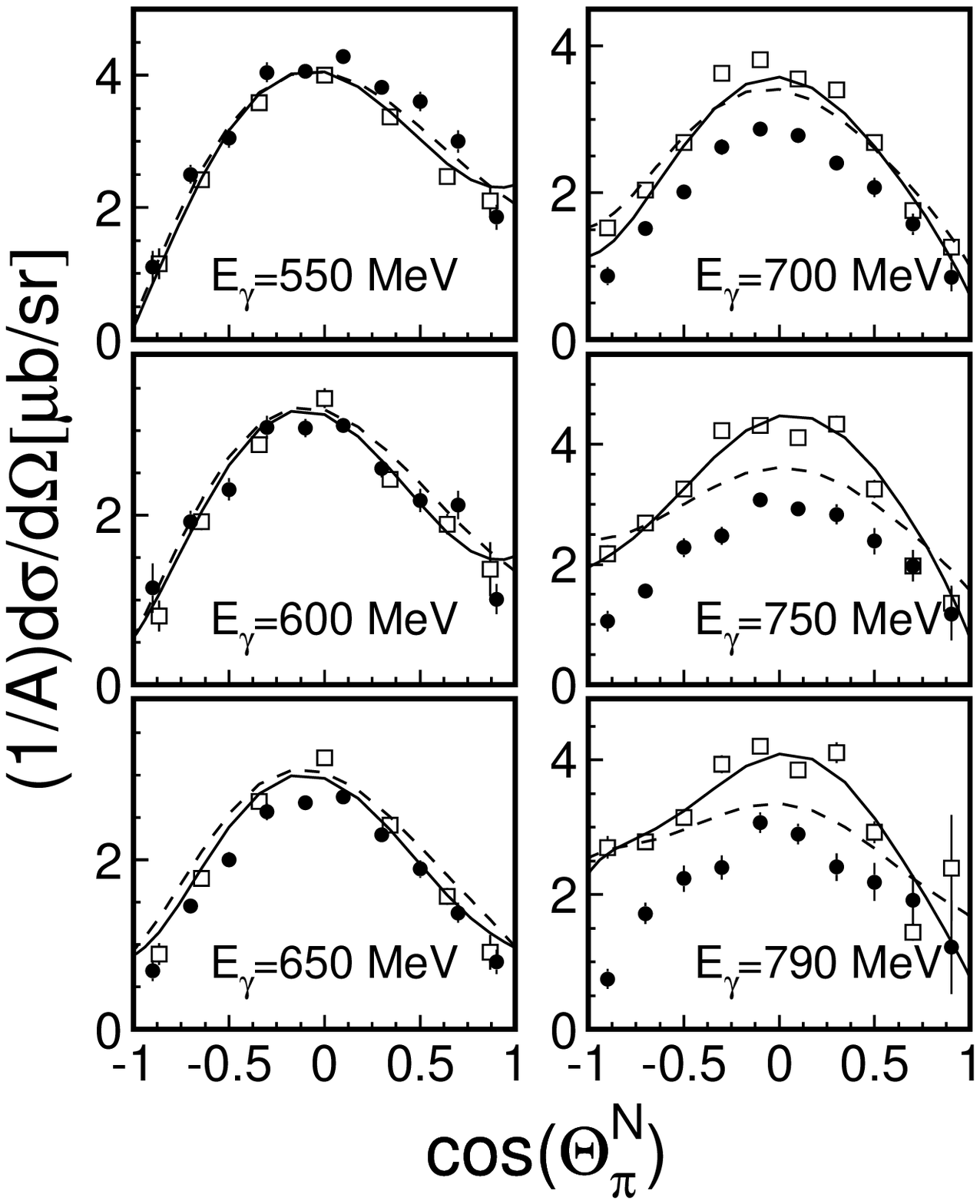}
}}
\caption{Single $\pi^0$ photoproduction off the free proton and the deuteron in
the second resonance region (note that $d(\gamma,\pi^0)X$ includes the $np\pi^0$
and $d\pi^0$ final states). Left hand side: total cross sections. 
Curves: results from the SAID analysis \cite{Said} (solid), and MAID-model 
\cite{Drechsel_99b} (dashed). For the deuteron from both models the sum of
proton and neutron cross section folded with nuclear Fermi motion is taken. 
Right hand side: angular distributions, solid curves: SAID proton, dashed curves:
Fermi smeared average of SAID proton and neutron.
}
\label{fig:secres}       
\end{figure} 
 
Photoproduction of neutral pions in the $\Delta$-resonance region is of course
quite well understood. But only a little bit higher in energy, in the second
resonance peak built from the P$_{11}$(1440), D$_{13}$(1520), and S$_{11}$(1535)
resonances, photoproduction of neutral pions off the deuteron is not well understood.  
This is demonstrated in Fig. \ref{fig:secres}. The results for the reaction 
$\gamma p \rightarrow p\pi^0$ for the free proton are reproduced by the SAID
partial wave analysis \cite{Said} and the MAID model \cite{Drechsel_99b} (which is
not surprising because the results are input to the model analyses). But the Fermi smeared 
sum of neutron and proton cross sections, which agrees well with the data in the tail
of the $\Delta$-resonance, significantly overestimates the second resonance peak.
In the absence of detailed model calculations for FSI effects and without exclusive 
data in coincidence with recoil nucleons, it is impossible to decide whether this 
indicates a problem in the model predictions for the free $\gamma n\rightarrow n\pi^0$
reaction or nuclear effects. Exclusive data in coincidence with recoil nucleons
could contribute a lot, because the comparison of the quasi-free cross section off
the proton to (the Fermi smeared) free proton cross section would allow a test of FSI
effects and such data would also allow a direct comparison of the neutron/proton
ratio to model results (assuming that FSI effects are not strongly different for 
protons and neutrons).  

The current experimental program for the investigation of the electromagnetic
excitation spectrum of the neutron therefore aims at detailed measurements of the 
quasi-free cross section off nucleons bound in the deuteron for several different
single ($\pi$, $\eta$, $\eta '$, $\omega$,...) and double ($\pi\pi$, $\pi\eta$...)
meson production channels. The comparison to other light target nuclei 
like $^3$He, $^4$He,... can add information about the importance of nuclear effects. 
This program will not only study differential cross sections but also single and double 
polarization observables.

\subsection{Meson-nucleus interactions}

The interaction of mesons with nuclei is a major source of information about 
the strong interaction. Elastic and inelastic reactions using secondary pion 
and kaon beams have revealed many details of the nucleon - meson potentials. 
However, secondary meson beams are only available for long-lived, charged mesons. 
Much less is known for short-lived mesons like the $\eta$, $\eta '$, and $\omega$. 
Their interactions with nuclei can be studied only in indirect ways. The general 
idea is to produce them by some initial reaction in a nucleus and then study their 
interaction with the same nucleus. 

In this section we will discuss three different lines of research for the interaction
of mesons with nuclei: (1) The study of the scaling of nuclear production cross 
sections with mass number, shedding light on nuclear absorption probabilities,
in-medium effects on meson-nucleon interaction probabilities, and meson in-medium 
widths, (2) the possible in-medium modification of the $\sigma$-meson related to 
partial restoration of chiral symmetry studied via double-meson production reactions,
and (3) the search for the formation of mesic nuclei via coherent meson production
reactions. 

\subsubsection{Meson absorption and in-medium widths}

\begin{figure}[thb]
\centerline{
\resizebox{0.75\textwidth}{!}{%
  \includegraphics{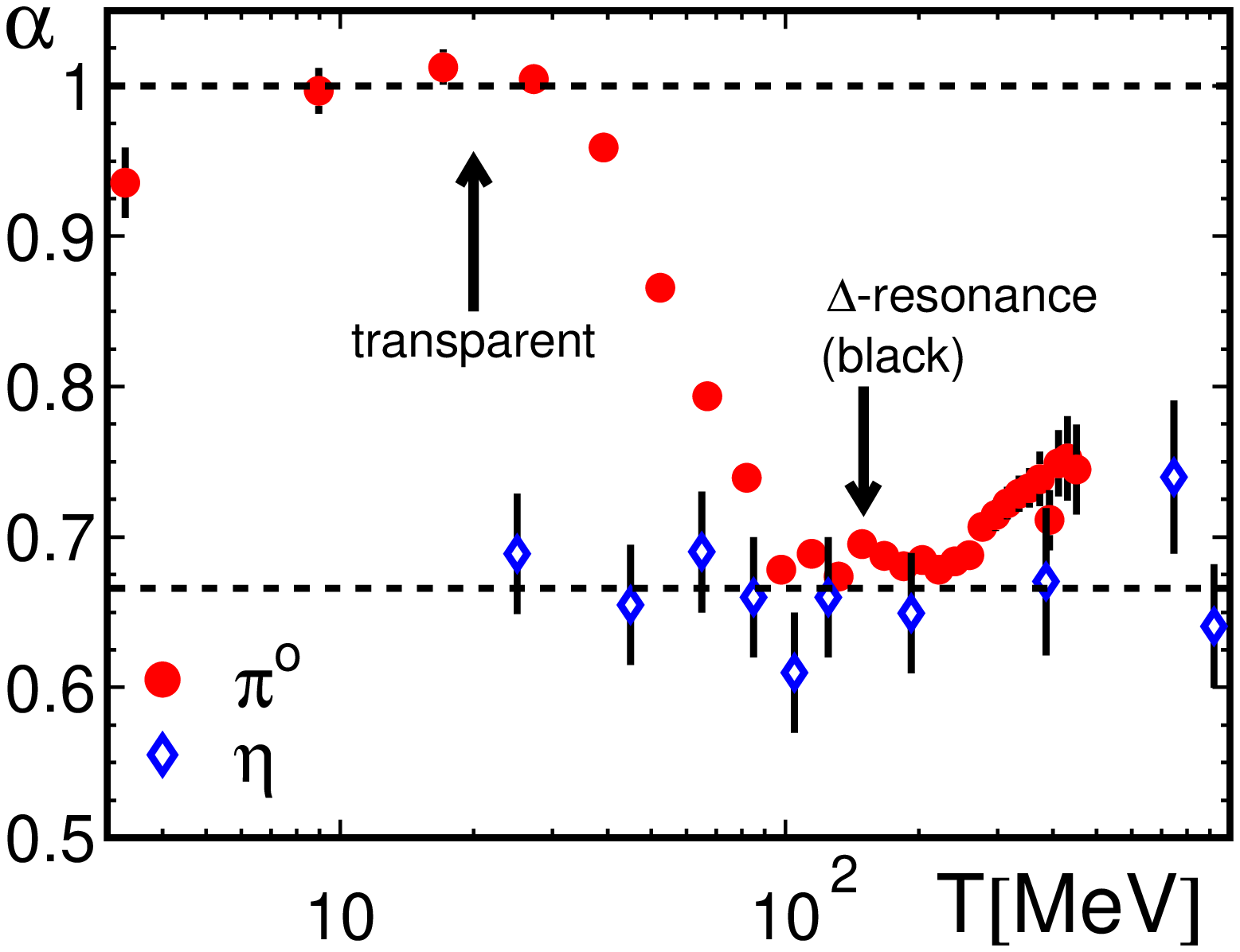}
}}
\vspace*{1cm}
\centerline{
\resizebox{0.9\textwidth}{!}{%
  \includegraphics{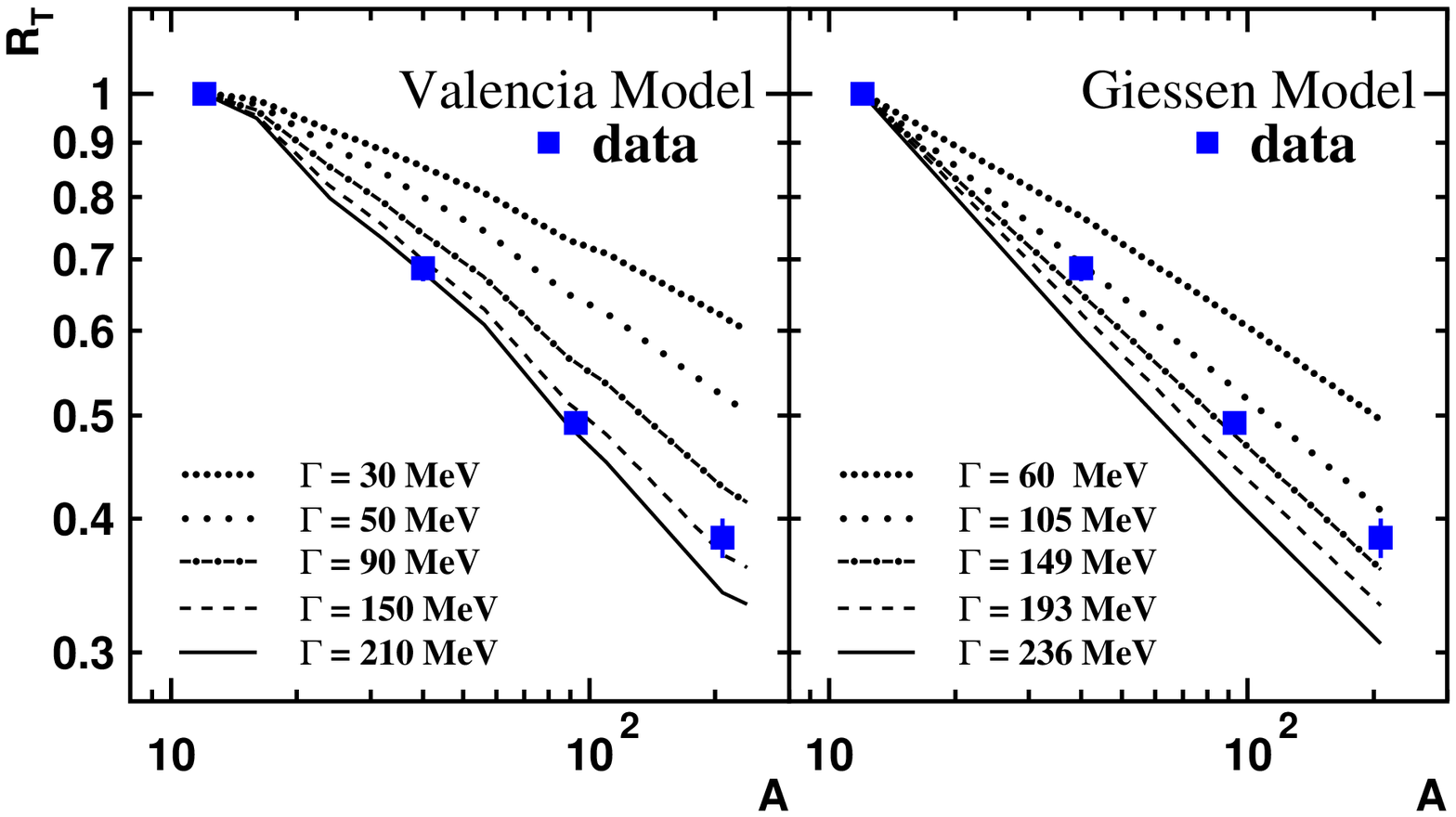}  
}}
\caption{Upper part: scaling parameter $\alpha$ as function of mesons 
kinetic energy for $\pi^o$ \cite{Krusche_04} and $\eta$ mesons 
\cite{Roebig_96,Mertens_08}. Bottom part: analysis of the transparency ratio
for photoproduction of $\omega$-mesons \cite{Kotulla_08}.
}
\label{fig:alpha}       
\end{figure}

The study of the scaling of meson production cross sections with the nuclear mass
number $A$ is a basic type of experiment which typically exhibits a 
\begin{equation}
\frac{d\sigma}{dT}(T)\propto A^{\alpha(T)}\;\;\; ,
\end{equation}
behavior, where $T$ is the kinetic energy of the pions. A value of $\alpha$ 
close to unity corresponds to a cross section scaling with the volume of 
the nucleus, i.e. with vanishing absorption, while a value of $\approx$2/3 
indicates surface proportionality, corresponding to strong absorption. 
These scaling coefficients can then be converted to absorption
cross sections using for example Glauber-type approximation models.  
Typical results for $\alpha$ as a function of $T$ are shown in Fig. \ref{fig:alpha}
for $\pi^0$ and $\eta$ mesons. Pions show strong FSI at kinetic energies large 
enough to excite the $\Delta$(1232), but are almost
undisturbed for energies below the $\Delta$ excitation threshold.
The situation is completely different for $\eta$-mesons, which show strong
absorption for all kinetic energies measured so far. The reason is the
overlap of the $s$-wave resonance S$_{11}$(1535), which couples strongly
to the $N\eta$-channel, with the production threshold. Eta-production
in the threshold region is therefore completely dominated by this resonance
\cite{Krusche_97}. The $\eta$-nucleon absorption cross section $\sigma_{\eta N}$
was determined from such data \cite{Roebig_96} to lie in the range of 30 mb, 
corresponding to a typical mean free path of $\lambda\approx$2 fm.  

A similar analysis is based on the so-called transparency ratio $R$, which compares 
the total production cross section $\sigma_{\gamma A\rightarrow VX}$ from a nucleus 
with mass number $A$ to $A$-times the elementary production cross section 
$\sigma_{\gamma N\rightarrow VX}$ on the nucleon:
\begin{equation}\label{ta}
R=\frac{\sigma_{\gamma A\rightarrow VX}}{A\;\sigma_{\gamma
N\rightarrow V X}}\; ,\;\;\;\;\;\;\;\;\;\;\;\;\;\;\;
R_A=\frac{12 \cdot \sigma_{\gamma A\rightarrow VX}}{A\;\sigma_{\gamma
C_{12}\rightarrow V X}}
\label{eq:transp}
\end{equation}
where $V$ denotes any meson. In order to account for the cross section difference 
between the proton and neutron and possible secondary production processes it is
better to normalize to an average nucleon cross section measured for a light target 
nucleus with equal proton and neutron number ($R_A$ defined in Eq. \ref{eq:transp} 
for the reference nucleus $^{12}$C). 
An example for such an analysis for the $\omega$ meson \cite{Kotulla_08} is shown at 
the right hand side of Fig. \ref{fig:alpha} where the 
transparency ratios are compared to model calculations assuming different in-medium
widths of the $\omega$ meson. As a result of the analysis, an in-medium width of the
$\omega$ meson on the order of 130 - 150 MeV in normal nuclear matter has been found
and, from a Glauber-type analysis, an absorption cross section of 70 mb was deduced.
This value is roughly a factor of three larger than the input used for
$\sigma_{\omega N}$ in the models. A similar result had been previously reported from 
the LEPS collaboration for the $\Phi$ meson \cite{Ishikawa_05}. In this case an
absorption cross section of $\approx$ 30 mb was deduced, which has to be compared to the
free nucleon absorption cross section of 7.7 - 8.7 mb. Both experiments have produced 
strong evidence for the much discussed in-medium modification of vector mesons.
A detailed review, including other related results for vector mesons, can be found in
\cite{Leupold_10}. 

\subsubsection{The $\sigma$-meson in matter and double-pion production}     

\begin{figure}[htb]
\centerline{
\resizebox{0.99\textwidth}{!}{%
  \includegraphics{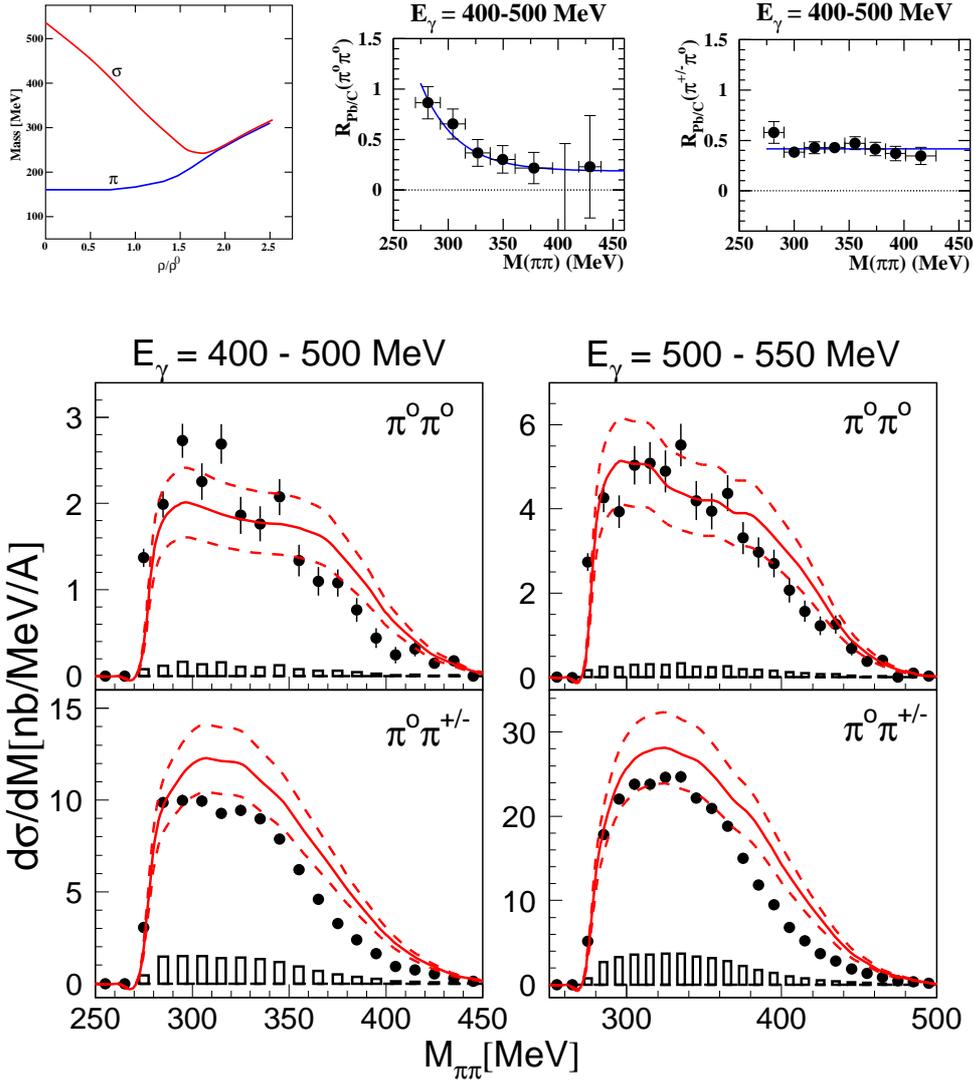}
}}
\caption{Upper row, left hand side: predicted density dependence of $\pi$ and $\sigma$
mass \cite{Bernard_87}, center and right: measured $R_{Pb/C}$ ratio for $\pi^o\pi^o$ 
and $\pi^{\pm}\pi^o$ \cite{Messchendorp_02}. Bottom row:
Pion-pion invariant mass distributions for $^{40}$Ca \cite{Bloch_07} compared to 
results of the BUU model \cite{Buss_06}. The bars at the bottom represent the 
systematic uncertainty of the data, the dashed lines represent the error band 
for the BUU calculation.
}
\label{fig:bloch}       
\end{figure}

A particularly interesting prediction for in-medium effects for scalar mesons
is the mass-split between the $J^{\pi}$=0$^{-}$ pion and its chiral partner,
the $J^{\pi}$=0$^{+}$ $\sigma$-meson. Their masses are very different in vacuum, 
which is a well-known manifestation of chiral symmetry breaking. 
The naive assumption that the two masses should become degenerate in the chiral 
limit is supported by model calculations. A typical result is the density 
dependence of the mass calculated in the Nambu-Jona-Lasino model by Bernard, 
Meissner and Zahed \cite{Bernard_87} (see Fig.~\ref{fig:bloch}, left hand side). 
The $\sigma$-mass drops as a function of nuclear density. The predicted effect is 
already large for normal nuclear matter density $\rho_o$, where the pion mass is
still stable. Since the $\sigma$ meson couples strongly to scalar-isoscalar
pion pairs the predicted effect should influence the invariant mass spectra of
such pairs in the nuclear medium, producing a downward shift 
of the invariant-mass distributions of such pairs. This prediction has been 
experimentally investigated with pion- and photon-induced double pion production 
reactions 
\cite{Bonutti_96,Bonutti_99,Bonutti_00,Camerini_04,Grion_05,Starostin_00,Messchendorp_02,Bloch_07}.
The CHAOS collaboration \cite{Bonutti_96,Bonutti_99,Bonutti_00,Camerini_04,Grion_05}
reported a downward shift for $\pi^+\pi^-$ pairs with respect to $\pi^+\pi^+$
pairs from pion induced reactions. The Crystal Ball collaboration at BNL
\cite{Starostin_00} observed a low-mass enhancement of strength for heavy nuclei in 
$\pi^-$-induced $\pi^0\pi^0$ production. In photon-induced reactions, a 
downward shift of the invariant-mass distributions of $\pi^0\pi^0$ pairs with 
respect to $\pi^0\pi^{\pm}$ pairs has been measured by the TAPS collaboration 
\cite{Messchendorp_02}. This is particularly clearly seen in the normalized ratios
of the invariant mass distributions defined by
\begin{equation}
R_{(Pb/C)}(\pi^o\pi^o) =   
\frac{d\sigma_{Pb}(\pi^0\pi^0)}{\sigma_{Pb}(\pi^0\pi^0) dM} \left/ 
\frac{d\sigma_{C}(\pi^0\pi^0)}{\sigma_{C}(\pi^0\pi^0) dM}
\right.
\end{equation}
and analogously for $R_{(Pb/C)}(\pi^o\pi^{\pm})$. The results are shown in Fig.
\ref{fig:bloch}. The ratio rises at small invariant masses for $\pi^0\pi^0$ pairs but 
not for $\pi^0\pi^{\pm}$ pairs, which do not couple to the $\sigma$. 
However, intricate final-state-interaction effects \cite{Buss_06,Bloch_07}
complicate the interpretation of the results. This is shown at the right hand side of
Fig. \ref{fig:bloch}, where precise invariant mass distributions for $^{40}$Ca
\cite{Bloch_07} are compared to calculation in the framework of the  
Boltzmann-Uehling-Uhlenbeck (BUU) model \cite{Buss_06}. The data have some excess
strength for $\pi^0\pi^0$ at small invariant masses, but the model results
show the same trend although they include only `trivial' FSI effects but no
$\sigma$-in-medium effects. In the model, the main effect arises from `side-feeding'.
Pions can undergo charge exchange scattering, re-distributing the strength between
different isospin states. Since the elementary production cross section for double
$\pi^0\pi^0$ pairs is much smaller than for double charged or mixed charge pairs 
(which are of comparable order) the main effect is a shift of strength from
$\pi^0\pi^{\pm}$ to $\pi^0\pi^0$. Inelastic scattering of pions tends to reduce their
kinetic energy, so that the pion pairs from `side-feeding' in average have smaller
invariant masses than those which have not undergone FSI. Therefore the net effect is 
an enhancement of $\pi^0\pi^0$ pairs at small invariant mass, just the same as 
predicted for the $\sigma$-in-medium effect. More detailed experiments trying to
disentangle the different contributions are required. Preliminary results from the
Crystal Ball experiment will be discussed below.  

\subsubsection{Mesic nuclei}

A much discussed question is, whether the strong interaction allows the formation 
of quasi-bound meson-nucleus states. So far all known meson-nucleus bound 
states involve at least partly the electromagnetic interaction. Pionic atoms 
are well established. Also deeply bound pionic states have been reported 
\cite{Geissel_02}, where the binding results from the superposition of the 
repulsive s-wave $\pi^-$-nucleus interaction with the attractive Coulomb force. 
Neutral mesons could form quasi-bound states only via the strong interaction. 
The meson-nucleus interaction for slow pions is too weak to form such states 
(cf the weak FSI effects for low energy pions in Fig. \ref{fig:alpha}). 
The possible formation of quasi-bound $\omega$-nucleus states and their
investigation via photon induced reactions has been discussed 
by Marco and Weise \cite{Marco_01}. However, so far, no conclusive experimental 
results are available.

The most promising case is the $\eta$-meson with the strong $s$-wave coupling to 
the S$_{11}$(1535) resonance. First hints for an attractive $\eta N$  $s$-wave 
interaction came from a coupled channel analysis of the $\eta N$ scattering 
length in 1985 by Bhalerao and Liu \cite{Bhalerao_85}. A short time later, 
Liu and Haider \cite{Liu_86} suggested the possible formation of quasi-bound 
$\eta$ - nucleus states for $A >$~10 nuclei. Experimental evidence has been 
sought in pion induced reactions \cite{Chrien_88,Johnson_93}, but those experiments 
did not produce conclusive evidence. More recently, Sokol and co-workers 
\cite{Sokol_99,Sokol_08} claimed evidence for the formation of $\eta$-mesic nuclei 
from bremsstrahlung induced reactions on $^{12}$C 
\begin{equation}
\gamma + ^{12}\mbox{C}\rightarrow 
p(n) + ^{11}_{\eta}\mbox{B}(^{11}_{\eta}\mbox{C})\rightarrow
\pi^+ + n + X
\end{equation}
where the $n\pi^+$ pairs were detected in the final state. The $\eta$-meson is 
produced in quasi-free kinematics on a nucleon (p,n) so that it is almost at rest 
in the residual $A=11$ nucleus. If a quasi-bound state is produced, the $\eta$-meson 
can be re-captured by a nucleon into the S$_{11}$ excitation, which may then decay 
into a pion-nucleon back-to-back pair. Sokol and Pavlyuchenko \cite{Sokol_08} claim 
an enhancement above background from quasi-free pion production for certain kinematic 
conditions.

After precise low-energy data for the photoproduction of $\eta$-mesons off the 
proton \cite{Krusche_95}, deuteron \cite{Krusche_95a,Hoffmann_97,Weiss_01,Weiss_03} 
and helium nuclei \cite{Hejny_99,Hejny_02} became available, refined model analyses 
of the scattering length were done by many groups (see \cite{Arndt_05} for a summary).
The results  for the imaginary part are rather stable and cluster between 0.2~fm and 
0.3~fm. The more important real part, which decides about the formation of bound
states, is still less constrained. It runs all the way 
from a negative value of $-$0.15~fm to numbers close to and even above $+$1~fm 
\cite{Arndt_05}. 
However, most of the more recent analyses prefer large values above 0.5~fm, 
which has raised discussions about very light mesic nuclei and prompted theoretical 
studies of the $\eta$-interaction with $^2$H, $^3$H, $^3$He, and $^4$He systems 
\cite{Ueda_91,Ueda_92,Wilkin_93,Rakityanski_95,Rakityanski_96,Green_96,Scoccola_98,Shevchenko_00,Grishina_00,Garcilazo_01} 
.

Experimental evidence for light $\eta$-mesic nuclei has mostly been
searched in the threshold behavior of $\eta$-production reactions.
The idea is that quasi-bound states in the vicinity of the production
threshold will give rise to an enhancement of the cross section relative to 
the expectation for phase space behavior. Many hadron induced reactions have 
been studied in view of such effects in particular: 
$pp\rightarrow pp\eta$ \cite{Calen_96,Smyrski_00,Moskal_04}, 
$np\rightarrow d\eta$ \cite{Plouin_90,Calen_98}, 
$pd\rightarrow\eta  ^3\mbox{He}$ \cite{Mayer_96},
$dp\rightarrow\eta  ^3\mbox{He}$ \cite{Smyrski_07,Mersmann_07,Rausmann_09}, 
$\vec{d}d\rightarrow \eta ^4\mbox{He}$ \cite{Willis_97}, 
and $pd\rightarrow pd\eta$ \cite{Hibou_00}. 
In particular the $pd\rightarrow\eta  ^3\mbox{He}$ \cite{Mayer_96} and
$dp\rightarrow\eta  ^3\mbox{He}$ reactions \cite{Smyrski_07,Mersmann_07,Rausmann_09}
show an extremely steep rise at threshold, implying a very large $\eta^3\mbox{He}$
scattering length. 

If such states do exist, they should show up independently of the initial state 
of the reaction. Threshold photoproduction of $\eta$-mesons from light nuclei is a  
clean tool for such experiments, but for that the $\eta$-mesons have to be produced 
coherently off the target nuclei. Since threshold photoproduction of the $\eta$ is 
dominated by an isovector spin-flip amplitude exciting the S$_{11}$(1535) 
(see \cite{Krusche_03} for a summary) the coherent cross section is very small for 
the isoscalar deuteron and practically forbidden for the isoscalar scalar $^4$He 
nucleus. Only the $I=1/2$, $J=1/2$ $^3$H and $^3$He nuclei have reasonably large cross 
sections for $\gamma A\rightarrow A\eta$.

The $^3$He system has been investigated with photon induced reactions 
by Pfeiffer et al. \cite{Pfeiffer_04}. Possible evidence for the formation of a 
quasi-bound state was reported from the behavior of two different reactions. 
The coherent $\eta$-photoproduction $\gamma ^3\mbox{He}\rightarrow \eta^3\mbox{He}$ 
showed a strong threshold enhancement with angular distributions much more isotropic
in the threshold region than expected from the nuclear form factor. Both are
indications at least for strong FSI processes. Furthermore, in an approach similar 
to the Sokol experiment \cite{Sokol_08}, the excitation function for $\pi^0$ - proton
pairs was investigated. After background subtraction, it showed a slight enhancement 
at the $\eta$-production threshold for pairs emitted back-to-back in the 
$\gamma ^3\mbox{He}$ cm-system. Preliminary results from a follow-up experiment with
much better statistical quality will be discussed below.

\section{Experimental facilities}
\label{sec:1}

In this section we will give a short overview of the main facilities involved in the 
study of photoproduction from nuclear targets. All experiments are based at electron
accelerators equipped with tagged photon beams, based on the bremsstrahlung technique
or Laser-backscattering. Most of them can be linearly and/or circularly polarized.
The additional availability of polarized protons and deuterons, using frozen-spin 
butanol or HD targets, allows the study of a full set of single and double polarization
observables for meson production reactions off the proton and the neutron. Since the 
detector systems at these facilities are optimized for different reaction types involving
different particles mostly from photoproduction off the free proton, they are more or
less well adapted for measurements off quasi-free neutrons. While some of them (in particular
the $4\pi$ electromagnetic calorimeters) can easily measure even all-neutral multiple-meson
final states like $\pi^0\pi^0 n$, $\pi^0\eta n$, devices relying more or less on magnetic
spectrometers need at least some charged particles in the final state and sometimes cannot
detect the recoil neutrons.

\subsection{The Crystal Ball/TAPS setup at the Mainz MAMI accelerator}

The MAMI accelerator in Mainz \cite{Herminghaus_83,Kaiser_08} now delivers electron beams 
with energies up to 1.5 GeV, typical longitudinal polarization of the beam reaches values 
above 80\% and is very stable. Photon beams are produced by the bremsstrahlung process and 
are tagged with the Glasgow magnetic spectrometer \cite{Anthony_91}; at maximum electron 
energy with a typical resolution of 4~MeV. Linearly polarized photon beams are produced 
using coherent bremsstrahlung from diamond lattices.
\begin{figure}[bth]
\centerline{
\resizebox{1.0\textwidth}{!}{%
  \includegraphics{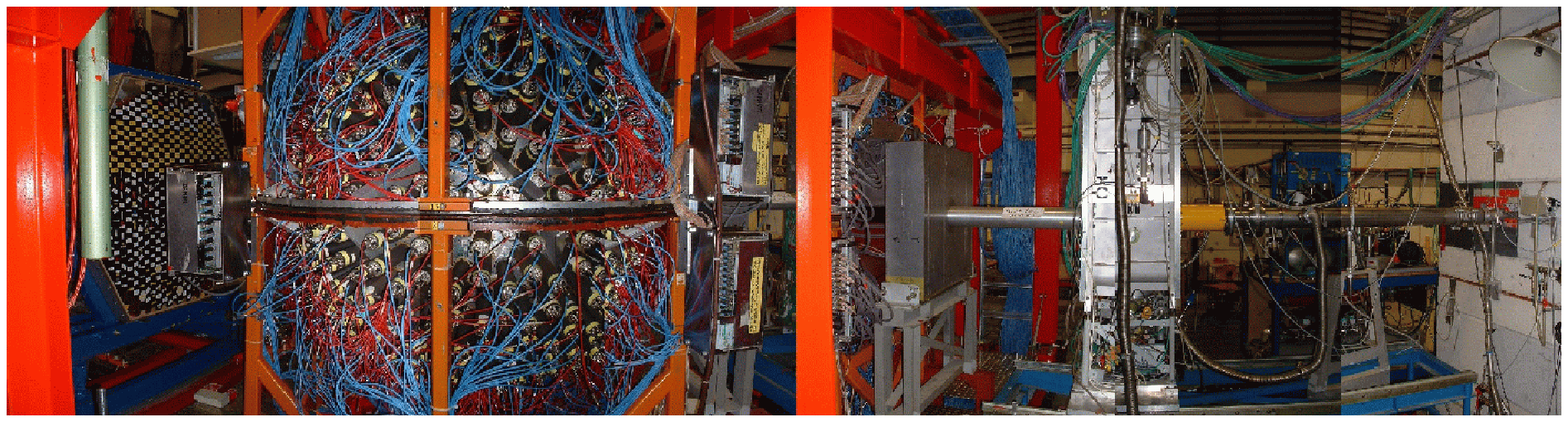}
}}
\centerline{
\resizebox{0.5\textwidth}{!}{%
  \includegraphics{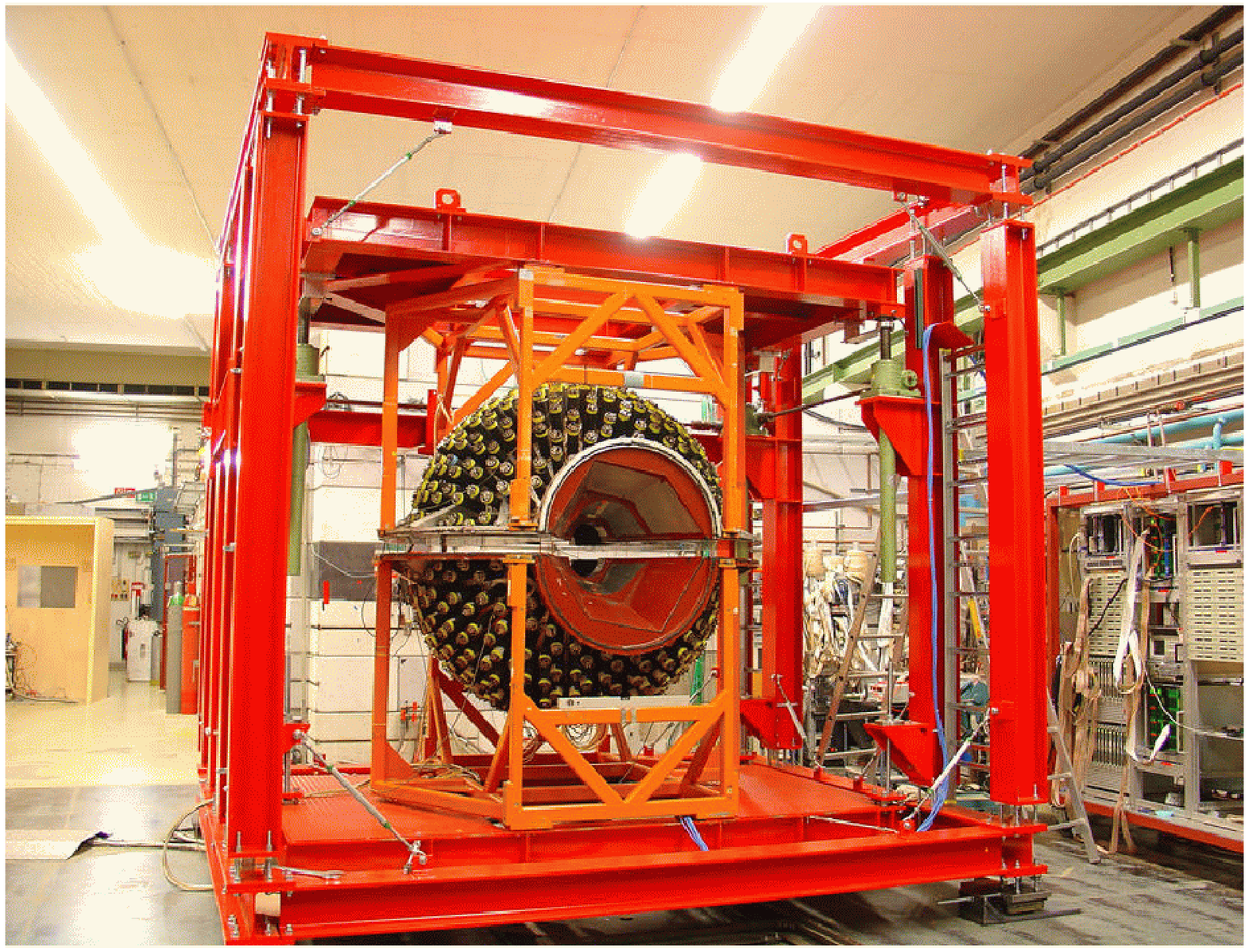}
 } 
\resizebox{0.5\textwidth}{!}{%
  \includegraphics{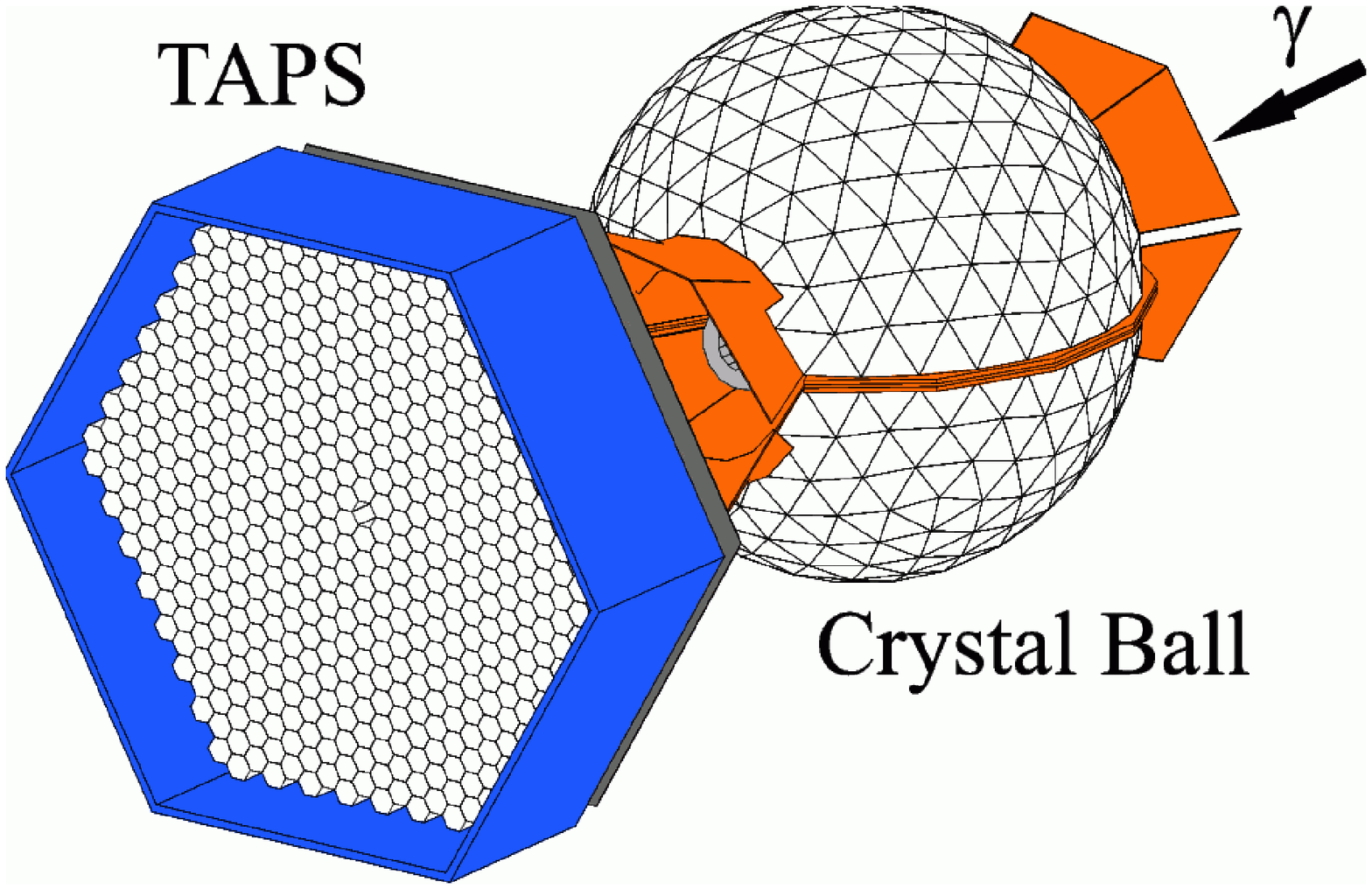}  
}}  
\caption{Overview of the MAMI setup. Upper part: beam coming from the
right, Crystal Ball at the left of the picture, TAPS forward wall at the very
left. 
Bottom part, left hand side: Crystal Ball in setup phase, right hand side:
schematic drawing of the calorimeter. 
\label{fig:cball}
}
\end{figure}
Since the accelerator is designed also for electron scattering experiments,
it can deliver electron beam intensities far larger than needed for tagged photon 
experiments ($\approx 100$ $\mu$A, compared to some tens of nA used for tagging). 
Liquid hydrogen, liquid deuterium, liquid $^{3,4}$He, gaseous polarized $^3$He, frozen-spin 
butanol (polarized protons), frozen-spin deuterated butanol (polarized deuterons), 
and solid targets can be and have been used at this facility. The typical detector 
equipment is summarized in Fig. \ref{fig:cball}. A $4\pi$ electromagnetic calorimeter 
has been set up, combining the Crystal Ball detector \cite{Starostin_01}
(672 NaI crystals covering the full azimuthal angle for polar angles from 20$^{\circ}$ 
to 160$^{\circ}$) with 384 BaF$_2$ crystals from the TAPS detector 
\cite{Novotny_91,Gabler_94}, configured as a forward wall. The forward wall is mounted
1.475 m downstream from the target and covers polar angles from $\approx$2$^{\circ}$ to 
21$^{\circ}$. The Crystal Ball is equipped with an additional 
{\bf P}article {\bf I}dentification {\bf D}etector (PID) \cite{Watts_04} and 
cylindrical multiple wire chambers surrounding the target and all 
modules of the TAPS detector are capped with individual plastic scintillators for charged
particle detection. This setup, and earlier simpler detector configurations based on the 
TAPS detector or the GDH-experiment, have been used to study 
photoproduction off nuclei from the deuteron to the most heavy targets. Currently, measurements 
of double polarization observables for meson photoproduction off the proton and the neutron 
are under way.

\subsection{The Crystal Barrel/TAPS setup at the Bonn ELSA accelerator}

The electron stretcher accelerator facility ELSA \cite{Husmann_88,Hillert_06} delivers 
electron beams with energies up to 3.5 GeV. 
\begin{figure}[thb]
\centerline{
\resizebox{1.0\textwidth}{!}{%
  \includegraphics{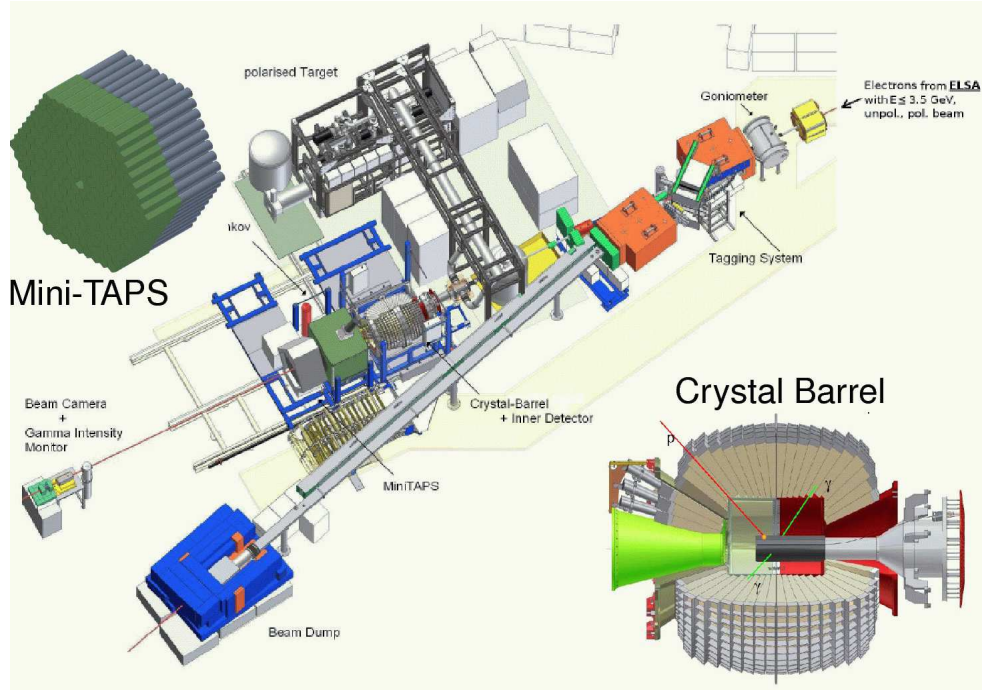}
}}
\caption{Experimental setup at the Bonn ELSA accelerator combining the Crystal Barrel
detector, part of the TAPS detector as forward wall, inner detectors, the photon tagging
facility and the frozen-spin polarized target.
}
\label{fig:cbarrel}       
\end{figure}
The intensity is lower (typically a few nA) than for MAMI. 
Circularly or linearly (see \cite{Elsner_09} for details) polarized beams are produced using 
the same techniques as at MAMI and a polarized frozen-spin butanol target, very similar to 
the one used at MAMI, is available. The whole setup is summarized in Fig. \ref{fig:cbarrel}.
The main detector is again a $4\pi$ electromagnetic calorimeter combining the Crystal Barrel 
detector \cite{Aker_92} (1290 CsI (Tl) crystals of 16 radiation lengths $X_o$ all mounted 
in a target pointing geometry) with part of the TAPS detector (216 BaF$_2$ crystals) as 
forward wall (`Mini-TAPS)'. In this case only `Mini-TAPS' and the most forward part of the
Barrel (90 CsI modules) are read out with photomultipliers. The rest of the Barrel is read
with photodiodes, so that it does not deliver timing information and cannot be used in the
first level trigger. This is not a significant disadvantage for measurements off the free
proton, where in addition to hits in the forward part of the calorimeter the proton can 
generate triggers in the `Inner-detector' \cite{Suft_05}, a three-layer scintillating fiber 
detector around the target, covering polar angles between 28$^{\circ}$ and 172$^{\circ}$.
It limits, however, the type of reactions that can be studied off quasi-free neutrons to
reactions with large photon multiplicity, which have reasonable trigger efficiencies from 
the forward part of the detector. Therefore, so far in particular channels like 
$\eta\rightarrow 3\pi^0\rightarrow 6\gamma$ \cite{Mertens_08,Jaegle_08}, 
$\eta '\rightarrow \pi^0\pi^0\eta\rightarrow 6\gamma$ \cite{Jaegle_11}, 
$\omega\rightarrow\pi^0\gamma\rightarrow 3\gamma$ \cite{Kotulla_08,Trnka_05,Nanova_10,Nanova_11},
$\pi^0\pi^0\rightarrow 4\gamma$ \cite{Jaegle_11c}, and $\pi^0\eta\rightarrow 4\gamma$
\cite{Jaegle_09} with three to six photons in the final state have been studied for nuclear
targets. It is, however, planned to upgrade the trigger capability of the detector
by a new readout system based on {\bf A}valanche {\bf P}hoto{\bf D}iodes (APDs), which
can deliver time information and trigger signals. Also at this facility a large
program to measure single and double polarization observables off the free proton and the 
quasi-free neutron is under way. 

\subsection{The CLAS detector at JLAB}

The CEBAF Large Acceptance spectrometer (CLAS) \cite{Mecking_03} is housed in Hall B of 
the Thomas Jefferson National Accelerator Facility in Newport News. The detector is
installed at a beam line which can deliver electron beams from 0.8 GeV to 6 GeV which may 
be polarized (polarization degrees up to 85\%). The photon tagging system can tag photons 
in the range from 20\% to 95\% of the incident electron energy (for electron energies up to
6.1 GeV). Circularly polarized photon beams are produced as usual with longitudinally
polarized electron beams and coherent bremsstrahlung from a diamond is used to obtain
linearly polarized beams. A butanol cryo-target of the frozen-spin type is available 
for polarized protons and deuterons. The core of the detector is a toroidal magnet built 
from six superconducting coils, symmetrically arranged around the beam line. All six 
sectors are equipped with 34 layers of tracking chambers. This setup allows the full 
reconstruction of the three-momentum of charged particles with high resolution. The 
tracking region is surrounded by plastic scintillators. Charged hadrons are identified 
by combining the momentum, time-of-flight, and path length information. In addition, 
for polar angles smaller than 70$^{\circ}$, photons and neutrons can be detected in an
electromagnetic calorimeter. The detector is often operated in the missing mass mode,
i.e. all particles except one are detected and the mass of the only missing particle
is than reconstructed from the reaction kinematics. In that way, all kinds of single meson
production reactions off free protons can be studied over a large range of incident photon
energies. This technique does not work for multiple production of neutral
mesons ($\pi^0\pi^0$, $\pi^0\eta$) or for photoproduction of neutral mesons off the 
neutron (unless the neutral mesons decay into long-lived charged mesons like 
K$^0\rightarrow \pi^+\pi^-$). Therefore, the programs aiming at the study of meson
photoproduction off neutrons at MAMI and ELSA on one side and at CLAS on the other side, 
are nicely complementary. The CLAS experiment is ideal for reactions with high charged
particle multiplicity in the final state like $\gamma n \rightarrow p\pi^-$ \cite{Chen_09},
or $\gamma n\rightarrow K^0\Lambda\rightarrow (\pi^+\pi^-)(\pi^- p)$, while the
$4\pi$ electromagnetic calorimeters are in advantage for the `all neutral' final states.      

Apart from the measurement of angular distributions and polarization observables for meson
photon production off quasi-free neutrons, the CLAS experiment has also contributed to the 
investigation of meson in-medium properties via a study of the 
$\gamma A\rightarrow e^+e^-X$ reaction \cite{Nasseripour_07,Wood_08}. The $\rho$, $\omega$,
and $\phi$ vector mesons have small branching ratios (order of 10$^{-5}$ to 10$^{-4}$) 
into the $e^+e^-$ channel and can be reconstructed from the invariant mass of the lepton 
pairs. The advantage of this type of experiment is that, both in the initial and final 
state, only electromagnetic interaction contributes, so that initial and final state 
interaction effects are minimized. Due to kinematical limitations only vector mesons with 
momenta above 1 GeV could be detected, so that the longer-lived $\omega$ and $\phi$
mesons decayed mostly outside the nuclei without significant in-medium modifications.
The short-lived $\rho$ in turn showed a clear collisional broadening of its width by
roughly 70 MeV, but no shift of its mass.    

\subsection{The GeV-$\gamma$ experiment at LNS at Tohoku University}

The third facility that uses a tagged bremsstrahlung beam is located at the 
Laboratory of Nuclear Science (LNS) at Tohoku University in Sendai, Japan. The 
primary electron beam with energies up to 1.2 GeV is produced by a synchrotron 
(Stretcher Booster Ring (STB)) \cite{Hinode_01}. The tagging system is described 
in detail in \cite{Yamazaki_05}. The system uses one of the magnets from the 
synchrotron (a C-shaped sector magnet) as magnetic spectrometer for photon tagging. 
The bremsstrahlung target is a thin carbon fiber string (diameter of 11 $\mu$m), which
can be moved step-wise into the internal beam to produce the desired intensity of
bremsstrahlung photons (typical intensities between 3$\times 10^{6}$ s$^{-1}$ and
2$\times 10^{7}$ s$^{-1}$). The tagging counters are two rows of plastic 
scintillators. The first row consists of 50 counters with a width of 5 mm, which
corresponds to an energy bin width of 6 MeV at $E_{\gamma}=$ 1 GeV. The second row
of scintillators has larger width, each module covering four modules of the first 
row (except that at highest energies which covers six modules) and reduces 
the background via a coincidence condition. 

The experimental setup for the detection of photons is described in 
\cite{Nakabayashi_06}. The main component are four blocks of in total 206 CsI 
detectors. The two forward blocks of 124 modules each cover polar angles from 
15$^{\circ}$ to 72$^{\circ}$ and azimuthal angles from -17$^{\circ}$ to 17$^{\circ}$
and the two backward blocks of 29 modules each cover polar angles of 
95$^{\circ}$ to 125$^{\circ}$ at azimuthal angles of -12$^{\circ}$ to 12$^{\circ}$.
Plastic scintillators in front of the CsI modules are used for charged particle
identification. The coverage of the solid angle was thus far below 4$\pi$.
More recently, a new detector setup, the electromagnetic calorimeter FOREST
\cite{Suzuki_09}, has been installed at LNS. It combines three electromagnetic 
calorimeters of CsI modules, Lead/SciFi blocks, and Lead Glass 
$\check{\mbox{C}}$erenkov counters, covering roughly 90\% of 4$\pi$.

For photoproduction reactions off the free proton or quasi-free nucleons from the 
deuteron solid hydrogen or deuterium targets are used. Polarization observables 
cannot be investigated.

This experiment has been for example used for the study of $\eta$ photoproduction
off the free proton \cite{Nakabayashi_06}, quasi-free of the deuteron \cite{Miyahara_07},
and also off heavy nuclei \cite{Kinoshita_06}. Due to the low acceptance and low
detection efficiency for recoil nucleons, the experiments off the nuclei supplied only
the inclusive cross section without detection of coincident recoil nucleons.
The quasi-free neutron cross section was extracted by subtracting the free-proton
cross section (after folding with the momentum distribution of protons bound in the 
deuteron) from the inclusive deuteron data.

\subsection{The LEPS facility at SPring-8 in Osaka}

The last two facilities are based on a different technique for the production of
tagged photon beams, namely the backscattering of laser light from high energy 
electrons. A certain advantage of this type of experiment is that linearly and
circularly polarized beams can be easily produced by using correspondingly polarized
laser light. The polarization characteristics are different from the bremsstrahlung 
technique, in particular for linearly polarized photons. The latter produces
polarization in the coherent peaks of the bremsstrahlung spectrum; reasonable
polarization degrees can only be obtained for photon energies up to 50\% of
the electron beam energy. Beams from laser backscattering reach the maximum
polarization degree at maximum photon beam energies.   

The {\bf L}aser-{\bf E}lectron {\bf P}hoton beam line (LEPS) experiment is located 
at the 8-GeV storage ring of the {\bf S}uper {\bf P}hoton {\bf ring} (SPring-8). 
The setup is described in
\cite{Sumihama_06}. Laser light from an Ar-ion laser (wavelengths between 333.6 
and 363.8 nm) is reflected into the electron beam line in a straight section. 
The photons backscattered from the 8 GeV electrons reach maximum energies of 
2.4 GeV. The scattered electrons are momentum analyzed by the last bending magnet
before the straight section of the beam line and then detected in a tagging counter.
The tagging detector combines two layers of scintillator hodoscopes with two layers
of silicon strip detectors. The scintillators of both hodoscopes are 5 mm wide and
arranged with 2.2 mm overlap. The hodoscopes reject accidental background and the
hit position on the strip detectors (strip width 0.1 mm) defines the energy of
the scattered electrons. The typical energy resolution is in the range of 15 MeV.

The LEPS detector system is quite different from the more-or-less 4$\pi$ covering
electromagnetic calorimeters discussed above or the large angle magnetic 
spectrometer CLAS. It is optimized for the detection of forward-going charged
hadrons. The two main components of the device are a dipole magnet at forward angles 
covering $\pm$0.4 rad in the horizontal and $\pm$0.2 rad in the vertical direction.
Charged particles deflected by the magnetic field are detected in a time-of-flight
wall made of 40 plastic scintillators each of 12 cm widths, 4 cm thickness and 
200 cm length mounted with an overlap of 1 cm. The start signal for the 
time-of-flight measurement is provided by the RF of the electron ring. 
The system is equipped with a start detector (for trigger purposes) and a 
$\check{\mbox{C}}$erenkov Aerogel counter for the rejection of electromagnetic background, 
both mounted in front of the magnet entrance. A silicon vertex detector also in front
of the magnet entrance and three multiple-wire drift chambers, one downstream and 
two upstream of the magnet, are used for track reconstruction of charged particles.

This setup is particularly well suited to study strangeness production via the
spectroscopy of forward going charged kaons (see e.g. \cite{Sumihama_06}).
It has significantly contributed to the study of the in-medium properties of vector
mesons with the measurements of transparency ratios for the photoproduction of
$\phi$ mesons off nuclei, where it found a substantial increase of the in-medium
$\phi N$ absorption cross section with respect to the $\phi N$ absorption in free
space. It was also strongly involved in experiments reporting tentative evidence
for the observation of pentaquark states in the $\gamma n\rightarrow K^+K^- n$
reaction using carbon targets \cite{Nakano_03}, which are, however, heavily disputed.

\subsection{The GRAAL facility at ESRF in Grenoble}

This overview would be incomplete without a short discussion of the 
{\bf GR}enoble {\bf A}nneau  {\bf A}c\-celerateur {\bf L}aser (GRAAL) at the 
{\bf E}uropean {\bf S}ynchrotron {\bf R}adiation {\bf F}acility (ESRF) in Grenoble,
although this experiment has been shut-down recently.  

The photon beam was produced similarly to the LEPS experiment by shining a laser 
on the 6 GeV electron beam of the synchrotron. Depending on the wavelength of the 
laser, Compton-edges at 1.1, 1.4, 1.47, and 1.53 GeV have been produced. An overview 
over the beam-line setup and tagger device is e.g. given in \cite{Bartalini_05}.
Also in this setup one of the synchrotron bending magnets was used for the deflection
of the backscattered electrons, which were then detected in a silicon microstrip
detector (128 strips with a pitch of 300 $\mu$m). Typical resolution for the photon
beam energy was 16 MeV (FWHM). Polarized beams were produced by the use of polarized
laser light. Polarized targets were not used.

The detection system {\bf L}arge {\bf A}cceptance {\bf GR}aal-beam {\bf A}pparatus 
(for) {\bf N}uclear {\bf G}amma {\bf E}xperiments (LA$\gamma$RANGE) combined
two main components \cite{Bartalini_05}. The target was surrounded by a BGO
electromagnetic calorimeter consisting of 488 crystals, 21 radiation lengths long,
which covered polar angles from 25$^{\circ}$ to 155$^{\circ}$ for
the full azimuthal angular range. Two cylindrical wire chambers and a barrel of
plastic scintillators in the BGO ball allowed the reconstruction of position
and energy loss of charged particles. The position of charged particles at 
forward angles up to 25$^{\circ}$ was measured by two planar wire chambers and
the energy was derived from a time-of-flight measurement by a hodoscope of plastic
scintillators placed 3 m away from the target. Photons and neutrons at forward angles
could be detected in a lead-scintillator sandwich wall behind the plastic hodoscope.
  
During the last few years, this experiment produced significant results for
meson photoproduction reactions of the quasi-free neutron bound in the deuteron, 
among them the first measurement of the beam asymmetry $\Sigma$ for
$\eta$ photoproduction of the neutron \cite{Fantini_08}, a measurement of total cross
sections, invariant mass distributions, and beam asymmetries for double $\pi^0$
production off the neutron \cite{Ajaka_07}, and the first experimental evidence
for the narrow structure observed recently in the excitation function of
$\eta$ photoproduction off the neutron \cite{Kuznetsov_07}.    

In the meantime, the experiment has been dismounted and the BGO ball is now being
installed at the Bonn ELSA accelerator in a new setup, combining it with a forward
magnetic spectrometer.

\section{Analysis methods}

The measurement of meson production reactions off the neutron via quasi-free reactions 
off bound neutrons requires in general the detection of the participant neutron. This
can be done with different techniques, electromagnetic calorimeters like the Crystal
Ball (NaI scintillators), Crystal Barrel (CsI scintillators), or TAPS (BaF$_2$
scintillators) are also capable of neutron detection, although with limited detection 
efficiency (typically on the 15\% - 30\% level) and without any energy resolution.
Kinetic energies of the recoil neutrons can be measured via time-of-flight provided
the target - detector distance is large enough, which is the case for the TAPS forward 
walls at MAMI and ELSA and was the case for the time-of-flight forward walls of the GRAAL
experiment. But this covers only the forward range of the recoil neutrons, corresponding to
backward angles of the produced mesons. Alternatively, as discussed below, the kinetic energy
may be reconstructed from the reaction kinematics. Detection efficiencies for the neutrons 
must either be simulated with Monte Carlo programs or can be measured with reactions
where direction and kinetic energy of the neutrons can be reconstructed from the
reaction kinematics (e.g. from reactions like $\gamma p\rightarrow n\pi^+$ or
$\gamma p \rightarrow n\pi^0\pi^+$), where the mesons are detected. The simulation of the
neutron detection efficiencies is notoriously difficult and must always be cross-checked
with experimental results. Nevertheless, as discussed in detail in \cite{Jaegle_11b}
the use of the GCALOR program package \cite{Zeitnitz_01} provides rather satisfactory results.     

\begin{figure}[thb]
\centerline{
\resizebox{0.63\textwidth}{!}{%
  \includegraphics{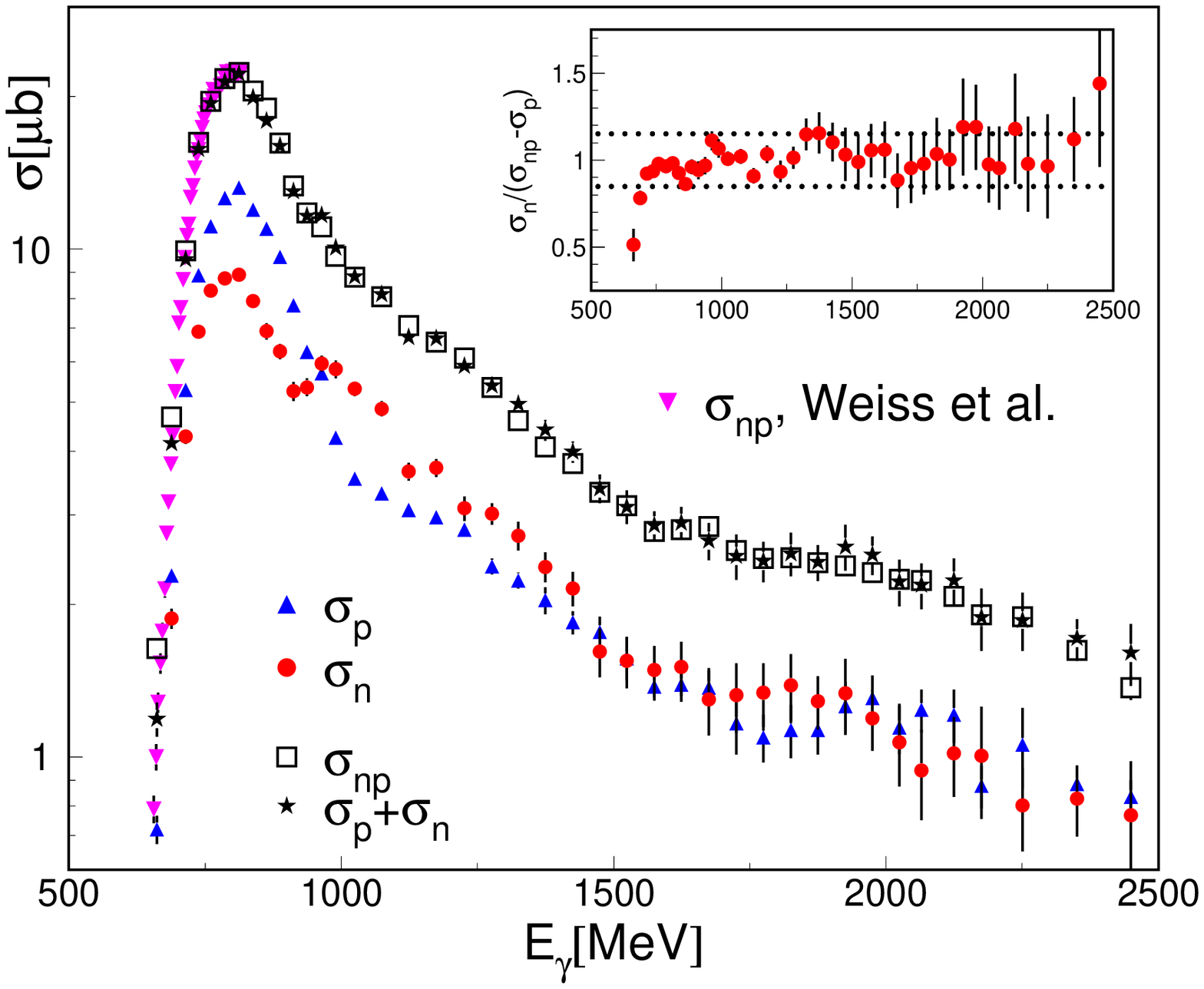}
}  
  \resizebox{0.37\textwidth}{!}{%
  \includegraphics{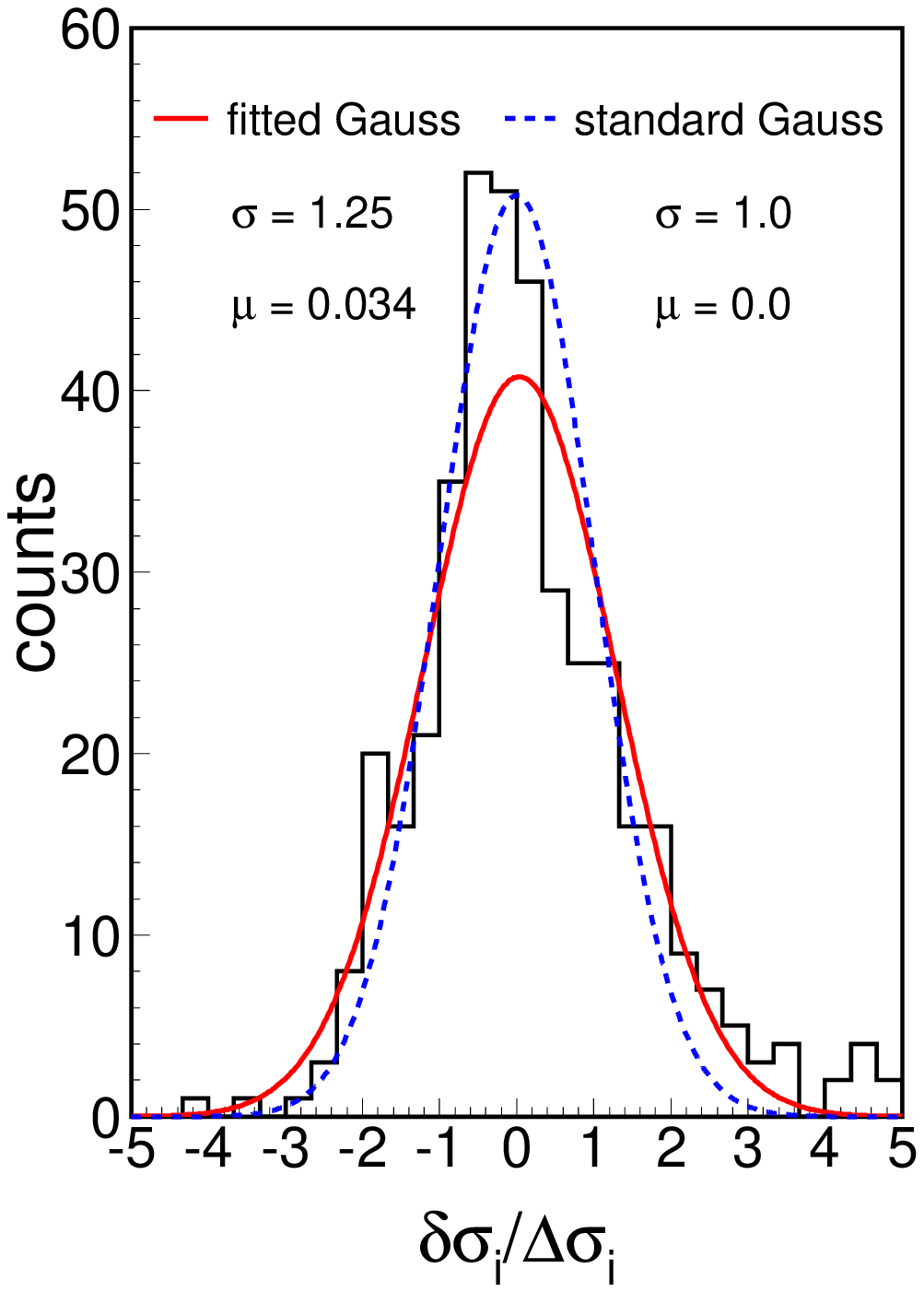}
}}
\caption{Comparison of cross sections for quasi-free $\eta$ photoproduction
\cite{Jaegle_11b}. 
Left hand side: total cross sections,
(Blue) upward triangles: quasi-free proton cross section $\sigma_p$, 
(red) dots: quasi-free neutron cross section $\sigma_n$, (black)
open squares: inclusive quasi-free cross section $\sigma_{np}$, 
(black) stars: $\sigma_n+\sigma_p$. 
Downward (magenta) triangles: inclusive quasi-free cross 
section from Weiss et al. \cite{Weiss_03}. Insert: ratio of neutron cross sections. 
Right hand side:
Distribution of deviations between $d\sigma_n/d\Omega$ and
$d\sigma'_n/d\Omega = d\sigma_{incl}/d\Omega-d\sigma_{p}/d\Omega$.
Solid (red) curve: fitted Gaussian distribution (width $\sigma$=1.25, 
mean $\mu$=0.034), dashed (blue) curve: standard Gauss: ($\sigma$=1, $\mu$=0).
}
\label{fig:eta_sum}       
\end{figure}

Fortunately, at least for some of the experiments, detection efficiency related problems
can be tracked in a simple way. The cross sections for meson production off the deuteron
(or other light nuclei) must obey the relation:
\begin{equation}
\sigma_{incl} = \sigma_p + \sigma_n + \sigma_{coh}\;\; ,
\label{eq:sumx}
\end{equation}   
where $\sigma_{incl}$ is the inclusive cross section without any condition for recoil baryons,
$\sigma_p$, $\sigma_n$ are the cross sections in coincidence with recoil protons and neutrons,
and $\sigma_{coh}$ is the coherent production cross section of the respective nucleus.
As an example we discuss the photoproduction of $\eta$-mesons off the deuteron. In this case,
(as in many others) the coherent production cross section $\sigma_{coh}$ is 
negligible, so that the quasi-free neutron and proton cross sections must add up to the
inclusive cross section. Only the detection efficiency of the $\eta$-meson enters into
the inclusive cross section (where events with and without recoil nucleons are excepted).
The completely different proton and neutron detection efficiencies enter into $\sigma_p$
and $\sigma_n$ so that Eq. \ref{eq:sumx} in general holds only when all detection efficiencies
are correct. This means that the neutron cross section can be measured in two different ways,
either as $\sigma_n$ by the detection of recoil neutrons or as difference 
$\sigma_{n'}=\sigma_{incl} - \sigma_p$. The result of such an analysis from the Crystal Barrel/TAPS
experiment \cite{Jaegle_11b} is shown in Fig. \ref{fig:eta_sum}. The left hand side of the
figure demonstrates the agreement of the total cross sections; the sum of quasi-free proton and
neutron cross section agrees with the independently constructed inclusive cross section,
and correspondingly the ratio of the neutron cross sections $\sigma_n$, $\sigma_{n'}$ is close 
to unity (see insert of Fig. \ref{fig:eta_sum}, left hand side).

As a further test (cf Fig. \ref{fig:eta_sum}, right hand side)
the distribution of the deviations $\delta\sigma_{i}$ normalized by the statistical
uncertainties $\Delta\sigma_{i}$    
\begin{equation}
\frac{\delta\sigma_i}{\Delta\sigma_i}\equiv 
\frac{d\sigma'_n/d\Omega-d\sigma_n/d\Omega}
{\sqrt{\Delta^2 (d\sigma'_n/d\Omega)+\Delta^2 (d\sigma_n/d\Omega)}}
\end{equation} 
for all data points (420 entries) of the angular distributions from production 
threshold to 2.5 GeV is compared to a Gaussian distribution. 
The fitted Gaussian distribution corresponds to a width of $\sigma=(1.25\pm0.10)$ 
and a mean of $\mu=(0.034\pm 0.110)$, fairly close to a standard Gaussian distribution.
In particular, the mean is not significantly different from zero so that no 
indication for a systematic deviation is indicated. This test provides stringent limitations 
for systematic uncertainties related to the detection of the recoil nucleons. In case of the
Crystal Barrel and Crystal Ball experiments, it has become a standard procedure. Some other
experiments, like for example LNS at Tohoku do not measure the recoil neutrons and rely
completely on $\sigma_{n'}$ constructed from inclusive deuteron and free (Fermi smeared) proton 
data (see e.g. \cite{Miyahara_07}). 

\begin{figure}[thb]
\centerline{
\resizebox{0.505\textwidth}{!}{%
  \includegraphics{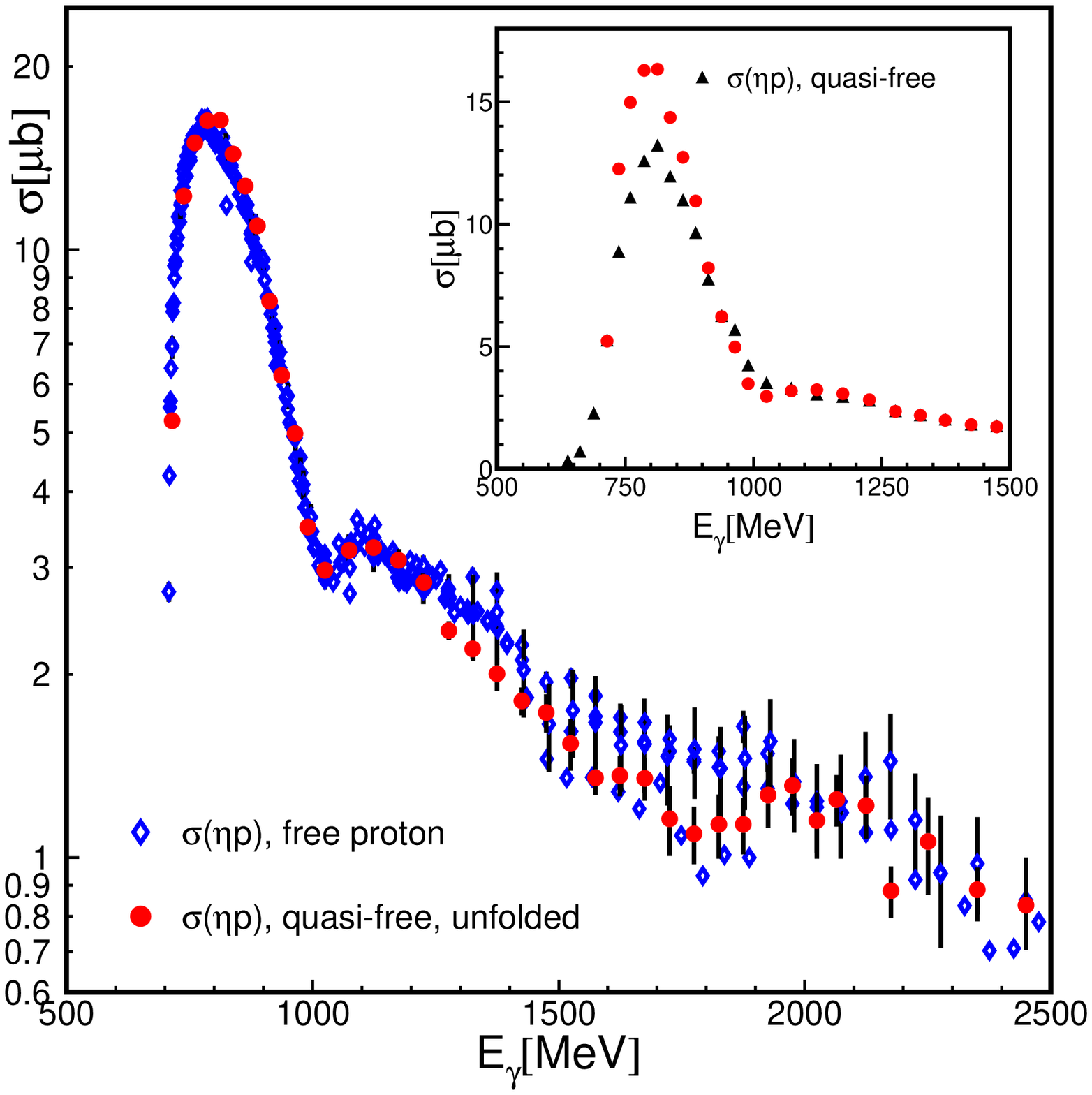}
}  
  \resizebox{0.495\textwidth}{!}{%
  \includegraphics{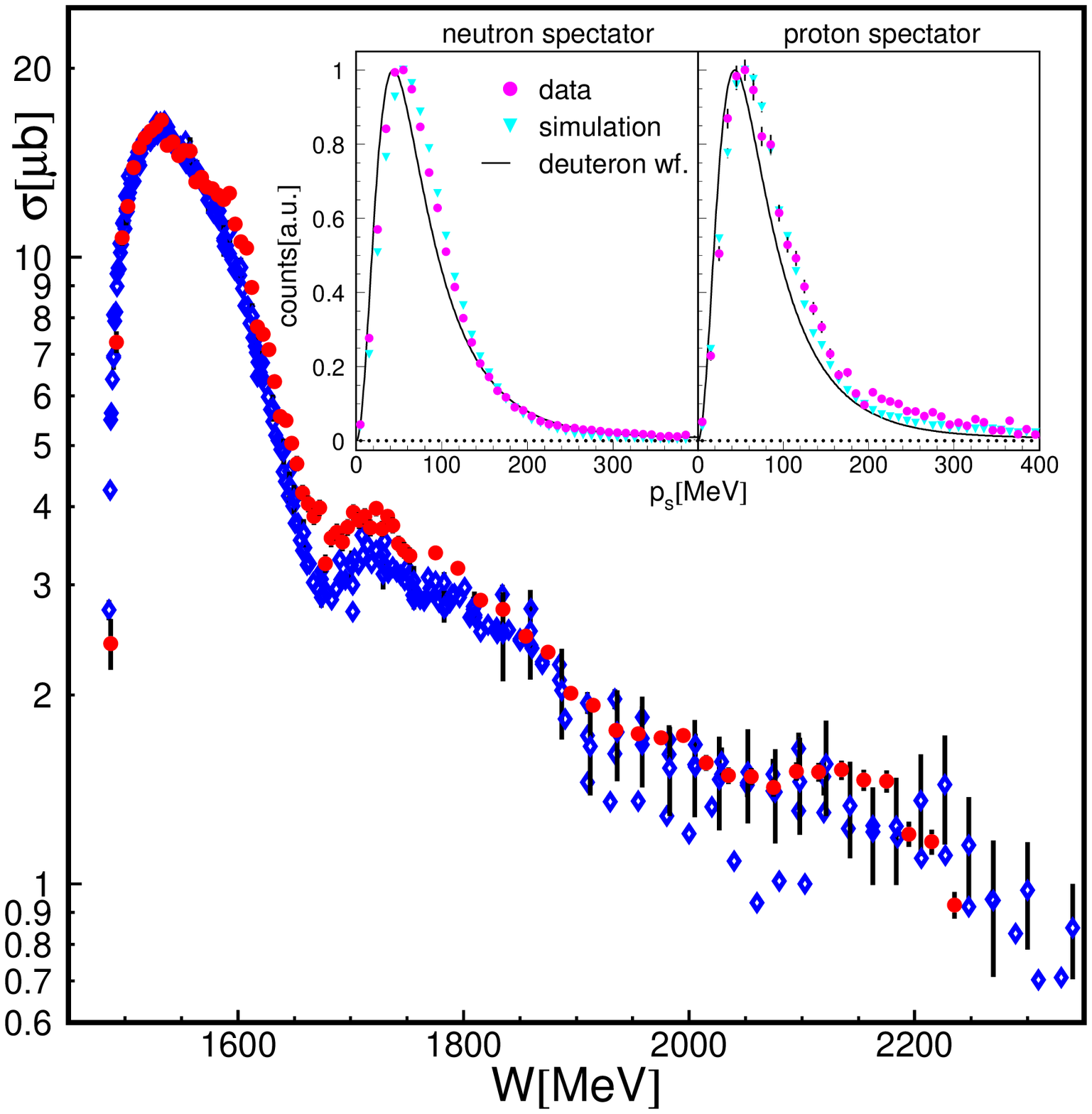}
}}
\caption{Comparison of free and quasi-free photoproduction of $\eta$-mesons off
the proton (see text). 
Left hand side: 
(Blue) diamonds world data base for
$\gamma p\rightarrow p\eta$ versus incident photon energy. 
(Red) dots: quasi-free $\gamma 'p'\rightarrow 'p'\eta$ reaction after applying 
correction factors for Fermi smearing. Insert: quasi-free data with (red dots)
and without (black triangles) correction for Fermi motion. 
Right hand side:
Free proton data (blue diamonds) versus $W$. (Red) dots: quasi-free data from 
kinematical reconstruction of $W$. Insert: reconstructed momenta of spectator
nucleons compared to momentum distributions from deuteron wave function
(neutron spectator corresponds to detection of recoil proton and vice versa). 
}
\label{fig:eta_compa}       
\end{figure}

The more important problem related to the use of the deuteron or other light nuclei as
target is the influence of nuclear effects, which may modify the elementary cross sections. 
The always present trivial effect from nuclear Fermi motion is quite well understood and, 
as discussed below, can be reliably accounted for. In the most simple approach, the 
results from model calculations can be folded with the momentum distribution of the
bound nucleons before comparison to the data. In a somewhat more involved procedure 
discussed below, in many cases it is also possible, to correct the effects in the data.

Much more difficult are nuclear FSI effects etc., which are not very well understood in 
models, and may have a very different influence on different reaction channels.
However, one can always cross-check the importance of such effects by a 
comparison of free proton data to the quasi-free proton data extracted from nuclear targets,
taking into account the effects of nuclear Fermi motion. As an example, we discuss again 
$\eta$-photoproduction, but other reactions, even with very small cross sections,
like Compton scattering \cite{Wissmann_99}, have been also studied in this way.  

A comparison of free and quasi-free proton data is shown in Fig. \ref{fig:eta_compa}.
The figure on the right hand side shows a comparison of the world data base for the
total cross section off the free proton to quasi-free proton data corrected for Fermi 
motion. In this case the correction was done by folding the free proton data with 
the Fermi motion, calculating as function of incident photon energy a correction
factor from the ratio of unfolded to folded cross section and applying these factors
to the quasi-free data (alternatively one could of course also compare directly the 
folded free-proton data to the quasi-free data). The agreement between the two data sets 
is quite good, demonstrating that nuclear effects beyond Fermi motion are not 
important in this case. The insert of the figure shows the effect of the Fermi smearing,
which is of course particularly important in the steep slopes of the cross sections. 

The above method does not assist in the extraction of an approximation of the free neutron 
cross section from quasi-free neutron data. In this case, the correction factors are unknown
and one could only try to compute them in an iterative way, but this will converge only 
as long as no narrow structures or steep slopes are involved. However, there is an
alternative possibility. Instead of analyzing the cross section as function of $E_{\gamma}$
(or $W$ computed from $E_{\gamma}$) one can directly compute the $W$ of the system decaying
into $N\eta$ from the four-vectors of the $\eta$ and the recoil nucleon. In this way,
the influence of Fermi motion is completely removed, at the price, that the experimental
resolution for the four-vectors enters into $W$. The result of such an analysis for
$\eta$-production off the proton is shown on the right hand side of Fig. \ref{fig:eta_compa}. 
Again, free proton data and reconstructed quasi-free data are in very good agreement.

Applying this method to neutron data involves the technical problem of measuring the neutron 
four-vectors with good resolution. The high granularity of the $4\pi$ calorimeters 
provides good resolution for the direction of the neutron. However, the kinetic energy 
cannot be reconstructed from the deposited energy and the use of time-of-flight methods
is restricted to the solid angle covered by the forward wall detectors 
(typically 20$^{\circ}$ of polar angle) since for the other detectors the 
time-of-flight path is too short. However, in the case of a deuteron target, a direct measurement
of the neutron kinetic energy in not necessary; it can be reconstructed from the reaction
kinematics. All kinematical observables of the initial state (photon and deuteron) are known.
In the final state, the mass of all particles is known and the three-momentum of the meson
is measured. When in addition the direction of the participant nucleon is measured, only
four kinematical observables are missing: the kinetic energy of the participant nucleon
and the three-momentum of the spectator nucleon. Since energy and momentum conservation
give four independent constraints, the system is completely determined. The insert of
Fig. \ref{fig:eta_compa} shows the momentum distributions of the undetected spectator
nucleons constructed this way. They are compared to the momentum distribution predicted
by the deuteron wave function, which, after taking care of detector resolution effects
with a Monte Carlo simulation, agrees very well with the data. With this method one can in 
addition, via cuts on the spectator momentum, remove those events which are not close to
quasi-free reaction kinematics and might suffer from stronger off-shell effects.  

\section{Final and preliminary results}

In this section we will summarize new results from the ELSA and MAMI 
experiments, which have just been published, are in press, or in an almost final 
state of analysis.   

\subsection{Meson photoproduction off the neutron - the $\eta '$ meson}

Photoproduction of $\eta '$ mesons has recently attracted quite some interest.
Due to its large mass of 958 MeV the production threshold on the free nucleon lies at 
1447 MeV, corresponding to $W\approx$1896 MeV. This is a range where many hitherto unknown 
N$^{\star}$ resonances are predicted by quark models. Like the $\eta$, the isoscalar $\eta '$ 
can only be emitted by N$^{\star}$ states and close to threshold only a few partial waves 
contribute. 

\begin{figure}[htb]
\centerline{
\resizebox{0.5\textwidth}{!}{%
  \includegraphics{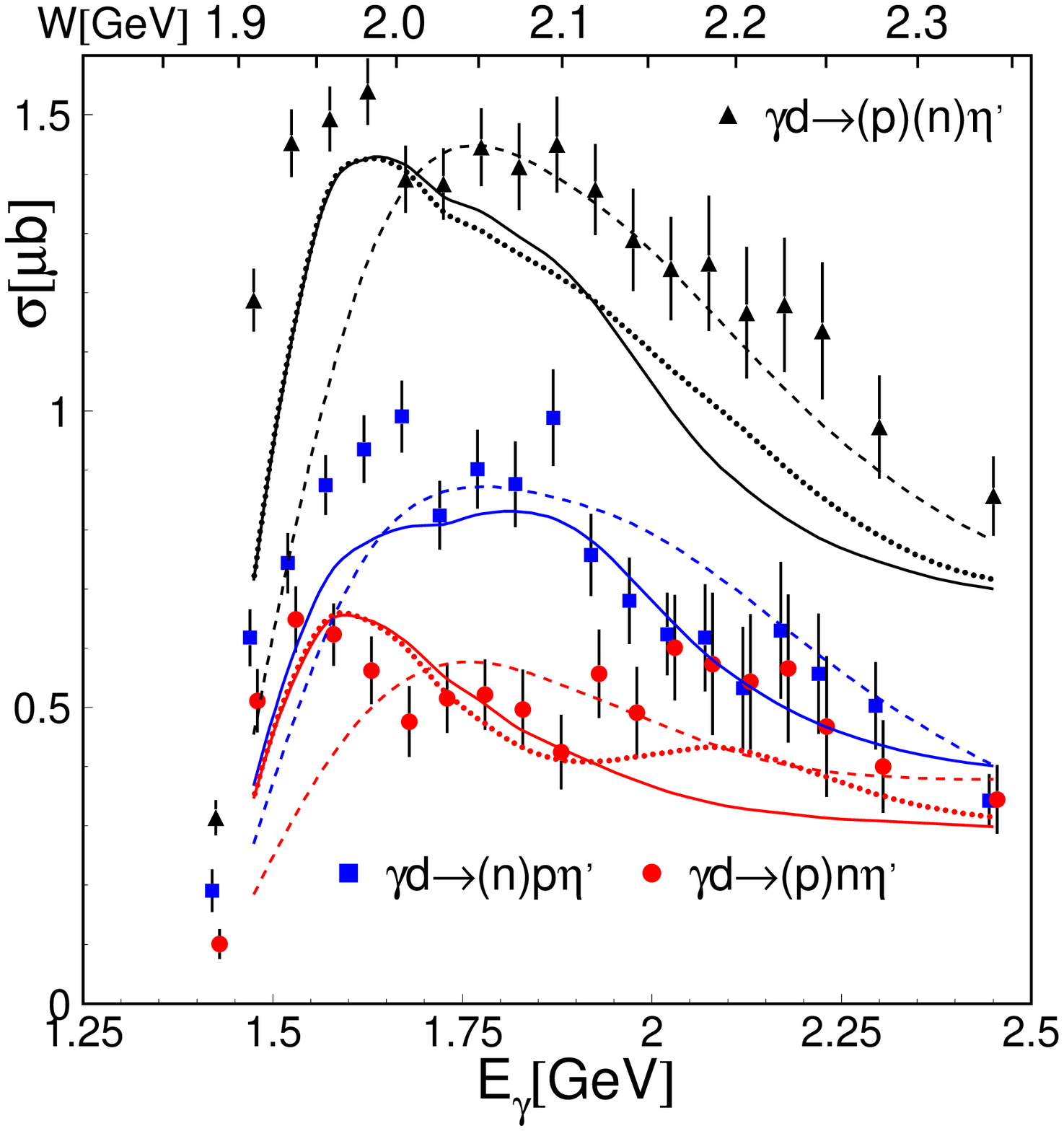}
}  
  \resizebox{0.5\textwidth}{!}{%
  \includegraphics{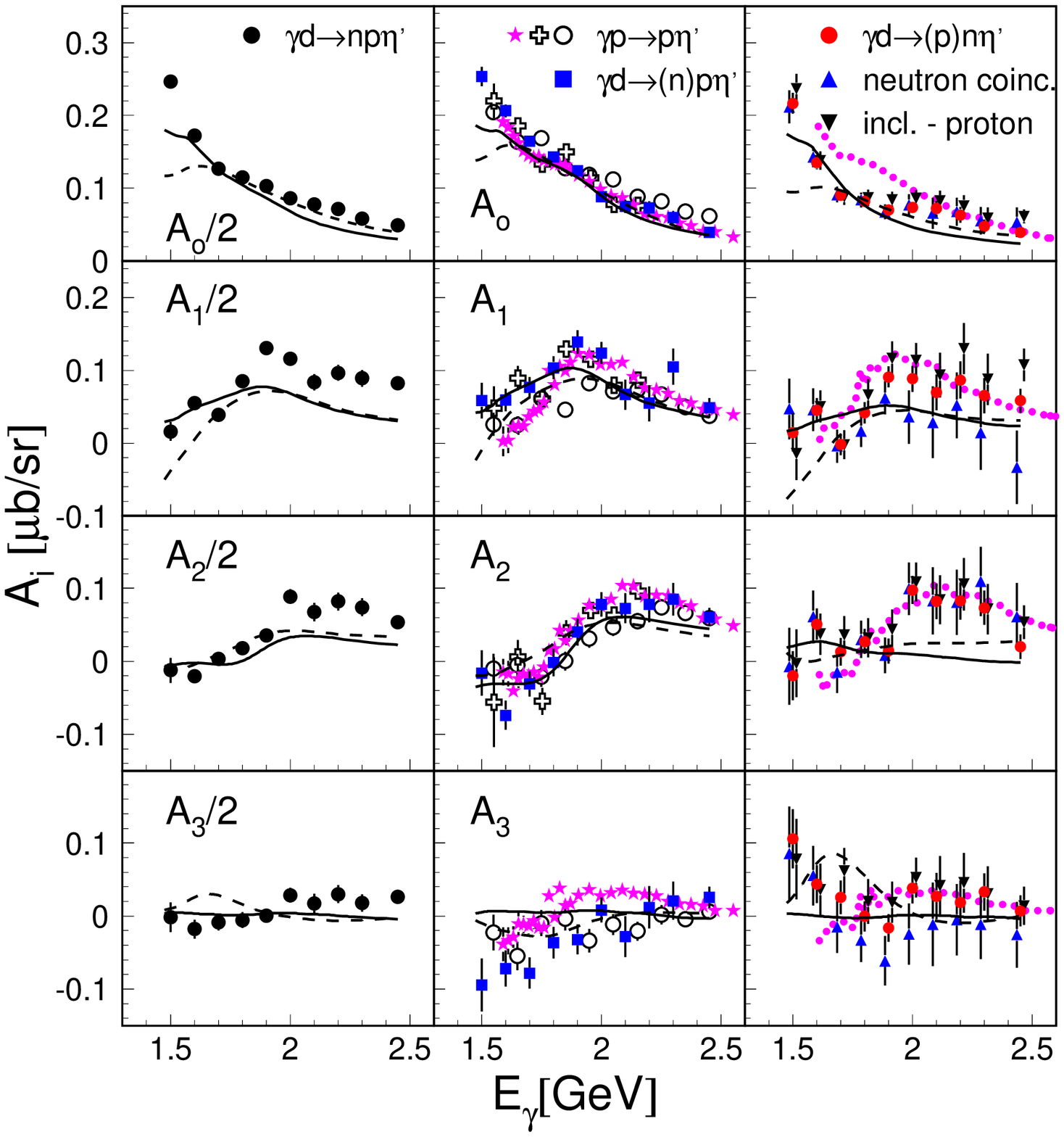}
}}
\caption{Left hand side:
Total cross section for $\eta '$ production off the deuteron. \cite{Jaegle_11}. 
Curves: fits with reaction models model, solid, dotted: different solutions 
from \cite{Nakayama_06}, dashed: $\eta '$-MAID \cite{Chiang_03}.  
Right hand side:
Coefficients of the Legendre Polynomials for the fitted angular distributions \cite{Jaegle_11}. 
Left hand column: inclusive reaction scaled down by factor of 2.
Center column (proton targets): quasi-free data (blue squares);
free proton data: open crosses \cite{Dugger_06}, open circles \cite{Crede_09}, 
(magenta) stars \cite{Williams_09}. 
Right hand column (quasi-free neutron data): (blue) upward triangles
from neutron coincidence, (black) downward triangles from difference of 
inclusive and proton data, (red) circles from averaged data. 
In all plots solid lines: NH model, dashed lines: $\eta '$-MAID;
for neutron: (magenta) dotted lines CLAS proton data.
}
\label{fig:igal_etapqf}       
\end{figure}

Rather precise data off the free
proton have been measured by the CLAS experiment \cite{Dugger_06,Williams_09} and at ELSA
\cite{Crede_09}. Unfortunately, it turned out \cite{Nakayama_06} that in this case 
even close to threshold differential cross section data alone cannot fix the model parameters.
Although most analyses agree on a strong contribution from a S$_{11}$ resonance,  
the position, width, and coupling strength of this state are not well constrained
and different solutions are possible for the contribution of further resonances. Part of the
problem results from a fairly large $t$-channel background contribution. Nevertheless,
recently also the first measurement off the quasi-free neutron was reported from ELSA
\cite{Jaegle_11}. The total cross section and the coefficients of Legendre polynomials fitted
to the angular distributions are compared to proton data and model results in Fig.
\ref{fig:igal_etapqf}. The results for proton and neutron are similar but indicate clearly
different contributions in particular in the intermediate energy range. Also for the neutron
final state, no unique solution could be obtained \cite{Jaegle_11}. The measurement of 
polarization observables in the near future will help to solve this problem. The measurement 
of the quasi-free reaction off the neutron has shown that, in this channel, no significant
nuclear effects will complicate such experiments. The quasi-free proton data are in excellent
agreement with free proton data, even effects from Fermi motion are not significant at the
current level of statistical precision \cite{Jaegle_11}.

\subsection{Meson photoproduction off the neutron - single $\pi^0$ production}

\begin{figure}[thb]
\centerline{
\resizebox{0.5\textwidth}{!}{%
  \includegraphics{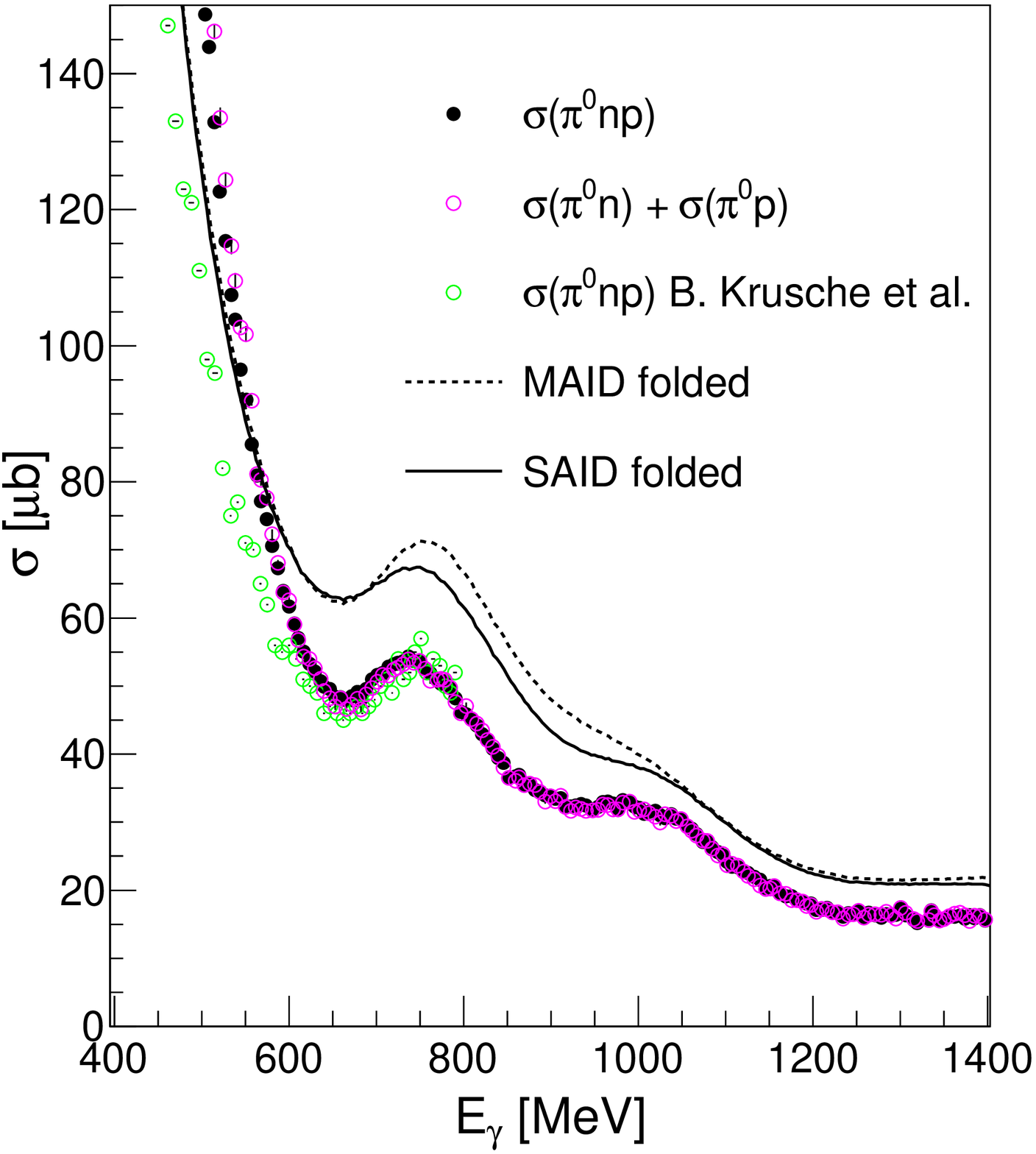}
}  
  \resizebox{0.5\textwidth}{!}{%
  \includegraphics{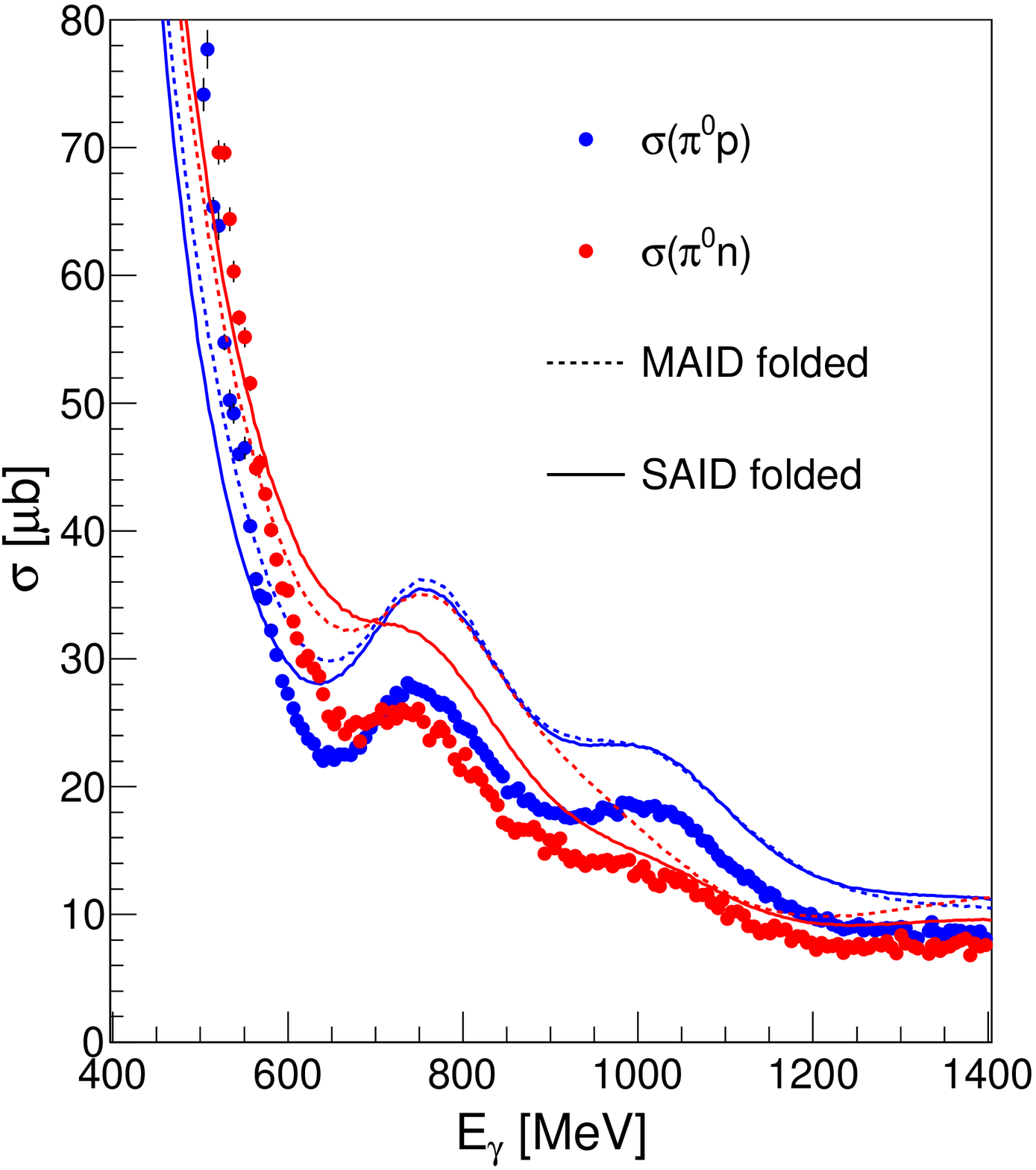}
}}
\caption{Preliminary results for quasi-free photoproduction of $\pi^0$ mesons off the
deuteron in the second and third resonance region \cite{Dieterle_11}. 
Left hand side: 
inclusive $\gamma d\rightarrow np\pi^0$ data compared to sum of
exclusive cross sections and results from MAID (dashed line) \cite{Drechsel_99b}
and SAID (solid line) \cite{Said}. Both model results are for the incoherent sum of 
proton/neutron cross sections folded with Fermi motion.
Right hand side:
Same for exclusive quasi-free $\gamma p\rightarrow p\pi^0$ and $\gamma n\rightarrow n\pi^0$
reactions. 
}
\label{fig:manuel}       
\end{figure}

The situation is completely different for the single $\pi^0$ channel, for which preliminary 
results \cite{Dieterle_11} are shown in Fig. \ref{fig:manuel}. We had already discussed in the 
introduction ( sec. \ref{sec:intro_res}) that quasi-free inclusive data off the deuteron are 
not in agreement with the results from the MAID \cite{Drechsel_99b} or SAID \cite{Said} analyses,
although the model results themselves are in fairly good agreement (as expected since they 
have been fitted to the same relatively large body of data available for the 
$\gamma p\rightarrow p\pi^0$, $\gamma p\rightarrow n\pi^+$, and $\gamma n\rightarrow p\pi^-$
reactions).

The left hand side of the figure demonstrates that the new preliminary quasi-free
data for $\gamma p\rightarrow p\pi^0$ and $\gamma n\rightarrow n\pi^0$
sum up correctly to the inclusive cross section (the coherent part must still be separated, 
but is small at photon energies above 500 MeV) and that the inclusive data agree with the 
previous measurement (within systematic uncertainties). However, they clearly disagree with the
MAID and SAID results (sum of proton and neutron cross section folded with Fermi motion). 
At the right hand side of the figure quasi-free proton and neutron cross sections are compared
to the MAID and SAID results. For this reaction, throughout the second and third resonance
region neither the proton nor the neutron data agree with the model results. Their energy
dependence is similar, but on an absolute scale the data are lower than expected by roughly
25\%. A similar result has been reported from the LNS experiment \cite{Shimizu_09}.
The model results agree very well with free proton data, to which they have been
fitted. This means that quasi-free and free proton data disagree by $\approx$25\% 
(and probably a similar effect occurs for the neutron). Interpretation of the quasi-free data
thus will require detailed studies of nuclear effects on this reaction. Assuming that these
effects are similar a comparison of the neutron/proton ratio from models and data might
give results which are easier to interpret. The total cross sections already indicate 
different resonance contributions to the proton and neutron excitation.   

\subsection{Meson photoproduction off the neutron - the $\eta$-meson}

Photoproduction of $\eta$ mesons off the free proton has become a very well studied reaction
over the last 15 years
\cite{Krusche_95,Dugger_02,Williams_09,Crede_09,Ajaka_98,Bock_98,Armstrong_99,Thompson_01,Renard_02,Crede_05,Nakabayashi_06,Bartholomy_07,Elsner_07,Denizli_07,Sumihama_09,McNicoll_10}.
It is particularly attractive due to the strong dominance of the S$_{11}$(1535) resonance and 
(at least at moderate incident photon energies) comparably small contributions of non-resonant
background terms. The main original motivation for the study of the 
$\gamma 'n'\rightarrow 'n'\eta$ reaction was related to the isospin structure of the 
S$_{11}$(1535) resonance (see \cite{Krusche_03} for a summary) and model predictions
\cite{Chiang_02} for a strong contribution of the D$_{15}$(1675) resonance to the neutron
excitations. However, when these experiments reached incident photon energies around 1 GeV
an unexpected result was found. Measurements at GRAAL \cite{Kuznetsov_07}, Tohoku-LNS 
\cite{Miyahara_07}, ELSA \cite{Jaegle_08,Jaegle_11b}, and MAMI \cite{Werthmueller_09} revealed 
a prominent peak-like structure around $W\approx$1.7 GeV, which has no counterpart for the proton.

\begin{figure}[thb]
\centerline{
\resizebox{0.37\textwidth}{!}{%
  \includegraphics{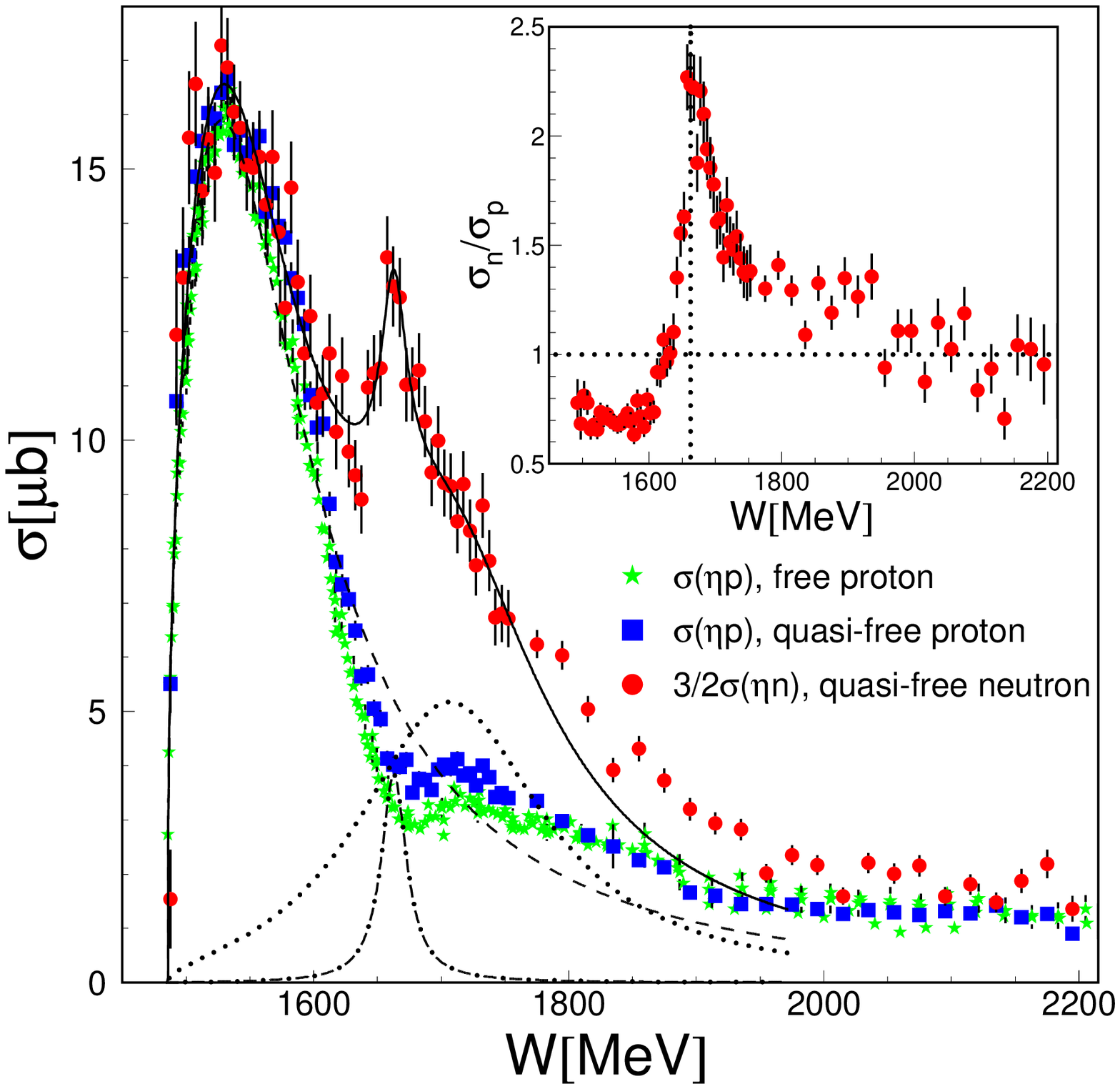}
}  
  \resizebox{0.63\textwidth}{!}{%
  \includegraphics{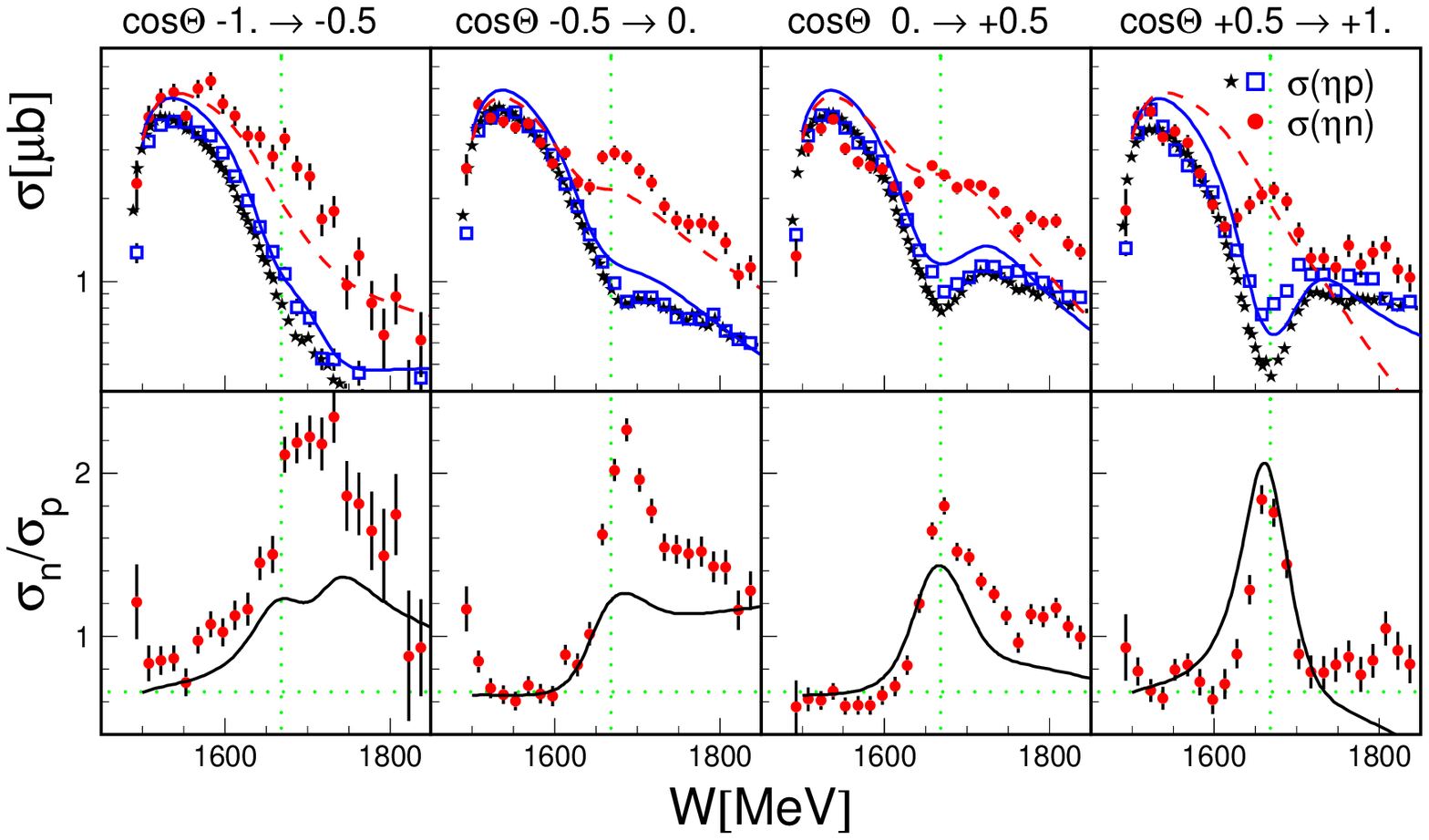}
}}
\caption{Quasi-free photoproduction of $\eta$-mesons off the proton and neutron \cite{Jaegle_11b}.
Left hand side: 
Total cross sections as function of final state invariant mass $W$
(Red) dots: quasi-free neutron, (blue) squares: quasi-free proton, 
(green) stars: free proton data.
Insert: ratio of quasi-free neutron - proton data. 
All curves for neutron data; dashed: fitted S$_{11}$ line shape, dotted:
broad Breit-Wigner resonance, dash-dotted: narrow Breit-Wigner, solid:
sum of all. 
Right hand side:
First row: excitation functions for different bins of $\eta$ cm polar angle.
(Blue) open squares: quasi-free proton data, (black) stars: free proton data 
from \cite{McNicoll_10}, (red) dots: quasi-free neutron data scaled up by 3/2. 
(Blue) solid lines: $\eta$-MAID \cite{Chiang_02} for the proton target, 
(red) dashed lines: $\eta$-MAID for the neutron target. 
Second row: ratio of neutron and proton cross section for data and $\eta$-MAID.
Vertical dotted lines: position of narrow peak in neutron data,
horizontal dotted lines: $\sigma_n/\sigma_p$=2/3.
}
\label{fig:igal_eta}       
\end{figure}

The result for the total cross section from a new analysis of the 
ELSA data \cite{Jaegle_11b}, using the kinematical reconstruction discussed above to eliminate 
the effects from Fermi motion, is shown at the right hand side of Fig. \ref{fig:igal_eta}. A
phenomenological fit of the data with the resonance shape of the S$_{11}$(1535) resonance,
a conventional Breit-Wigner that effectively subsumes the contributions from other broad 
resonances and non-resonant backgrounds, and an additional narrow Breit-Wigner curve results 
in a width (FWHM) of the narrow structure of only (25$\pm$12) MeV; close to the experimental
resolution. Also all other experiments have reported widths lower than $\approx$40 MeV.

The nature of this structure is not yet understood. Different scenarios involving contributions
from specific resonances as well as interference patterns in the same partial wave have been 
discussed in the literature. Fix, Tiator, and Polyakov \cite{Fix_07} have investigated whether
the data could be consistent with the excitation of a narrow P$_{11}$-state. This work was 
motivated by the idea that the P$_{11}$-state of the proposed anti-decuplet of pentaquark 
states should be relatively narrow (width of the order of 10 MeV), have a strong
electromagnetic coupling to the neutron, and a large $\eta N$ decay branching ratio 
\cite{Polyakov_03,Arndt_04}. Due to Fermi smearing, they found solutions not only for 
narrow but also conventionally broad states. Similarly, an analysis performed in the 
framework of the Bonn-Gatchina model (BoGa) \cite{Anisovich_09} can reproduce the data 
by either adding a `conventionally' broad P$_{11}$ resonance, a very narrow P$_{11}$ state, 
or even by a careful adjustment of the interference pattern for the $s$-wave amplitudes. 
Shklyar, Lenske, and Mosel \cite{Shklyar_07} found solutions just from coupled-channel 
effects in the S$_{11}$ - P$_{11}$ sector, without introducing any additional resonances. 
A similar result from a coupled-channel K-matrix approach was presented by Shyam and Scholten
\cite{Shyam_08}. Finally, D\"oring and Nakayama \cite{Doering_10}, using an $s$-wave coupled 
channel model, find a `dip-bump' structure in the neutron cross section related to the 
opening strangeness thresholds of $K\Lambda$ and $K\Sigma$ photoproduction around 900 MeV 
and 1050 MeV. 

The scenarios with a contribution from one single conventionally broad resonance
without intricate interference effects have been ruled out in the meantime by the data shown
in Fig. \ref{fig:igal_eta}. Also shown at the right hand side of Fig. \ref{fig:igal_eta}
are excitation functions and neutron/proton ratios for four different ranges of the $\eta$
cm polar angle. These data show that the structure is particularly strong for forward
angles. A comparison to the most precise, recent free-proton data \cite{McNicoll_10} reveals, 
that the peak-like
structure for the neutron corresponds to a dip-like structure at the same $W$ and with comparable
width in the proton excitation functions. The similarity in the neutron-proton ratio between
the data and the MAID prediction \cite{Chiang_02} results only from this proton dip, the neutron
peak is not seen in the model results. This also causes some doubts, as to whether the proton dip
is interpreted by the correct mechanism in the model.  

\begin{figure}[thb]
\centerline{
\resizebox{0.5\textwidth}{!}{%
  \includegraphics{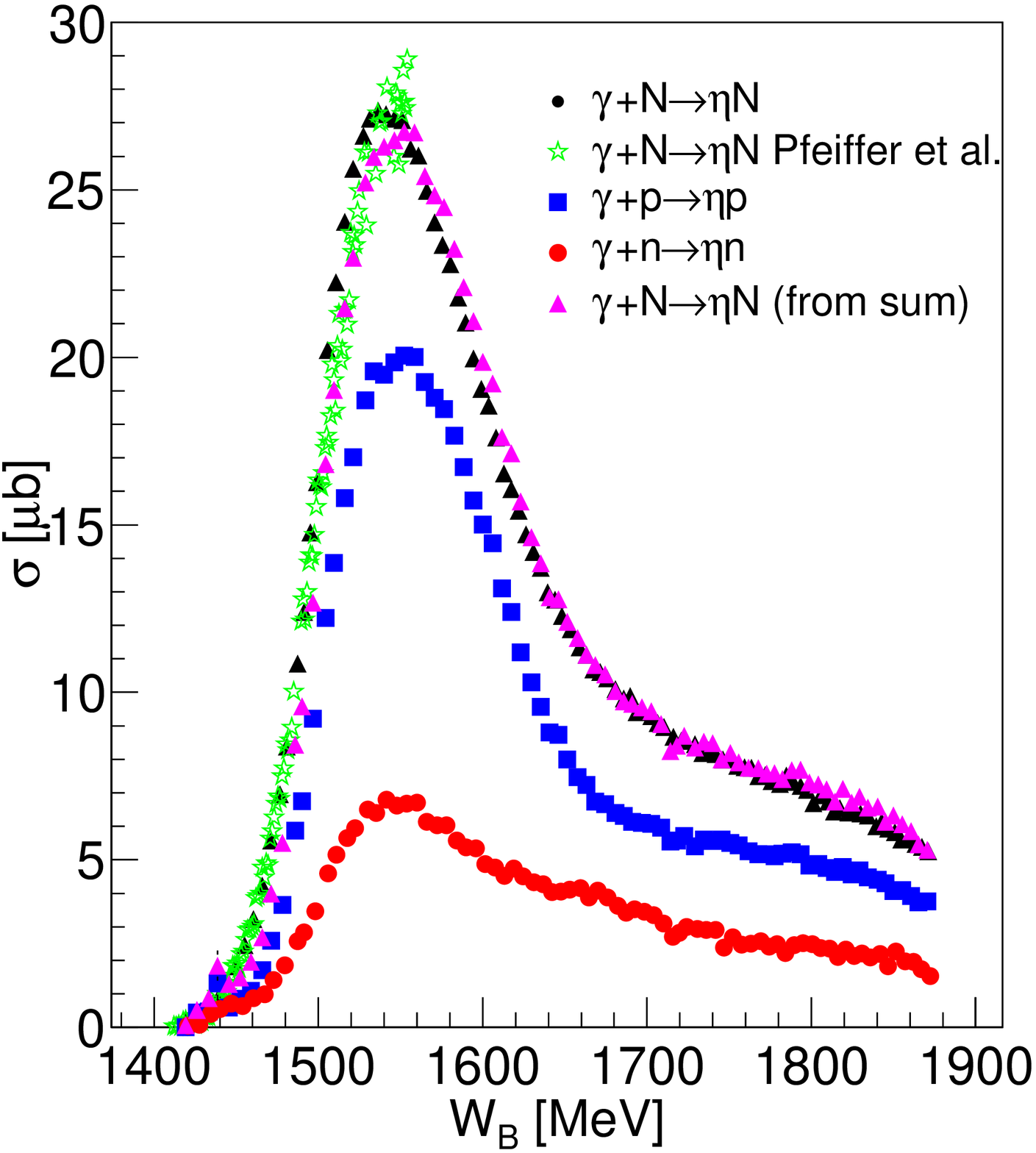}
}  
  \resizebox{0.5\textwidth}{!}{%
  \includegraphics{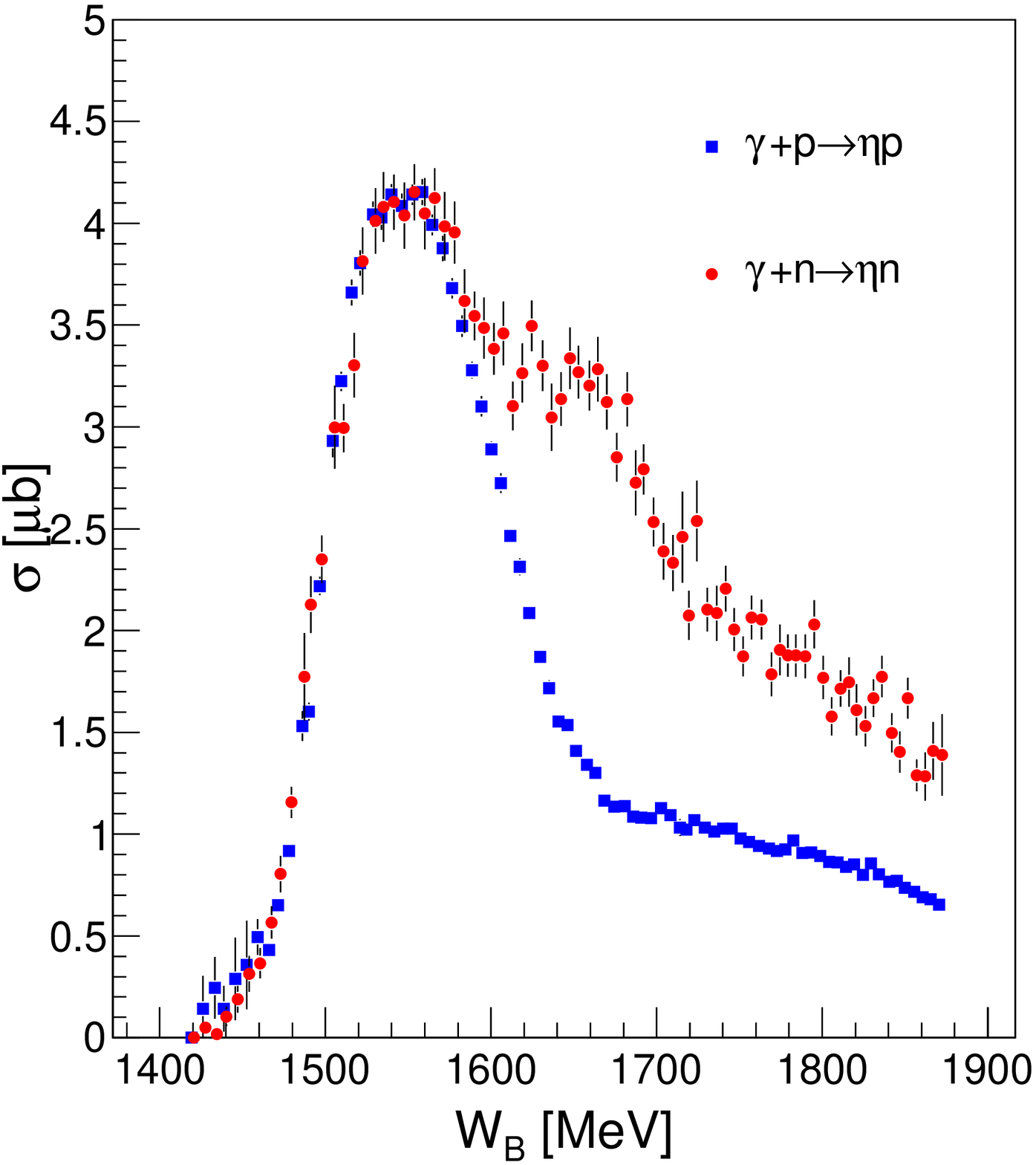}
}} 
\centerline{
  \resizebox{0.5\textwidth}{!}{%
  \includegraphics{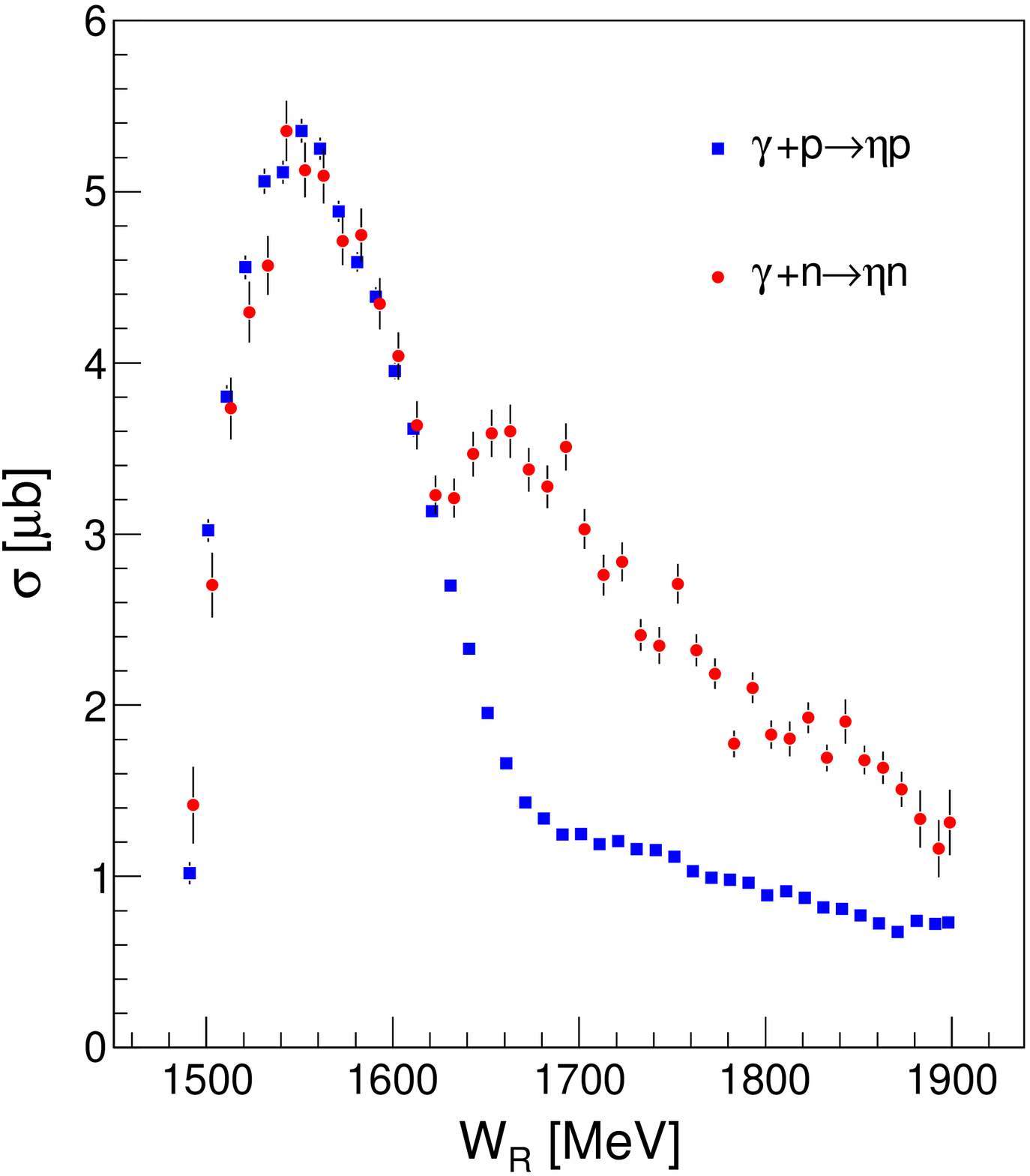}
}  
  \resizebox{0.5\textwidth}{!}{%
  \includegraphics{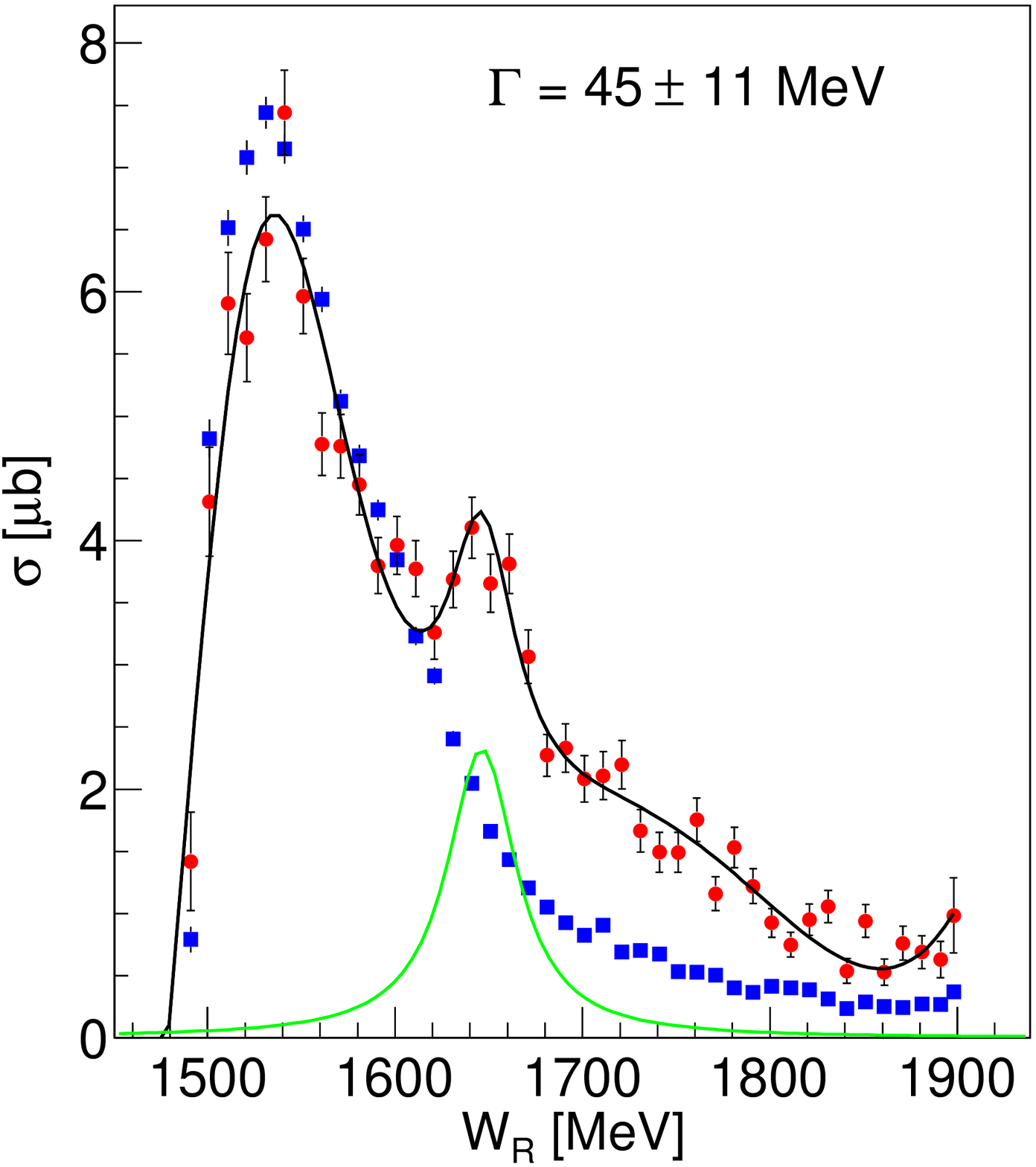}    
}}
\caption{Preliminary results for quasi-free $\eta$-photoproduction off $^3$He \cite{Witthauer_10}. 
Upper left: comparison of quasi-free cross sections off protons and neutrons with
inclusive cross section. Previous data for inclusive reaction from Pfeiffer et al.
\cite{Pfeiffer_04}. 
Upper right: quasi-free proton and neutron cross sections (normalized in S$_{11}$ peak)
as function of $W$ from $E_{\gamma}$ ($W_B$).
Lower left: same as function of $W$ reconstructed from final state kinematics ($W_{R}$)
(kinetic energy of neutron from reconstructed kinematics).
Lower right: same as function of $W_R$ for kinetic energy of neutron from time-of-flight
measurement (only for $\eta$ backward angles). 
}
\label{fig:lilian}       
\end{figure}

In the meantime, an experiment at MAMI \cite{Witthauer_10} has identified the same structure
in quasi-free photoproduction off $^3$He, i.e. in a different nuclear environment. This is shown
in Fig. \ref{fig:lilian}. The kinematic reconstruction of the neutron works less well
for the helium target nucleus; one has to make the approximation that the pair of spectator
nucleons in the four-body final state has no relative momentum. Due to this approximation, the
resolution for the $W$ excitation function is less good. However, the peak is still clearly 
visible and a more narrow width (close to the experimental resolution) can be recovered for the 
angular range where the neutron kinetic energy can be directly determined by the time-of-flight 
method (Fig. \ref{fig:lilian}, lower right corner).

Finally, it should be mentioned that recently Kuznetsov and co-workers \cite{Kuznetsov_11} 
have reported from the GRAAL experiment a similar structure in Compton scattering off the
quasi-free neutron. If this structure is confirmed, it will impose significantly more stringent 
conditions on explanations with complicated interference patterns etc. which are not very likely 
to be identical for such different reaction channels.  

Up to now, two important questions are open for this structure: what is its width on the free neutron
and in which partial wave amplitude(s) does it appear. So far we have only upper limits for the
widths. New data from MAMI \cite{Werthmueller_09} for quasi-free production off neutrons bound 
in the deuteron, which are still under analysis, have much better statistical quality than the
previous experiments. This may help to further constrain the width by more detailed comparisons to
the experimental resolution. The identification of the relevant amplitudes, and thus the quantum
numbers of a prospective nucleon resonance, requires the measurement of polarization observables.
This program has began in parallel in Mainz and Bonn, using deuterated frozen-spin butanol
targets. First data have been taken in Bonn for the helicity dependence of the cross section 
(observable $E$, circularly polarized beam on longitudinally polarized target), and in Mainz for 
the target asymmetry (transversely polarized target), and the double polarization observable $F$ 
(transversely polarized target, circularly polarized beam).

\subsection{Meson photoproduction off the neutron - double pion production}

As already discussed in the introduction, photoproduction of pion pairs may allow the investigation of
nucleon resonances which have only small branching ratios for decays to the nucleon ground state but
decay dominantly to intermediate N$^{\star}$ or $\Delta^{\star}$ resonances. One expects  
that this scenario becomes more important the higher the excitation energy of the states. At the 
same time, for selected isospin channels, double pion production also gives access to $\rho$
and $\sigma$ meson decay channels. A large body of data has become available in particular for 
reactions off the free proton. For incident photon energies up to the second resonance region 
total cross sections and invariant-mass distributions of the $\pi\pi$- and the $\pi N$-pairs have 
been measured with the DAPHNE and TAPS detectors at the MAMI accelerator in Mainz 
\cite{Braghieri_95,Haerter_97,Zabrodin_97,Zabrodin_99,Wolf_00,Langgaertner_01,Kotulla_04,Sarantsev_08},
and at higher incident photon energies at GRAAL in Grenoble (also the linearly polarized beam asymmetry)
\cite{Assafiri_03} and at ELSA in Bonn \cite{Sarantsev_08,Thoma_08}. Also the first double polarization 
observables ($E$, longitudinally polarized target, circularly polarized beam) have been measured
in the context of the Gerasimov-Drell-Hearn project \cite{Ahrens_03,Ahrens_05,Ahrens_07}.
However, in spite of all these efforts even for the proton target in the low energy region the
interpretation of the data is still strongly controversial. All models agree that the production 
of neutral pions involves smaller contributions from non-resonant terms like pion-pole diagrams
or terms of the $\Delta$-Kroll Rudermann type. But the reaction models do not even agree for the
dominant contributions to the $\pi^0\pi^0$ channel. In the second resonance region, the model of 
the Valencia group \cite{Gomez_96,Nacher_01,Nacher_02} is dominated by the  
D$_{13}(1520)\rightarrow \Delta\pi^0\rightarrow p\pi^0\pi^0$ reaction chain, Laget and co-workers 
\cite{Assafiri_03} find a much more prominent contribution from the
P$_{11}(1440)\rightarrow N\sigma$ decay,and the recent Bonn-Gatchina analysis \cite{Sarantsev_08}
claims a large contribution from the $D_{33}(1700)$ state, which is almost negligible in the other 
models. 

\begin{figure}[thb]
\centerline{
\resizebox{0.50\textwidth}{!}{%
  \includegraphics{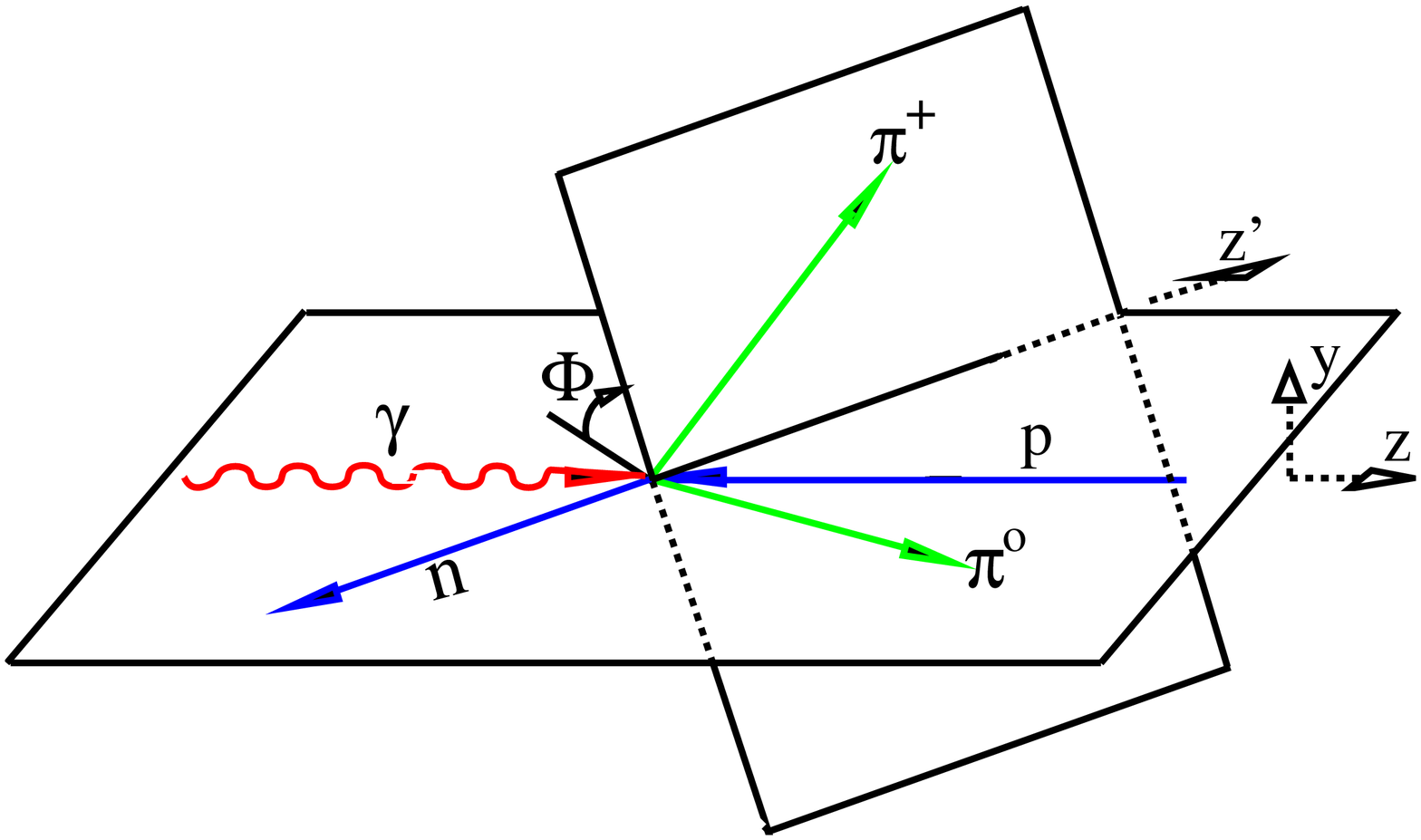}
}
\resizebox{0.50\textwidth}{!}{%
  \includegraphics{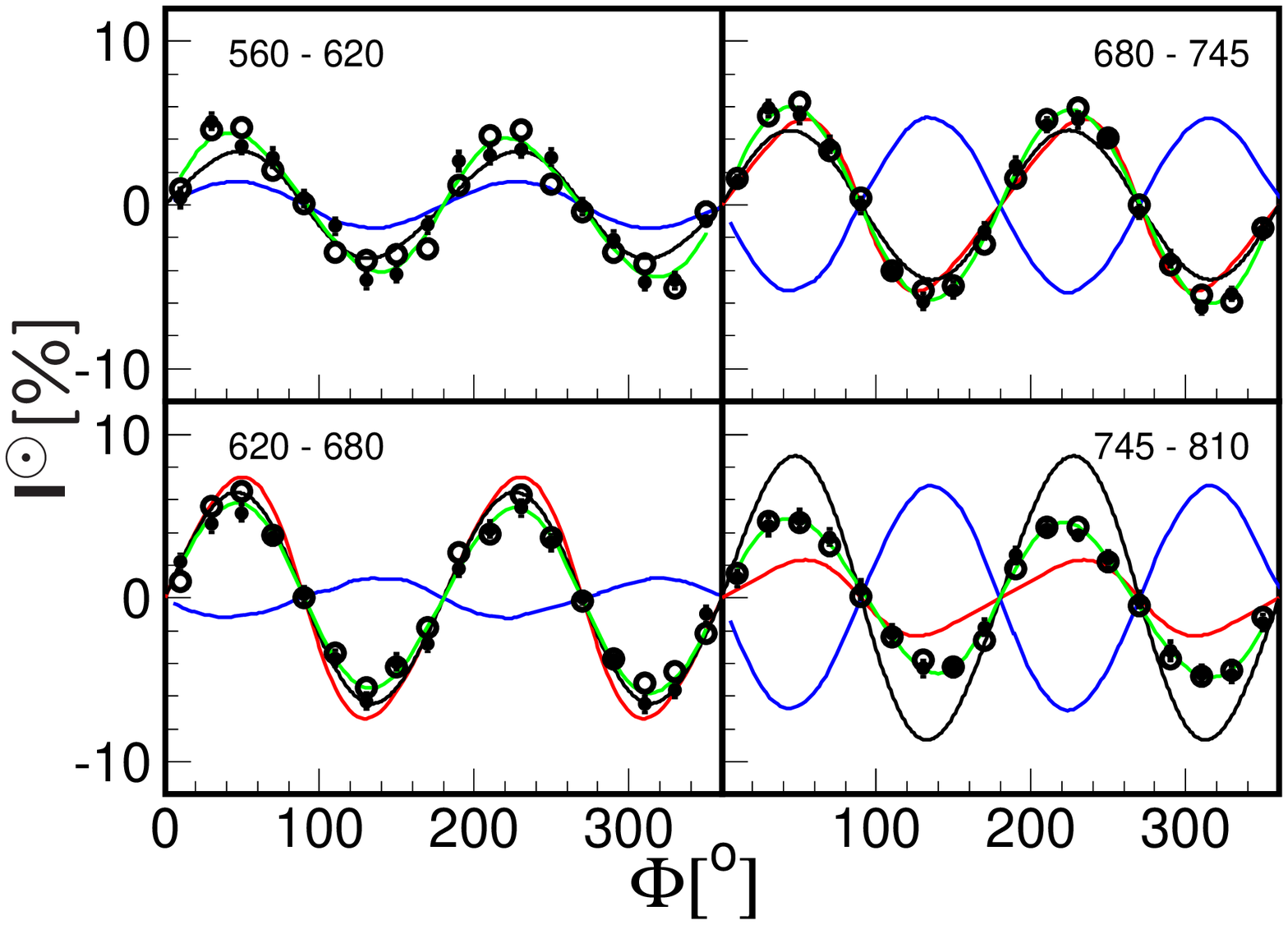}  
}}
\caption{Beam-helicity asymmetry for the double pion production off the nucleon
\cite{Krambrich_09}. 
Left hand side: definition of the angle $\Phi$.
Right hand side: results for the free proton compared to models.
Curves: green: fit to data, black: Bonn-Gatchina model \cite{Thoma_08,Sarantsev_08},
red: Fix and Arenh\"ovel \cite{Fix_05}, blue:  Roca \cite{Roca_05}.
}
\label{fig:markus_def}       
\end{figure}

Several model predictions \cite{Nacher_02,Fix_05,Roberts_05,Roca_05} indicated that polarization 
observables are extremely sensitive to the reaction mechanisms, but it came as a surprise when the 
first measurement of the beam helicity asymmetry of the $\gamma p\rightarrow p\pi^+\pi^-$ 
reaction at the CLAS facility \cite{Strauch_05} produced results that could not be reproduced by 
any reaction models. This observable can be measured for three-body final states with a circularly
polarized photon beam on an unpolarized target and is defined by:  
\begin{equation}
\label{eq:asym}
I^{\odot}(\Phi)=\frac{1}{P_{\gamma}}
	        \frac{d\sigma^{+}-d\sigma^{-}}{d\sigma^{+}+d\sigma^{-}}
	       =\frac{1}{P_{\gamma}}
                \frac{N^{+}-N^{-}}{N^{+}+N^{-}}\;\;,
\label{eq:circ}		
\end{equation}
where $d\sigma^{\pm}$ is the differential cross section for each of the two
photon helicity states, and $P_{\gamma}$ is the degree of circular polarization 
of the photons. The definition of the angle $\Phi$ is shown in Fig. \ref{fig:markus_def}.

In the meantime \cite{Krambrich_09} it has been found that, also for the $n\pi^0\pi^+$ final 
state all reaction models fail. For the $p\pi^0\pi^0$ final state at least some analyses 
(cf. Fig. \ref{fig:markus_def}) are close to the data, but also in this case the Valencia model 
\cite{Roca_05}, which is in good agreement with all other observables, is out of phase with 
the data.  

The data base for reactions off the neutron is still much more sparse. At low incident photon energies
total cross section data, invariant mass distributions, and the helicity dependence of the cross 
section for $\gamma n\rightarrow p\pi^-$are available from Mainz \cite{Zabrodin_99,Ahrens_11}. 
Results for the $n\pi^0\pi^0$ final state have been reported from Mainz \cite{Kleber_00} and GRAAL
\cite{Ajaka_07}. 

New, precise results for total cross sections and invariant mass distributions
for the quasi-free $\gamma n\rightarrow n\pi^0\pi^0$ reaction have been measured by the CBELSA 
experiment and are in preparation for publication \cite{Jaegle_11c}. Data for the same channel and
for the $p\pi^-\pi^0$ final state have been also measured with the Crystal Ball/TAPS setup at 
MAMI up to photon energies of 1.4 GeV and are currently under analysis \cite{Oberle_11}. 
\begin{figure}[thb]
\centerline{
\resizebox{1.\textwidth}{!}{%
  \includegraphics{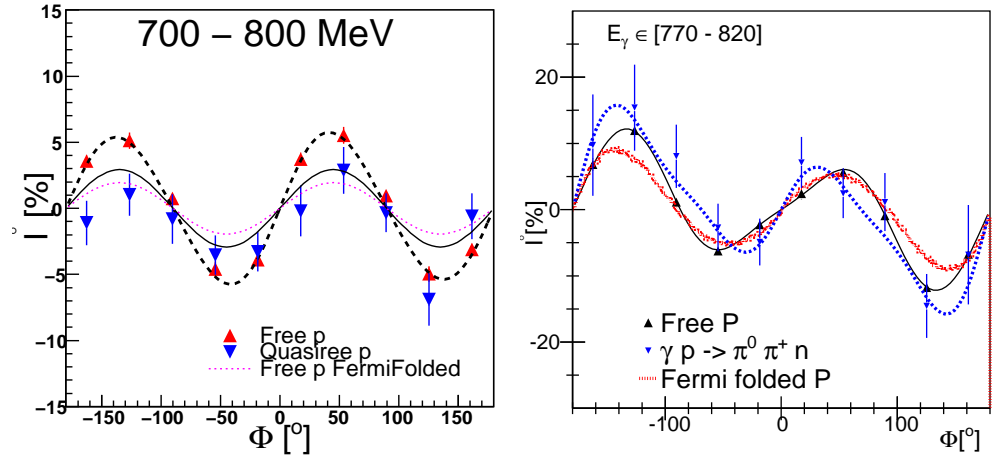}
}}
\caption{Preliminary results for the beam-helicity asymmetry for quasi-free protons
for the $p\pi^0\pi^0$ (left hand side) and the $n\pi^0\pi^+$ (right hand side) final states 
\cite{Oberle_11} compared to free proton results \cite{Krambrich_09}. 
}
\label{fig:markus_asym}       
\end{figure}
These data have been measured with circularly polarized photons, so that the above discussed
beam-helicity asymmetry can be extracted. It was first shown that the results for this observable 
for quasi-free production off the bound deuteron also reproduce the free proton 
data. As demonstrated in Fig. \ref{fig:markus_asym} this is the case for both isospin channels.
The final analysis of the quasi-free neutron data, using kinematical reconstruction to
eliminate the Fermi motion effects, is under way. Initial, very preliminary results seem to indicate,
that for the $\pi^0\pi^0$ channel the proton and neutron asymmetries are very similar, while they 
differ considerable for the mixed charged channel. 

\subsection{Coherent photoproduction off the deuteron - $\pi^0\eta$-pairs}

Coherent photoproduction of mesons can give valuable additional information about the isospin
structure of electromagnetic excitations, however, up to now, it has been little explored due
to the, in general, very small cross sections. One such channel is the $\eta$ photoproduction
off light nuclei discussed in the next section in the context of $\eta$-mesic nuclei. 

\begin{figure}[htb]
\centerline{
\resizebox{0.95\textwidth}{!}{%
  \includegraphics{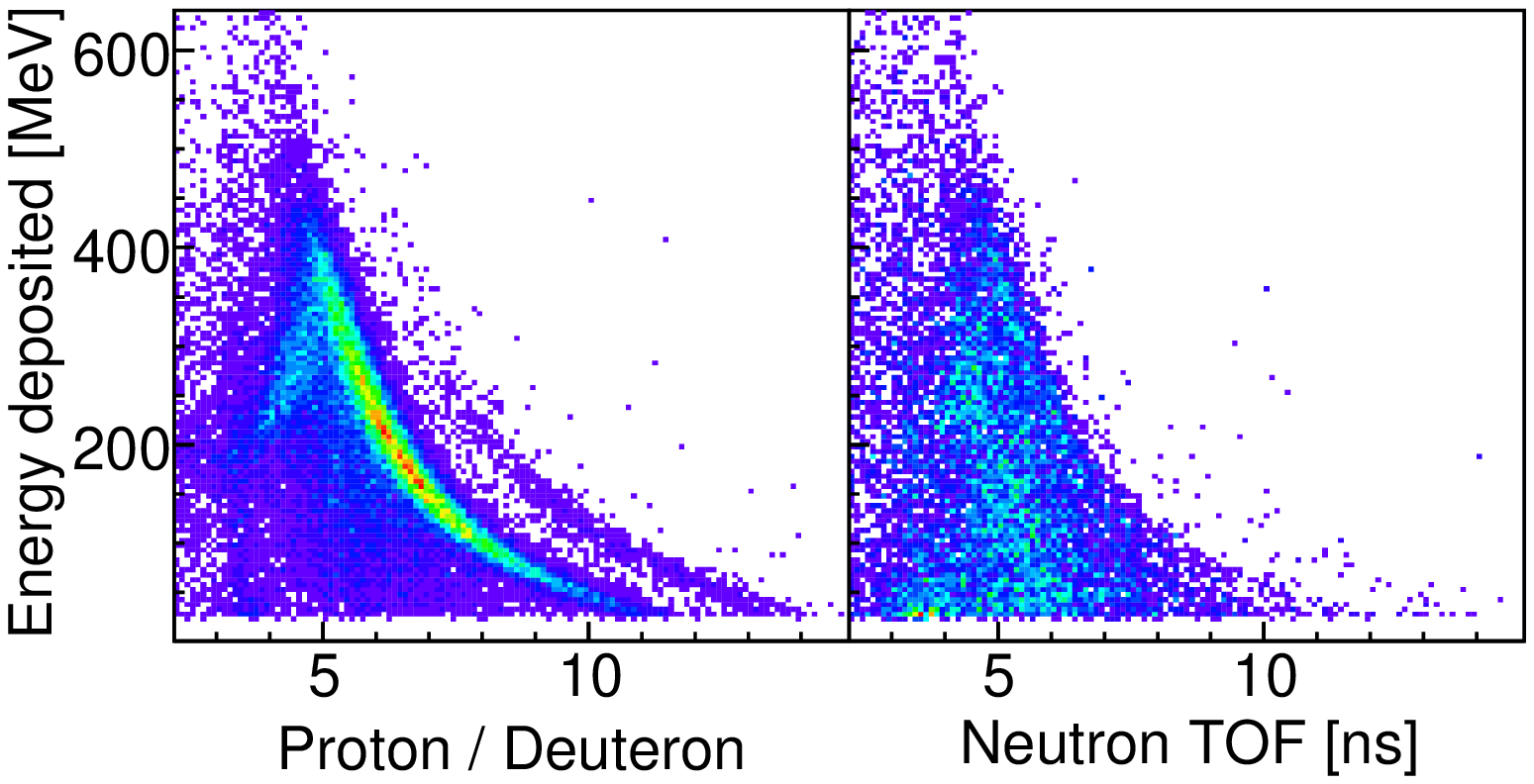}
}}
\centerline{
\resizebox{0.95\textwidth}{!}{%
  \includegraphics{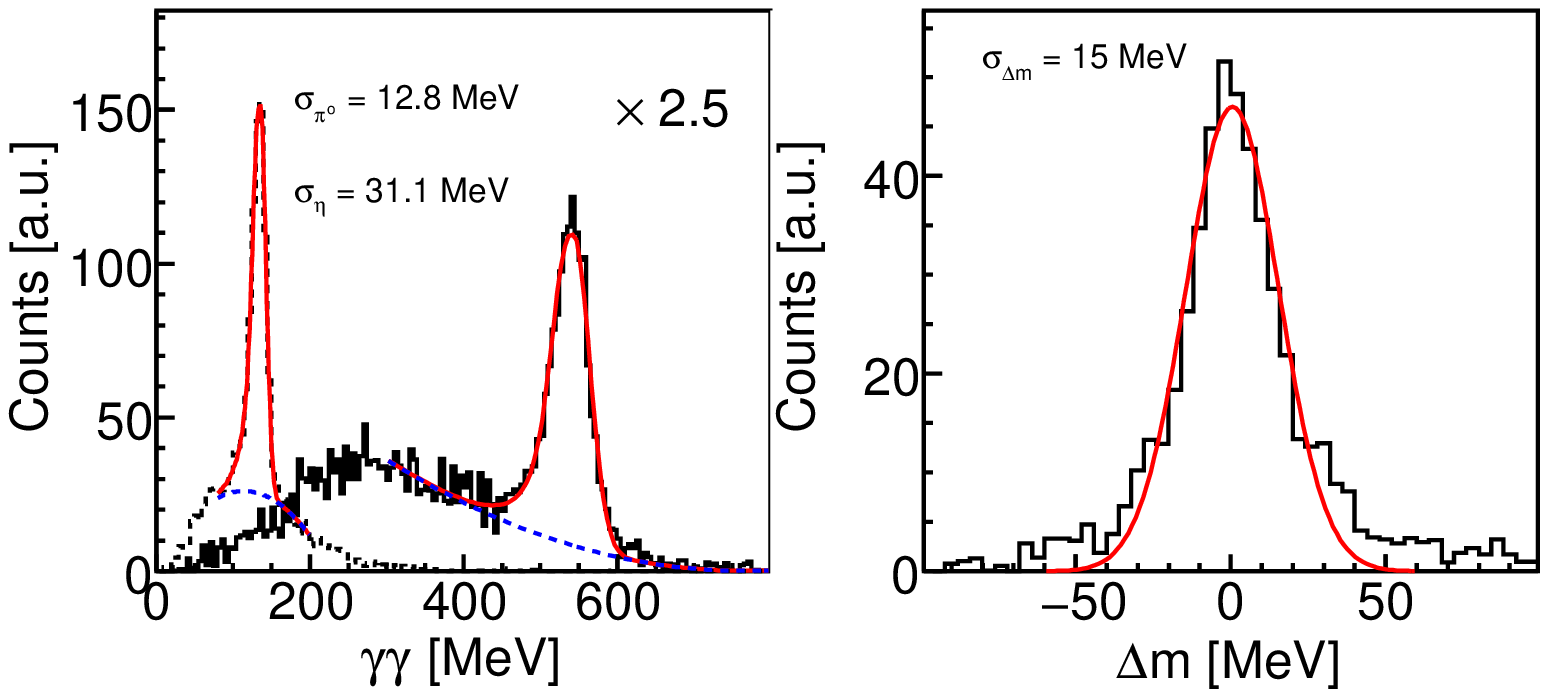}  
}}
\caption{Identification of the reaction $\gamma d\rightarrow d\pi^0\eta$
\cite{Jaegle_09}.
Upper part: time-of-flight versus energy for charged (left hand side) and neutral
(right hand side) recoil particles. For the charged particles a prominent band for
protons and a weaker band for neutrons is visible.
Bottom part: 
Left hand side: invariant mass spectra for $\pi^0$ and $\eta$ mesons
corresponding to the deuteron band in ToF-versus-E.
Right hand side: missing mass spectrum after cut on invariant mass. 
}
\label{fig:pieta_iden}       
\end{figure}

On the deuteron, this channel is suppressed by more than three orders of magnitude with respect 
to the elementary cross section off the free nucleon. One might therefore think, that the study 
of such a production process for multiple meson production reactions is hopeless. The contrary 
is the case, double $\pi^0$, $\eta\pi^0$, and even triple $\pi^0$ off the deuteron (and partly 
also off $^3$He) have been identified recently. Here we will briefly discuss the 
$\gamma d\rightarrow d\eta\pi^0$ reaction.

Production of $\eta\pi^0$ pairs of the free proton was studied intensively during the last 
few years at GRAAL \cite{Ajaka_08}, ELSA \cite{Horn_08a,Horn_08b,Gutz_10}, and MAMI 
\cite{Kashevarov_09,Kashevarov_10}. Different model analyses of the data 
\cite{Horn_08a,Kashevarov_10} found that the threshold region is strongly dominated by the 
decay of one single nucleon resonance, namely the 
D$_{33}(1700)\rightarrow\mbox{P}_{33}(1232)\eta\rightarrow\mbox{N}\eta\pi^0$ decay chain.
The dominance is so strong that, so far, after the pion-decay of the $\Delta$ resonance and
the $\eta$-decay of the S$_{11}$(1535), this is the third case, where a resonance 
can be studied in great detail in one specific decay channel with low background contributions. 

The identification of the coherent reaction \cite{Jaegle_09} in data from the CBELSA experiment
is summarized in Fig. \ref{fig:pieta_iden}. The upper part of the figure shows 
energy-versus-time-of-flight spectra for recoil particles detected in TAPS in coincidence
with $\pi^0\eta$ pairs. For charged recoil particles clear proton and deuteron bands are visible.
The lower part of the figure shows at the left hand side the invariant mass signals for $\pi^0$
and $\eta$ after a cut on the deuteron band and at the right hand side the missing mass spectrum
after cuts on the pion and eta invariant masses. This spectrum is not only almost background free,
the narrow width of the peak also proves that only the coherent reaction has been selected
(the quasi-free reaction has broader peaks due to Fermi smearing).

In view of resonance excitations, the coherent reaction adds additional isospin selectivity.
The final $d\eta\pi^0$ state has isospin $I=1$. This means that the resonance excitation must have
been isospin changing from the $I=0$ deuteron to an $I=1$ $d^{\star}$, which is only possible
for the excitation of a $\Delta$ resonance. Thus processes like 
$\gamma N\rightarrow N^{\star}\rightarrow S_{11}(1535)\pi\rightarrow N\eta\pi^0$, which 
contribute to the elementary reaction are isospin forbidden. 

\begin{figure}[thb]
\centerline{
\resizebox{1.\textwidth}{!}{%
  \includegraphics{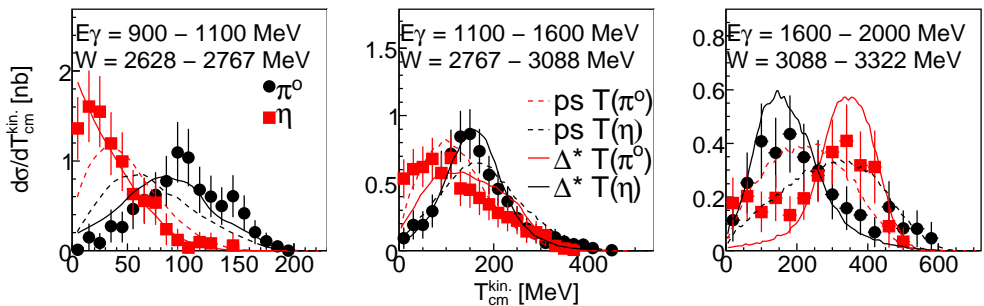}
}}
\caption{Distribution of kinetic energies for $\pi^0$ (red squares) and $\eta$ 
(black dots) mesons from $\gamma d\rightarrow d\pi^0\eta$ \cite{Jaegle_09}
for different bins of incident photon energy.
The dashed curves are simulations assuming phase space, the solid curves assuming
sequential decays of a $\Delta^{\star}$ via $\eta$ emission to the $\Delta$(1232).  
}
\label{fig:pieta_dis}       
\end{figure}

Kinetic energy spectra of the pions and $\eta$ mesons from the coherent reaction are shown in 
Fig. \ref{fig:pieta_dis}. They support the sequential decay scenario of the D$_{33}$(1700)
resonance. The kinetic energies of the pions always peak at energies corresponding to the decay
of the $\Delta$ resonance, while the kinetic energies of the $\eta$ mesons shift upwards with
increasing incident photon energies, scanning the line shape of the D$_{33}$(1700) resonance.

\subsection{Meson photoproduction off nuclei}

\subsubsection{$\eta$-mesic nuclei}
\begin{figure}[thb]
\centerline{
\resizebox{0.64\textwidth}{!}{%
  \includegraphics{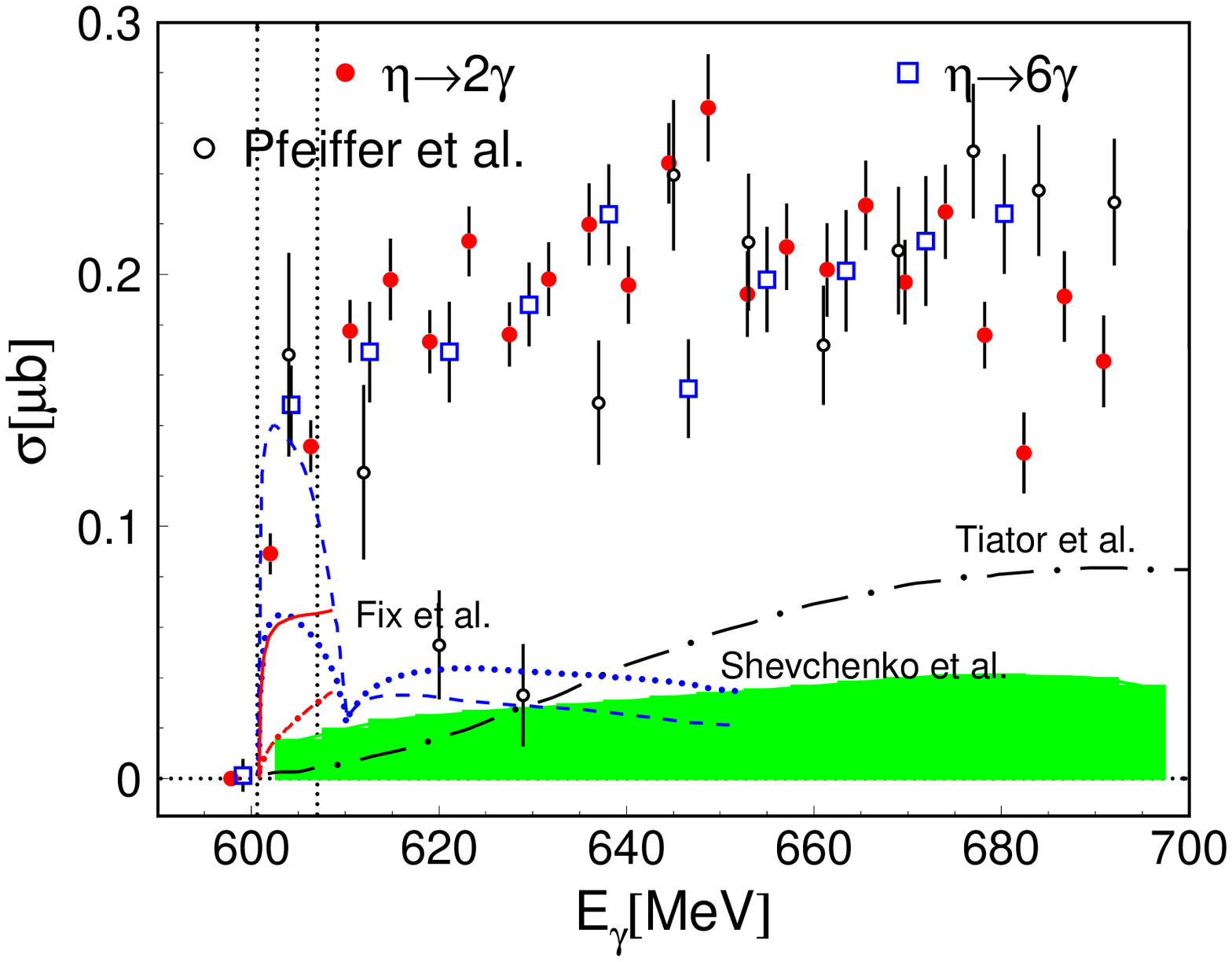}
}  
  \resizebox{0.36\textwidth}{!}{%
  \includegraphics{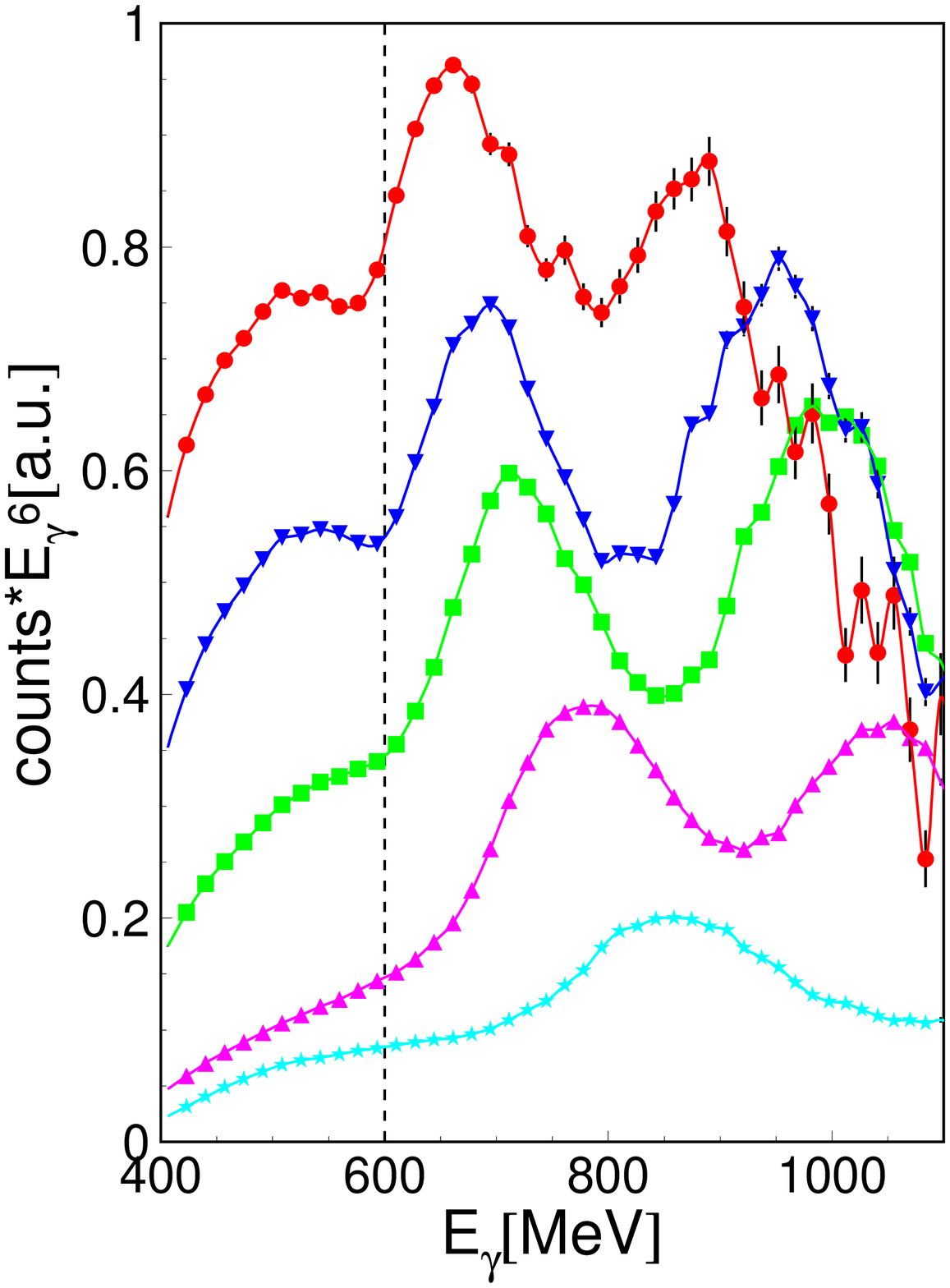}
}}
\caption{Left hand side:
Preliminary total cross section for the $\gamma ^{3}\mbox{He}\rightarrow ^3\mbox{He}\eta$
reaction from $\eta\rightarrow 2\gamma$ and $\eta\rightarrow 6\gamma$ decays \cite{Pheron_10}.
The shaded band at the bottom indicates the systematic uncertainty. The two vertical
lines indicate coherent and breakup threshold. Previous data from Pfeiffer et al.
\cite{Pfeiffer_04}.
Theory curves: (blue) dotted and dashed
from Shevchenko et al. \cite{Shevchenko_03} for two different versions of elastic
$\eta N$ scattering, (red) solid (dash-dotted): Fix and Arenh\"ovel \cite{Fix_03}
full model (plane wave), (black) long dash-dotted: Tiator et al. \cite{Tiator_94}.   
Right hand side:
excitation functions of $\pi^0-p$ back-to-back pairs 
for different ranges of the opening angle $\Psi_{\pi p}$ after removal of the overall 
energy dependence $\propto E_{\gamma}^{-6}$. From top to bottom opening angle ranges of:
165$^{\circ}$ - 180$^{\circ}$, 150$^{\circ}$ - 165$^{\circ}$, 140$^{\circ}$ - 150$^{\circ}$,
130$^{\circ}$ - 140$^{\circ}$, and 120$^{\circ}$ - 130$^{\circ}$.
Vertical lines: $\eta$-production threshold.
}
\label{fig:francis}       
\end{figure}

We have discussed in the introduction in detail the search for $\eta$-nucleus (quasi)-bound
states, so-called $\eta$-mesic nuclei. Tentative evidence, although at low statistical quality,
for such states had been reported from a measurement of meson photoproduction off $^3$He by 
Pfeiffer and co-workers \cite{Pfeiffer_04} with the TAPS detector at MAMI. This experiment has 
been repeated with much higher statistical quality and a larger incident photon energy coverage 
with the Crystal Ball/TAPS experiment at MAMI \cite{Pheron_10}. The most 
important experimental findings are summarized in Fig. \ref{fig:francis}.  

The left hand side of the figure shows the total cross section for the coherent 
$\gamma ^3\mbox{He}\rightarrow \eta^3\mbox{He}$ reaction extracted from the two-photon
and three-pion decays of the $\eta$-meson. Both data sets are in good agreement and
confirm the extremely sharp rise of the excitation function at the kinematic threshold 
for coherent $\eta$ production. The data are also in reasonable agreement with the previous
experiment by Pfeiffer et al. \cite{Pfeiffer_04}, although they do not show the dip-like
structure in the previous data at incident photon energies around 620 MeV.
Here, one should note that, in this range, the systematic uncertainty in the previous data
was comparably large because the coherent signal had to be extracted by fitting a small
coherent contribution in the missing energy spectra, which were dominated by the quasi-free 
reaction. At lower incident photon energies, the quasi-free contribution is relatively much 
smaller and at higher incident energies it is better separated from the coherent component.
The present experiment profited not only from the better statistics but also from the almost
$4\pi$ coverage of the detector system, which allowed the removal of a significant fraction of the
quasi-free background by the detection of associated recoil nucleons. Altogether, the
extraordinary threshold behavior of this reaction, which is very similar to the hadron induced
reactions \cite{Smyrski_07,Mersmann_07,Rausmann_09}, was confirmed and supports the assumption 
of a resonant behavior at threshold. In the figure, the data are also compared to the results 
of different models. The plane-wave models (Tiator and co-workers
\cite{Tiator_94}, Fix and Arenh\"ovel \cite{Fix_03}) have a completely different energy
dependence in the threshold region. The model of Fix and Arenh\"ovel, including final state
interactions, reproduces a very steep slope at threshold, but does not agree with the
magnitude of the cross section in that region. Also the models by Shevchenko et al.
\cite{Shevchenko_03} for two different versions of elastic $\eta N$ re-scattering show a steep
rise at threshold; but can reproduce neither energy dependence nor magnitude of the data
over a larger energy range. The differential cross section data (not shown), 
confirm also the previous finding. The angular dependence at higher incident photon energies is
dominated by forward peaking related to the nuclear form factor. But in the threshold region they 
are much more isotropic and very close to threshold they show a behavior which is opposite to 
what is expected from the form factor. This is evidence for at least strong FSI effects in the
threshold region.  

The other tentative signature for the formation of a quasi-bound state reported by Pfeiffer
et al. \cite{Pfeiffer_04} was a small, narrow peak-like structure at the coherent 
$\eta$ production threshold in the excitation function of proton-$\pi^0$ pairs emitted 
back-to-back in the photon-$^3$He cm system. The interpretation of this structure is related to 
the following picture. When a quasi-bound $\eta$-nucleus state forms, the $\eta$ may be re-captured
by a nucleon which is then excited to the S$_{11}$(1535) resonance with a large probability.
This resonance has very similar decay branching ratios into $N\pi$ and $N\eta$ (both in the range
between 45\% and 55\% \cite{PDG}). That means the S$_{11}$ can either re-emit the $\eta$ or
a pion. Due to the small mass of the pion, its emission is also possible below the 
coherent $\eta$ production threshold, where the decay into $N\eta$ necessarily leads again to the
formation of a bound system. Therefore, while in coherent $\eta$ production we observe only the 
abrupt rise of the cross section at threshold, in the pion channel a peak may occur, which is
indicative for the width of the resonant structure. There is, however, even after appropriate
kinematical cuts, a large background from quasi-free single $\pi^0$ production. In the work of 
Pfeiffer et al. a large part of this background was subtracted with a side-band analysis: 
the excitation function for $\pi^0 - p$ pairs with opening angles between 150$^{\circ}$ -170$^{\circ}$
was subtracted from the one between  170$^{\circ}$ -180$^{\circ}$ after proper normalization.
The idea was, that the background from quasi-free $\pi^0$ production changes only slowly with the
opening angles. An analysis of the same type applied to the new high statistics data reproduces the
peak observed in \cite{Pfeiffer_04} with much higher statistical significance. However, the new data
allow also a more detailed study of the quasi-free single $\pi^0$ background. This is shown in Fig.
\ref{fig:francis}, right hand side. The excitation functions for different pion - proton opening angle
$\Psi$ have been constructed after cuts on the reaction kinematics, which enhance the possible 
signal from $\eta$-mesic nuclei with respect to the quasi-free background. The overall energy 
dependence in the range of interest on the order of $E_{\gamma}^{-6}$ has been removed. The figure 
shows clearly remnant structures from the bumps of the second and third resonance region in 
quasi-free pion production. These structures move unfortunately with decreasing opening angle to 
higher incident photon energies. This is a trivial kinematic effect because changing opening angles 
require different combinations of incident photon energy and nuclear Fermi momentum to produce the 
same value of $W$. Since the nucleon resonance bumps are at fixed positions in $W$, they must 
shift with $E_{\gamma}$ as function of $\Psi$. The peak reported by Pfeiffer et al. \cite{Pfeiffer_04}
occurs exactly at the coherent $\eta$ production threshold. But the back-to-back pairs 
(165$^{\circ}$ - 180$^{\circ}$) do not show a peak at this position, rather the rising slope
of the low energy tail of the second resonance region. At the same position the 
150$^{\circ}$ - 165$^{\circ}$ data are not yet in this rising slope but by chance in a little dip
just before it. Subtraction of the scaled background thus produces a narrow peak-like structure at
threshold, which is however artificially caused by the dip in the background. A true signal from
an $\eta$-mesic state would be somewhere hidden in these complicated structures, and there is
almost no hope to extract such a signal.

A final remark should be made to the search for $\eta$-mesic states in other light nuclei. 
The use of photon induced $\eta$-production restricts possible target nuclei to spin $J\neq$0
and isospin $I\neq$0 nuclei like $^3$He because the dominant amplitude is an isovector spin-flip.
The next best candidate would be $^7$Li. Such data from the Crystal Ball/TAPS experiment at MAMI 
are in preparation for publication \cite{Yasser_10}. Also in this case a significant enhancement
of the cross section in the threshold region was observed, although not as dramatic as for the 
$^3$He nucleus. 

It would be even more interesting, to search for $\eta$-mesic states in the much more strongly 
bound $^4$He. However, as discussed above, in this case coherent $\eta$-production is not a possible
entrance channel. Nevertheless, in this case the coherent production of $\pi^0\eta$ pairs discussed
in the previous section could possibly be explored. Since this reaction is dominated near threshold
by the process 
$\gamma \mbox{N}\rightarrow \mbox{D}_{33}(1700)\rightarrow\mbox{P}_{33}(1232)\eta\rightarrow\mbox{N}\eta\pi^0$ 
the electromagnetic excitation must go via the $V3$ amplitude, which is identical for protons and
neutrons so that no cancellation can occur. Since D$_{33}$ and P$_{33}$ have opposite parity
but the same total angular momentum, the D$_{33}\rightarrow\eta \mbox{P}_{33}$ transition
is $s$-wave ($J=0$ $\eta$ meson with negative intrinsic parity). The subsequent 
P$_{33}\rightarrow \mbox{N}$ transition is the usual magnetic $p$-wave decay of the $\Delta$
resonance known from single pion production. Consequently, coherent production of $\eta\pi^0$
pairs is not suppressed for the scalar isoscalar $^4$He nucleus.  
An additional advantage of this reaction is, that one can select events, where the pion carries
away most of the momentum, so that the $\eta$-meson has only a small momentum relative to the
nucleus, as needed for the formation of a quasi-bound state. 
   
\subsubsection{Photoproduction of pion pairs off nuclei}

New, precise data have been obtained with Crystal Ball/TAPS at MAMI for the photoproduction of
neutral and mixed charged pion pairs of a series of nuclei ($^2$H, $^7$Li, $^{12}$C, $^{40}$Ca,
$^{93}$Nb, and $^{nat}$Pb. The main purpose of this experiment was to collect further data
which give some clues to distinguish between trivial FSI effects and genuine in-medium modifications
of the $\sigma$ meson in the pion-pion invariant mass distributions. Compared to the results by
Messchendorp et al. \cite{Messchendorp_02} and Bloch and coworkers \cite{Bloch_07} the present
data \cite{Yasser_10} improve the situation in two aspects. The much better statistical quality
does not only provide more significant signals in the invariant mass distributions but also allows 
the study of these effects more systematically to low incident photon energies. The addition of the
light nuclei ($^2$H, $^7$Li) gives better reference points for low average nuclear density, where
in-medium effects of the $\sigma$ should not play an important role.

\begin{figure}[thb]
\centerline{
\resizebox{0.50\textwidth}{!}{%
  \includegraphics{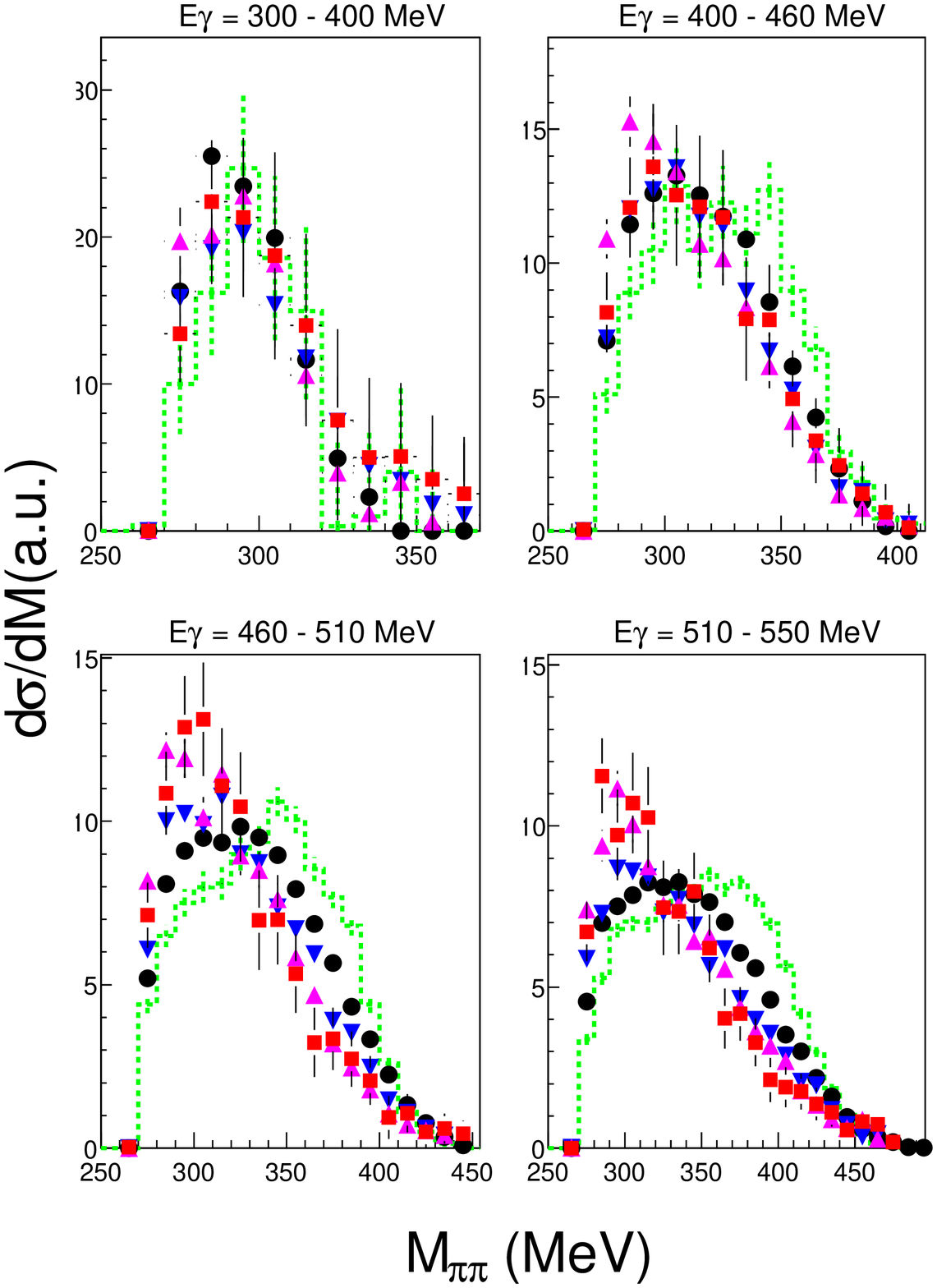}
}
\resizebox{0.50\textwidth}{!}{%
  \includegraphics{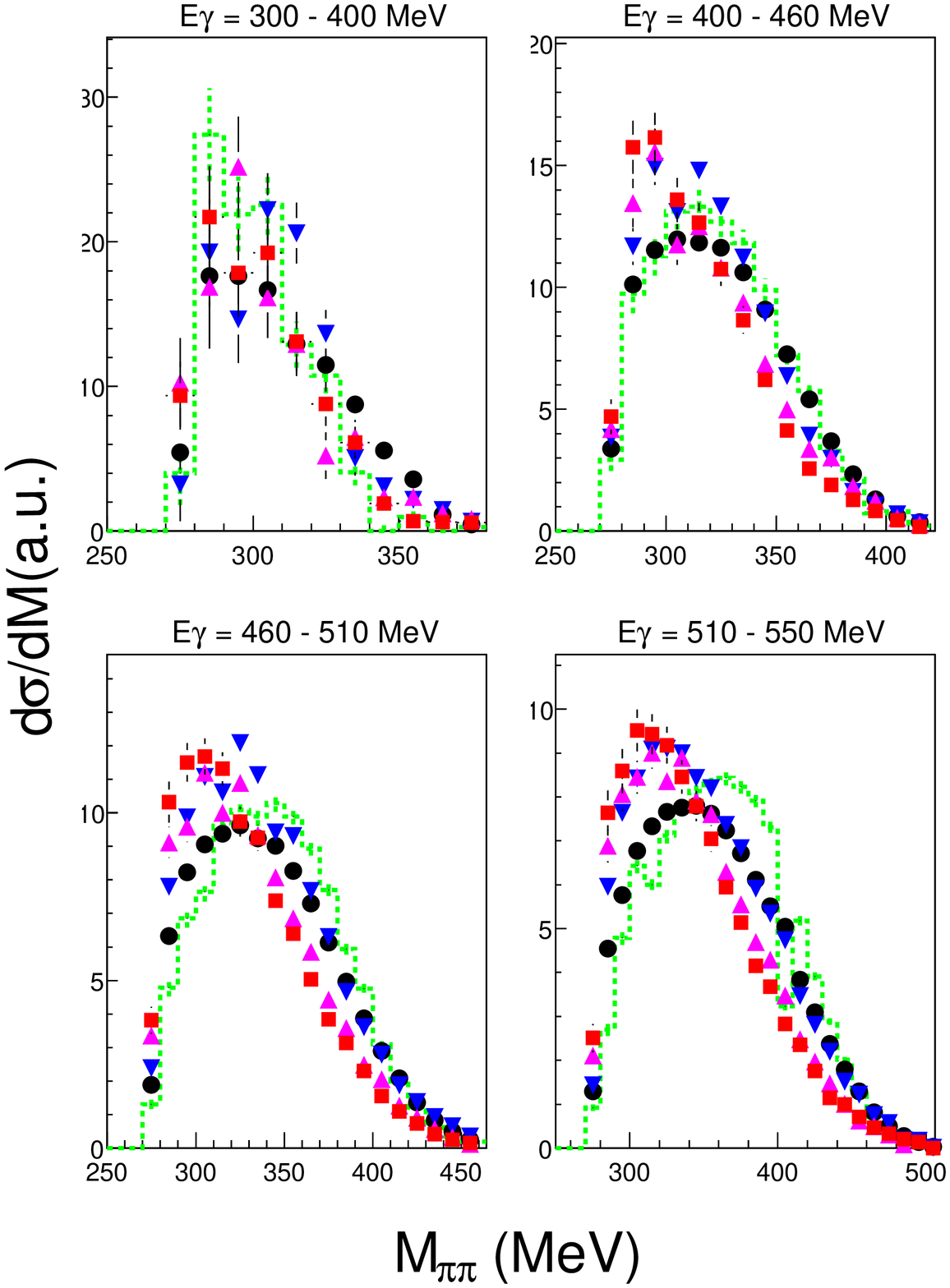}  
}}
\caption{Preliminary invariant mass distributions of $\pi^0\pi^0$ pairs (left hand side)
and $\pi^0\pi^+$ pairs (right hand side) off nuclei for different ranges of incident photon 
energy (all distributions normalized to total cross section \cite{Yasser_10}. 
(Green) histograms: deuterium, (black) dots: $^7$Li, (blue) downward triangles: $^{12}$C,
(magenta) upward triangles: $^{40}$Ca, (red) squares: $^{nat}$Pb.
}
\label{fig:yasser}       
\end{figure}

Preliminary results for the invariant mass distributions are compared in Fig. \ref{fig:yasser}
for the $\pi^0\pi^0$ and the $\pi^0\pi^{\pm}$ final states for four different ranges of incident
photon energies from the immediate threshold region at 300 MeV up to 500 MeV. They can be summarized
in the following way. For both isospin channels the invariant mass distributions of different target
nuclei are very similar; there is not even much difference between the deuteron (basically
proton/neutron average of elementary cross sections) and lead nuclei. At these very low incident
photon energies, the produced pions must have small momenta, below the threshold where they could 
excite the $\Delta$-resonance. This means, that as discussed in the introduction (cf Fig.
\ref{fig:alpha}), they are almost unaffected by FSI effects. Nevertheless, possible in-medium
effects of the $\sigma$ meson should also occur in this region. Towards higher photon energies, the
picture changes completely. In particular the $\pi^0\pi^0$ invariant mass distributions for heavy
nuclei develop a very significant low mass enhancement with respect to the light nuclei. 
This effect is also visible for the mixed charge channel, although less pronounced. This picture 
is consistent with the FSI effects, in particular from quasi-elastic charge-exchange scattering 
of the pions, discussed in the introduction. They increase with incident photon energy,
are more important for $\pi^0\pi^0$ than for $\pi^0\pi^{\pm}$ (because of the relative
size of the cross sections of feeding and fed channels), and give rise to enhancements at
low invariant masses. Although a detailed comparison to models must still be done, the
preliminary results seem to indicate that the observed effects are mainly related to pion-FSI.

\section{Conclusions and Outlook}

We have discussed in this paper the two main aspects of a world-wide active program for
the study of photon induced meson production reactions off nuclear targets: the study
of the electromagnetic excitations of the neutron and the investigation of meson-nucleus
interactions and meson in-medium properties. Many new results have become available during 
the last few years and much more is to come in the near future.

The measurement of meson production reactions off quasi-free nucleons has been put on a much 
more systematic basis, in particular by the detailed study of nuclear effects on the 
quasi-free proton cross sections which can be compared to their free counter parts. Effects 
from nuclear Fermi motion can be reliably removed by kinematic reconstruction techniques.
The picture for nuclear effects beyond Fermi motion is not unique. For some reactions like 
$\eta$- and $\eta '$-production they do not seem to play any role. For other reactions,
as in example single $\pi^0$ photoproduction, they seem to be substantial, so that in
this case it is probably better to compare the less affected neutron/proton ratios to 
model predictions. Already measured data from the CBELSA (Bonn) and Crystal Ball (Mainz)
experiments for many further final states ($n\pi^0$, $n\omega$, $n\pi^0\pi^0$, 
$p\pi^0\pi^-$, $n\pi^0\eta$, $p\pi^-\eta$) from photoproduction off the neutron are 
currently under analysis and will soon become available. Furthermore, currently, at both
experiments, first beam times with a polarized deuterated butanol target for the 
measurements of (double) polarization experiments are under way and will continue in
2011/2012. These data will greatly extend the data base for photoproduction reactions
on the neutron and should help to solve urgent questions like the nature of the narrow
structure in the excitation function of $\eta$-photoproduction off the neutron.

New data for the measurement of coherent photoproduction off light nuclei serving as 
isospin filters will also become available soon. These reaction channels have recently 
been investigated for the first time for the production of meson pairs and triples like
$\pi^0\pi^0$, $\pi^0\eta$, and $\pi^0\pi^0\pi^0$. 

The measurement of meson-nucleus interactions is also continuing. New results for
the interaction of $\eta '$ mesons in nuclei will be published soon. The question of
the existence of $\eta$-mesic nuclei is not yet finally solved. The new precise results
for coherent $\eta$ photoproduction off $^3$He nuclei have confirmed the extraordinary
threshold behavior of this reaction, but the existence of a related structure in the
excitation function of $\pi^0$-proton back-to-back pairs could not be confirmed.
New results for $^7$Li will become available soon, and experiments are planned to
search for $\eta$-mesic $^4$He using the $\pi^0\eta$ production reaction. Furthermore, 
data are under analysis to study for the first time the threshold behavior of 
$\eta '$ production off the deuteron and off $^3$He. Finally, detailed investigations 
of the production of pion pairs off light and heavy nuclei, although not yet complete, 
seem to indicate that the previously observed effects in the invariant mass spectra of 
the meson pairs are due to final state interaction effects, rather than to in-medium 
modifications of the $\sigma$ meson.    

\section{Acknowledgments}
The preliminary results discussed in this paper have been obtained by the Crystal Ball/TAPS
and CBELSA/TAPS collaborations. They are part of the PhD theses of I.~Jaegle, F.~Pheron,
and Y.~Magrhbi, and of the Master theses of M.~Dieterle, M.~Oberle, and L.~Witthauer.
This work was supported by Schweizerischer Nationalfonds and Deutsche 
For\-schungs\-ge\-mein\-schaft (SFB 443, SFB/TR-16.)

\end{document}